\documentclass[1p,12pt]{elsarticle}
\pdfoutput=1

\usepackage{graphicx}
\usepackage{amssymb}

\usepackage{booktabs}
\usepackage{epstopdf}
\usepackage{pgf}
\usepackage[caption=false]{subfig}
\usepackage[section]{placeins} 
\usepackage[xindy]{glossaries}

\usepackage[titletoc]{appendix}

\usepackage[section]{algorithm}
\usepackage[noend]{algpseudocode}

\makeatletter
\def\BState{\State\hskip-\ALG@thistlm}
\makeatother

\journal{Journal of Computational Physics}

\graphicspath{ {./fig-ab/} {./fig-dc/} {./fig-eps2pdf/} {./fig-tg/} {./fig-golf/} {./fig-perf/} }
\newcommand{\pluseq}{\mathrel{+}=}

\newcommand{\forloop}[3]{\ensuremath{{#1} \in \lbrace {#2},\ldots,{#3} \rbrace}}

\newcommand\CONDITION[2]%
  {\begin{tabular}[t]{@{}l@{}l@{}}
     #1&#2
   \end{tabular}%
  }
\algdef{SE}[WHILE]{While}{EndWhile}[1]%
  {\algorithmicwhile\ \CONDITION{#1}{\ \algorithmicdo}}%
  {\algorithmicend\ \algorithmicwhile}
\algdef{SE}[FOR]{For}{EndFor}[1]%
  {\algorithmicfor\ \CONDITION{#1}{\ \algorithmicdo}}%
  {\algorithmicend\ \algorithmicfor}
\algdef{S}[FOR]{ForAll}[1]%
  {\algorithmicforall\ \CONDITION{#1}{\ \algorithmicdo}}
\algdef{SE}[REPEAT]{Repeat}{Until}{\algorithmicrepeat}[1]%
  {\algorithmicuntil\ \CONDITION{#1}{}}
\algdef{SE}[IF]{If}{EndIf}[1]%
  {\algorithmicif\ \CONDITION{#1}{\ \algorithmicthen}}%
  {\algorithmicend\ \algorithmicif}%
\algdef{C}[IF]{IF}{ElsIf}[1]%
  {\algorithmicelse\ \algorithmicif\ \CONDITION{#1}{\ \algorithmicthen}}

\algnewcommand{\Commentx}[1]{\Statex \(\triangleright\) #1}

\algtext*{EndFunction} 

\begin{document}

\begin{frontmatter}

\title{A Parallel Direct Cut Algorithm for High-Order Overset Methods with Application to a Spinning Golf Ball}

\author{J. Crabill}
\author{F. D. Witherden}
\author{A. Jameson}

\address{Dept. of Aeronautics and Astronautics, Stanford, California 94305}

\begin{abstract}
Overset methods are commonly employed to enable the effective simulation of problems involving complex geometries and moving objects such as rotorcraft. This paper presents a novel overset domain connectivity algorithm based upon the direct cut approach suitable for use with GPU-accelerated solvers on high-order curved grids.  In contrast to previous methods it is capable of exploiting the highly data-parallel nature of modern accelerators.  Further, the approach is also substantially more efficient at handling the curved grids which arise within the context of high-order methods.  An implementation of this new algorithm is presented and combined with a high-order fluid dynamics code.  The algorithm is validated against several benchmark problems, including flow over a spinning golf ball at a Reynolds number of 150,000.
\end{abstract}

\begin{keyword}
Overset \sep Chimera \sep CFD \sep Flux Reconstruction \sep Discontinuous Galerkin \sep GPU \sep Golf Ball
\end{keyword}

\end{frontmatter}



\section{Introduction}
\label{S:intro}

Overset grids are a commonly used technique within the field of computational fluid dynamics (CFD) to enable the simulation of cases which involve the relative motion of several distinct bodies \cite{handbookgrid}.  The idea is to generate a distinct mesh around each body and then overlay these meshes on top of a background grid.  The relevant system of governing equations are then solved on each grid, with a hole cutting algorithm being used to transfer data between grids as required.  Also known as the Chimera grid approach, this technique has enabled the accurate simulation of complex problems of engineering interest, including entire rotorcraft configurations with relative motion between one or more rotors in addition to a fixed fuselage.

One of the many advantages of this approach, compared with say deforming grids, is that it is possible to employ multiple solvers: an unstructured near-body solver to allow for easier mesh generation around each body, and a high-order Cartesian off-body solver with adaptive mesh refinement (AMR) for speed and simplicity.  In most production codes used for complex geometries, such as the CREATE-AV HELIOS software \cite{Helios}, the near-body solvers are typically 2nd or 3rd order accurate in space.  Higher-order near-body solvers are available, but only for (multi-block) structured grids currently, such as with the OVERFLOW code using 5th order accurate finite differences \cite{pulliam2011,buning2016}.  For the best possible performance, accuracy, and applicability to a wide range of complex geometries, however, a higher-order unstructured near-body solver is preferable.  As shown by Wissink \cite{HeliosStrand} and by Nastase et al. \cite{OversetDG2011}, without a high-order near body solver, vortical structures and other flow features incur high amounts of diffusion before (or while) passing into the high-order off-body solver, causing a large increase in error.

High order methods have a number of attractive attributes.  Not only are they less dissipative, enabling the simulation of vortex-dominated flows with fewer degrees of freedom than a lower-order method, but they have also been shown to give remarkably good results when used to perform implicit large eddy simulations (ILES) and direct numerical simulations (DNS) \cite{pypeta,pyfrstar}.  Popular examples of high-order schemes for unstructured grids include the discontinuous Galerkin finite element method, first introduced by Reed and Hill \cite{reedhill1973}, and spectral difference (SD) methods originally proposed under the moniker `staggered-grid Chebyshev multidomain methods' by Kopriva and Kolias in 1996 \cite{kopriva96} and later popularized by Sun et al. \cite{sun07}. In 2007 Huynh proposed the flux reconstruction (FR) approach \cite{Huynh}; a unifying framework for high-order schemes for unstructured grids that encompasses both the nodal DG schemes of Hesthaven and Warburton \cite{HesthavenWarburton} and, at least for a linear flux function, any SD scheme. More recently, Romero et al. \cite{romero2017direct} proposed a simplified formulation of the FR approach, direct flux reconstruction (DFR), which exactly recovers the nodal DG version of FR.

Within the context of overset grids, high-order methods of the discontinuous variety are particularly attractive.  By their nature, they involve a large amount of structured computation within each element, while relatively little data is needed from surrounding elements.  In addition to providing high-order accuracy on unstructured grids, this compactness provides a significant advantage over finite difference and finite volume methods for overset grids.  As described in detail by Galbraith \cite{Galbraith13}, the artificial boundary (AB) approach is an entirely natural extension of these methods to handle element interfaces which may overlap another grid.  For codes which utilize the message passing interface (MPI) paradigm for parallelization, only relatively minor modifications must be made to handle overset grids, since the AB faces appear identical to inter-processor boundaries. 

A number of high-order solvers have been integrated into overset-grid frameworks in recent years, using both the traditional volume-interpolation approach \cite{OversetDG2011,OversetDG2015,Brazell17} and the surface-based AB approach \cite{Galbraith13,Galbraith14,crabill16,Wang17}.  Research has focused on the DG and FR methods, with studies focusing on their use as either a near-body solver, an off-body solver, or both.  Brazell and Kirby et al. \cite{Brazell17,kirby17} have developed an advanced overset-capable DG solver, with the focus on an off-body AMR solver capable of both $h$- and $p$-adaptivity to track flow features of interest and match the resolution of the near-body grid.  Crabill et al. \cite{crabill16} previously showed an unstructured FR solver suitable for use on both near- and off-body moving grids, and Duan and Wang \cite{Wang17} more recently developed a similar capability for both mixed-element unstructured and semi-structured strand grids.

These studies and others indicate that high-order methods can maintain their order of accuracy on overset grids with a relatively small increase in absolute error.  However, to date no group has presented an overset method capable of solving large-scale, dynamic Navier--Stokes problems on moving overset grids which uses high-order unstructured solvers throughout the domain.  For static cases, the only requirement is that the overset interpolation process should add only a small amount of overhead to the underlying solver.  In the case of moving grids, however, the entire overset connectivity must constantly be recomputed, so both the connectivity and interpolation together should not add a significant amount of overhead.  This requires novel algorithms to perform the high-order overset connectivity quickly.

Furthermore, modern hardware is shifting towards the use of highly parallel accelerators such as Intel's line of Xeon Phi co-processors and NVIDIA's line of Graphical Processing Units (GPUs).  These accelerators provide the potential for an order of magnitude improvement in terms of performance-per-Watt over conventional CPUs, but require algorithms that map well to their hardware architecture.  In particular, algorithms which conserve memory bandwidth are essential since, on these platforms, memory bandwidth is typically the limiting factor.  Fortunately, high-order finite element-type methods (such as DG and FR) map extremely well onto these architectures, but limited work has been done to map overset methods onto them \cite{chandar12}.

In this work, we outline the development of a high-order GPU-accelerated solver capable of running on moving overset grids with curved elements.  The underlying numerical solver, ZEFR, is able to run on tensor-product elements using the DFR scheme.  ZEFR was developed in the Aerospace Computing Laboratory as a simple but high-performance CFD code for the purpose of developing new algorithms and applying them to useful test cases.  The code is written in a combination of C++ and CUDA and can effectively target NVIDIA GPUs.  Distributed memory systems are handled using the MPI standard.  All overset-related connectivity and interpolation is handled within the external Topology Independent Overset Grid Assembler (TIOGA) library \cite{tioga}, to which modifications have been made for the present work.  TIOGA handles the construction of binary search trees, the inter-process communication map, and approximate geometry representations, as well as performing donor searches and interpolation of solution data between grids.
Our key advancement lies in a novel overset domain connectivity algorithm inspired by the work of Galbraith \cite{Galbraith13,Galbraith14}, Sitaraman \cite{pundit}, and others, and has been developed specifically to leverage the considerable speedup possible with modern  hardware.  The proposed method is simple, robust, and fast for arbitrary configurations of curved-element grids.

The remainder of this paper is as follows.  In Section \ref{S:oga}, the application of the AB overset method to DFR will be discussed, along with a review of existing overset grid assembly methods.  Section \ref{S:motion} will discuss the additional work necessary to handle moving grids within the AB framework.  In Section \ref{S:pconn}, existing direct cut methods will be reviewed, and our novel Parallel Direct Cut hole-cutting method will be presented.  In Section \ref{S:results} we present the results of several test cases run with our new method, including the challenging problem of a spinning golf ball at a realistic Reynolds number of $150\,000$.  The performance and robustness of the method are also considered. Finally, in Section \ref{S:conclusion} conclusions are drawn.

\section{Overset Domain Connectivity}
\label{S:oga}

\subsection{Artificial Boundaries for Discontinuous Finite Elements}

The starting point for the present study is the AB method, initially developed by Galbraith for the DG method \cite{Galbraith13}, although it can be applied to any discontinuous finite element method.  In the AB method, all data transfer between overset grids occurs at interfaces (surfaces) tagged as overset boundaries.  Thus, each overset grid in an AB system is essentially a self-contained grid with modified boundary conditions. This is in contrast to traditional overset methods, which rely on volume interpolation to the interior degrees of freedom of each domain.  For high-order discontinuous finite element methods this approach is quite natural, as all elements operate essentially independently with coupling only at inter-element boundaries.  Not only does this simplify its implementation---since the artificial boundaries can be handled identically to standard inter-processor (MPI) boundaries---it also is much less expensive than volume interpolation, because volume interpolation requires interpolation to all of the interior solution points of an element, whereas the AB approach requires interpolation only to the points on the surface (face) of an element.  For a nominally $p+1$th order scheme in 2D or 3D, this reduces the amount of overset connectivity processing by a factor of $p+1$.  The amount of solution data transferred between grids is also reduced by a factor proportional to $p+1$. Viscous cases will in general require gradients to be interpolated in addition to the solution. Recent work by Duan and Wang \cite{Wang17} has also shown that the volume-interpolation approach incurs more error than the AB approach.  Further, the AB method requires less overlap between grids, with the consequence that more elements can be removed from each grid in the overset system, reducing its computational cost.

\begin{figure}
\centering
\subfloat[]{
  \includegraphics[width=.45\textwidth]{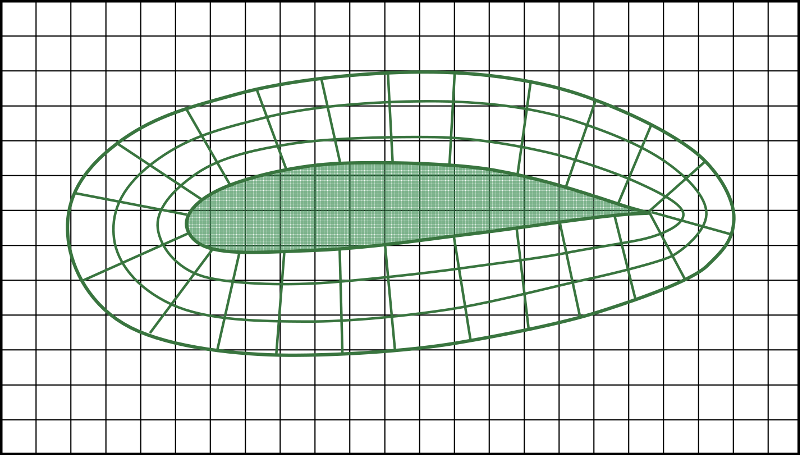}
} 
\hspace{5mm}
\subfloat[] {
  \includegraphics[width=.45\textwidth]{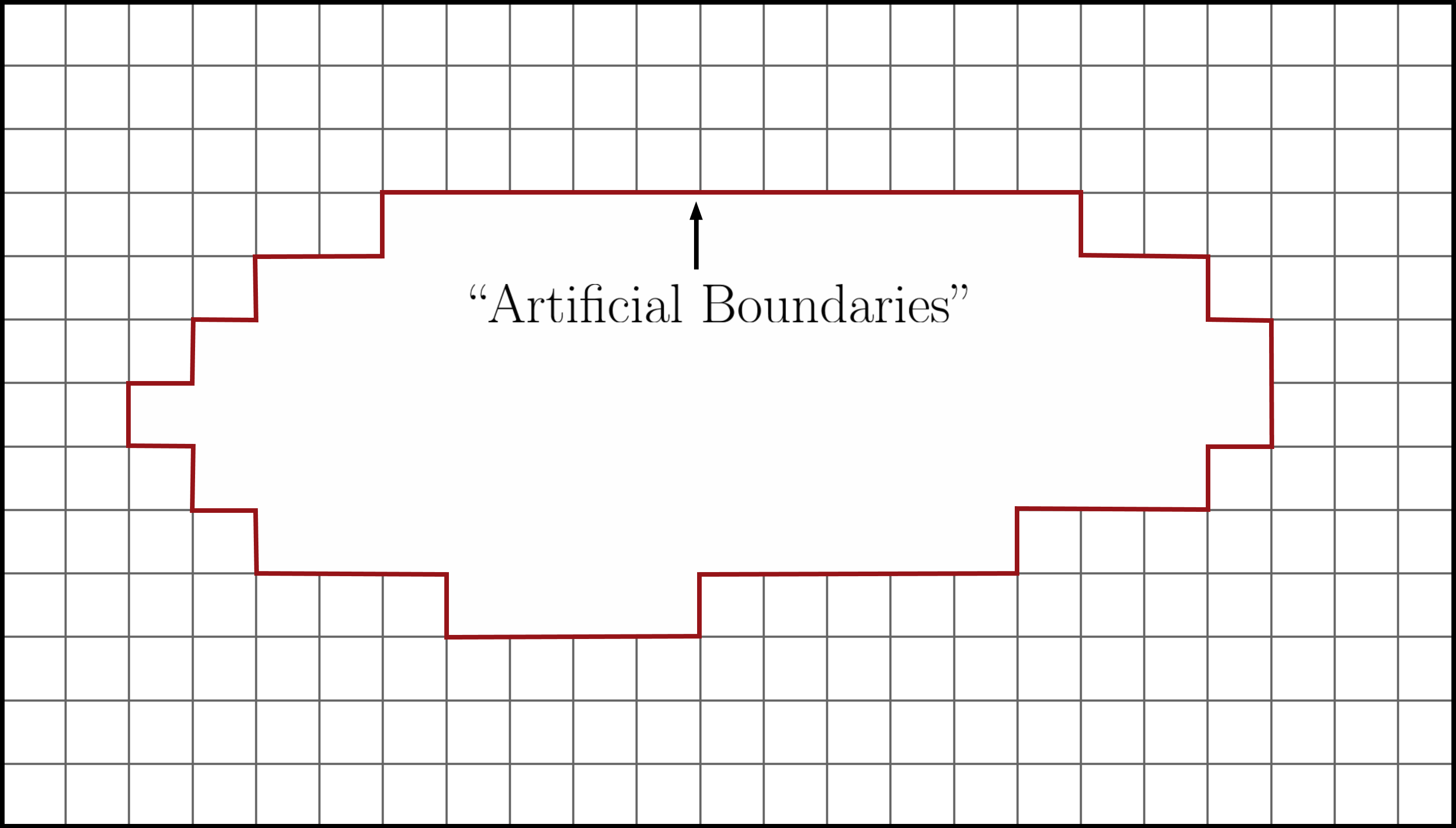}
}
\caption{Example of the basic hole cutting and artificial boundary creation process on a simple airfoil grid. (a) Simple body grid/background grid overset system. (b) The artificial boundaries which would be created from the grid system.}
\label{fig:blanking}
\end{figure}

\begin{figure}[t]
\centering
\subfloat[] {
  \includegraphics[width=.7\textwidth]{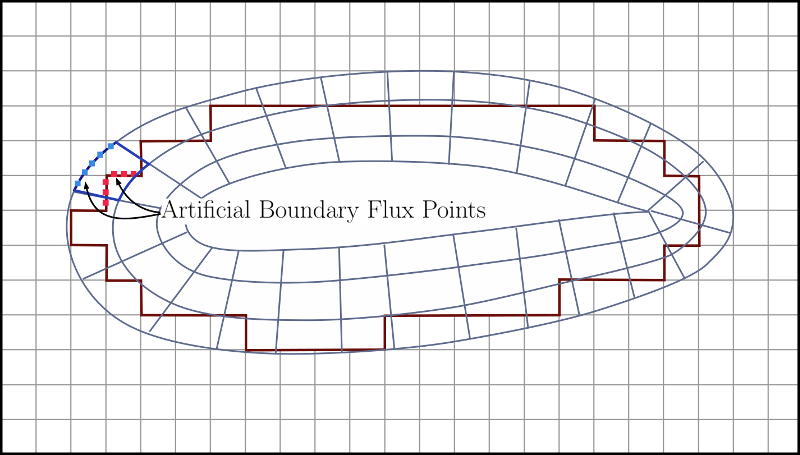}
}
\\ \vspace{6pt}
\subfloat[] {
  \includegraphics[width=.4\textwidth]{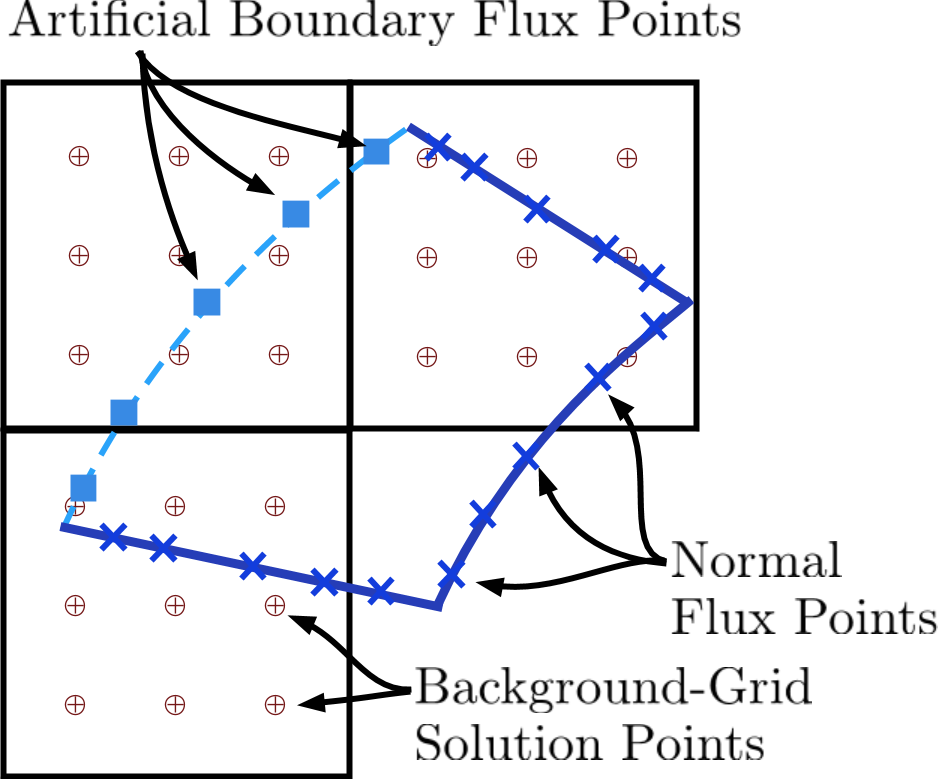}
  \label{fig:ab_a}
}
\hspace{5mm}
\subfloat[] {
  \includegraphics[width=.35\textwidth]{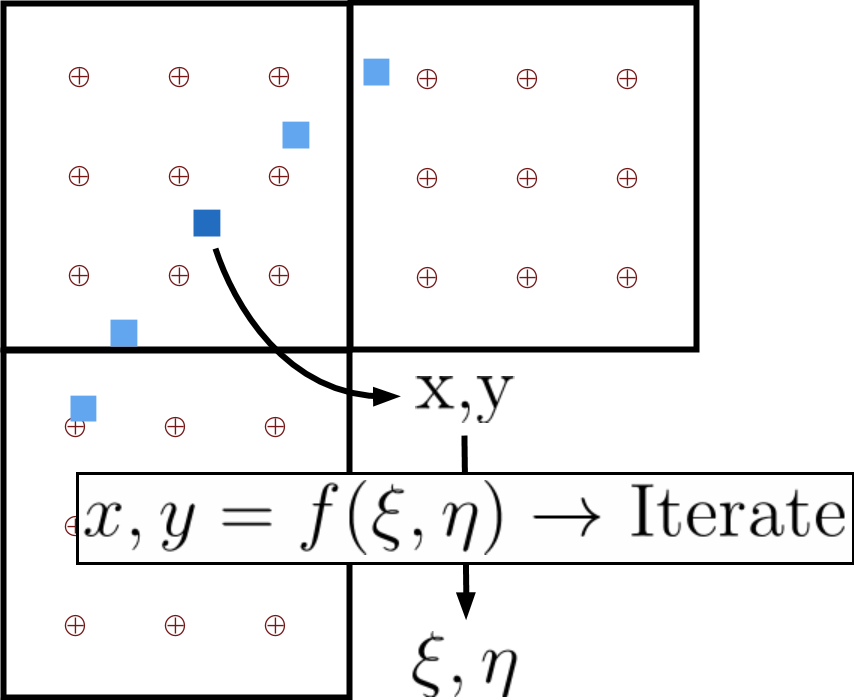}
  \label{fig:ab_b}
}
\caption{(a) Schematic representation of the resultant AB nodes which require interpolated data from the grid system in Figure \ref{fig:blanking}. (b) Several overlapping elements from the grid system, showing the flux points on the artificial boundary face where data must be interpolated to. (c) Schematic of the inverse-isoparametric mapping iterative method required to find the reference location for each fringe point.}
\label{fig:ab_method}
\end{figure}

\begin{figure}[t]
\centering
\subfloat[Two overlapping 1D grids.] {
 \includegraphics[width=.75\textwidth]{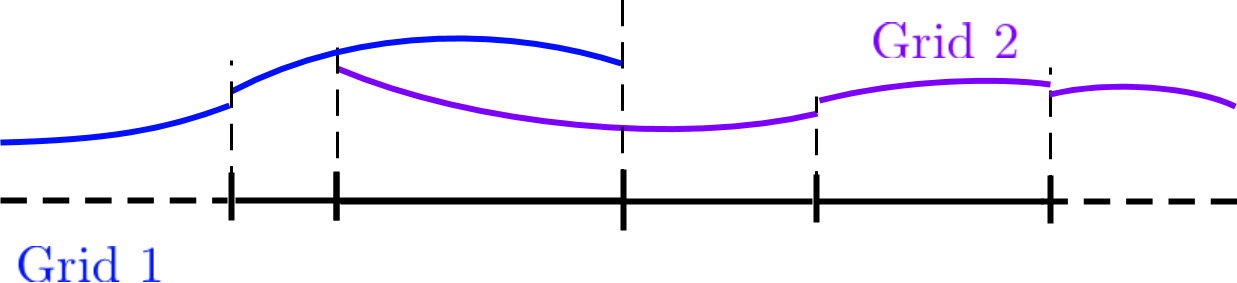} \label{fig:1d_a}
} \\ \vspace{6pt}
\subfloat[AB interpolation procedure for the blue grid] {
  \includegraphics[width=.75\textwidth]{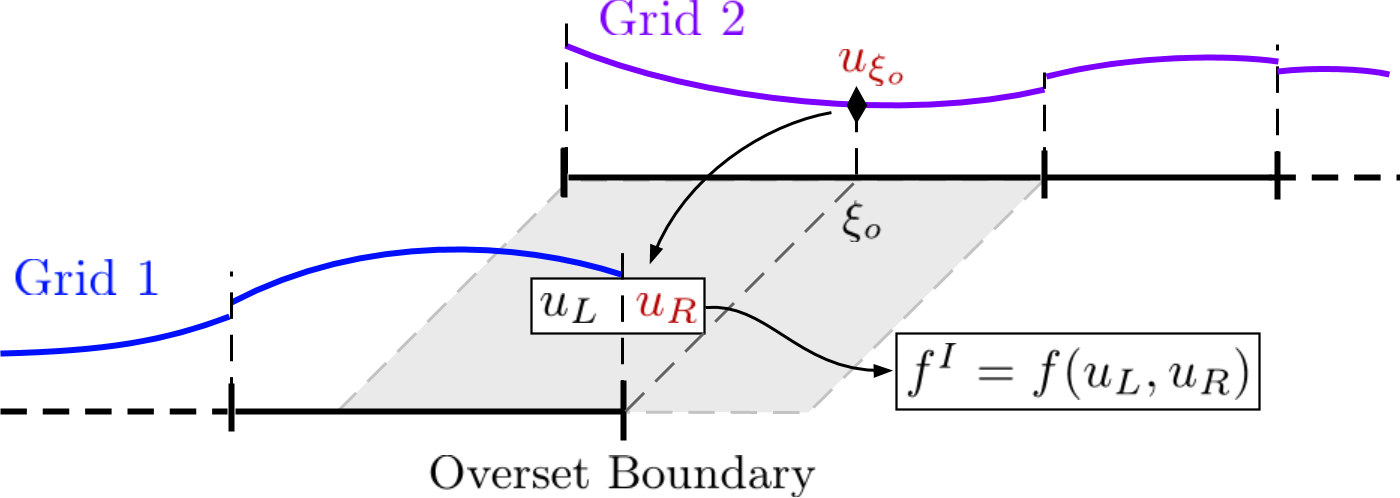} \label{fig:1d_b}
}
\caption{Schematic of AB method in 1D. $\xi_0$ is the reference location in Grid 2's donor element corresponding to the physical position of Grid 1's overset boundary. $u(\xi_0)$ is the interpolated solution value from Grid 2 at $\xi_0$, and is placed into Grid 1's memory as $u_R$.  The overset boundary of Grid 1 then acts as a normal inter-element interface and computes a common interface flux.}
\label{fig:overset1d}
\end{figure}

An AB overset simulation be broken down into three main steps:

\begin{enumerate}
\item \textit{Hole Cutting:}  All elements not required for computation on the combined overset grid system are `cut' (or blanked) from each grid.  This includes not only any elements lying within a solid boundary, but potentially any elements overlapping with a body (geometry-containing) grid.  As discussed in \cite{pundit}, the grid with the highest resolution in any given region may be kept and all others discarded, but for simplicity here, we will simply assume that any element enclosed within the overset boundary of another grid may safely be removed.  Wherever elements are removed from a grid, the leftover inter-element interfaces are subsequently tagged as being artificial boundary interfaces.  A schematic of the process is shown in Figure \ref{fig:blanking}; the full details of our hole-cutting algorithm are presented in Section \ref{S:pconn}.

\item \textit{Point Connectivity:}  The flux points lying on the artificial boundary faces of each grid are gathered and sent to all other grids which the points may lie within.  Each grid then attempts to find donor elements for the points it has been given (receptor points); this includes finding the reference location within the donor which maps to the physical position of each receptor point, as schematically shown in Figure \ref{fig:ab_method}.  This is typically accomplished by placing all potential donor elements into binary search trees which can quickly be searched for each receptor point.  The physical-to-reference position calculation is accomplished by performing Newton iterations on the isoparametric mapping for the donor element.

\item \textit{Data Interpolation:}  The computed reference coordinates are used to query the solution polynomial in the donor element for each receptor point which has a donor.  The interpolated conservative variables are then sent back to the receptor grid and unpacked into the proper location in the receptor grid's memory.  Each grid then carries on with its residual calculation and time advancement as normal, with the influence of the overset grids felt only in the computation of the common normal interface flux at the artificial boundaries.  A 1D schematic of the interpolation process is shown in Figure \ref{fig:overset1d}.
\end{enumerate}

Many algorithms exist for each of these steps, adapted to a variety of solver paradigms for many applications, from static analysis of a multi-element airfoil to dynamic analysis of a maneuvering helicopter.  The first step, hole cutting, is the main focus of our work here, and specific details of our approach will be presented in a subsequent section.  The second step, point connectivity, is required by all overset methods.  The third step, data interpolation, is a trivial polynomial interpolation procedure for discontinuous finite element-type methods.

\subsection{Review of Connectivity Methods}

Most of the work required for overset domain assembly is involved in the hole cutting and point connectivity procedures, though in some cases the steps overlap or are combined, such as when the hole cutting procedure is synchronous with the final point connectivity, or the point connectivity directly yields interpolation weights.  As discussed by Noack \cite{noack07}, automatic hole cutting methods for overset domain connectivity can be roughly be broken down into three broad classes of methods (as opposed to non-automatic hole cutting using user-defined analytic functions or shapes).

The first class, search-based methods, are characterized by their use of fast search methods, such as a KD-tree or an Alternating Digital Tree (ADT) \cite{ADT}.  Typically, a search-based method attempts to find a donor cell in another grid for each node in all grids.  Nodes without donors are potentially hole cells, and all nodes which located a possible donor cell lie in the overlap region between two or more grids, and hence are potential fringe (also called receptor) points.  Some form of approximate geometry representation, such as a Cartesian approximate map, can be used to distinguish nodes inside of solid boundaries from nodes completely outside the extents of other grids.  Alternately, some methods rely on each grid storing a wall distance at each node in the grid, then using that distance to find the median line between boundaries.  An example of this approach is the implicit hole cutting method introduced by Lee and Baeder \cite{lee03}, currently a very popular method, which combines the hole cutting and donor/receptor connectivity into a single process.

The next class of methods, query-based, are characterized by the use of an approximate geometry representation.  The approximate geometry must be easily queried to determine if a node or cell of one grid lies within a solid boundary of another grid; this can lead to a very fast and efficient hole-cutting method as long as the approximate geometry is easy to construct.  When constructing the approximate geometry, care is required if small features are present, such as a small gap between two bodies or between a wall boundary and its enclosing overset boundary. The approximate geometry must be adapted properly so that the error in the representation of the geometry is less than the size of such small features.  Examples here include Cartesian grid approximations, quad/octree grid representations, and other binary tree representations.  The difficulties associated with applying this approach to complex geometries with small features can be removed by using additional information and data structures in addition to the approximate geometry.  In PUNDIT \cite{pundit,pundit-2}, an approximate map back to the grid's elements is included in the approximate geometry, leading to a fast search method for performing donor/receptor point connectivity as well as sidestepping the limitations of the approximate geometry representation.

The last class of automatic hole-cutting methods are direct cut methods, in which the boundary surfaces of the grids are used directly to `cut' the connectivity of the other grids and remove all cells within them.  These methods still rely once again on tools such as fast binary search trees to locate the cells to be cut.  Few 3D implementations of this approach exist with the only other implementations known to the authors are the works of Galbraith \cite{Galbraith13} and of Noack \cite{noack07}.

Often implementations employ some combination of the above approaches and tools to perform the hole cutting.  For instance, the state-of-the-art connectivity package PUNDIT \cite{pundit} uses a hybrid application of the implicit hole-cutting approach, in which query-type structures are combined to speed up the implicit search method.

\section{Extension to Moving Grids}
\label{S:motion}

\begin{figure}
\centering
\subfloat[Overset grids at time $t$ with blanking $\mathcal{I}_{t}$]{
  \includegraphics[width=.5\textwidth]{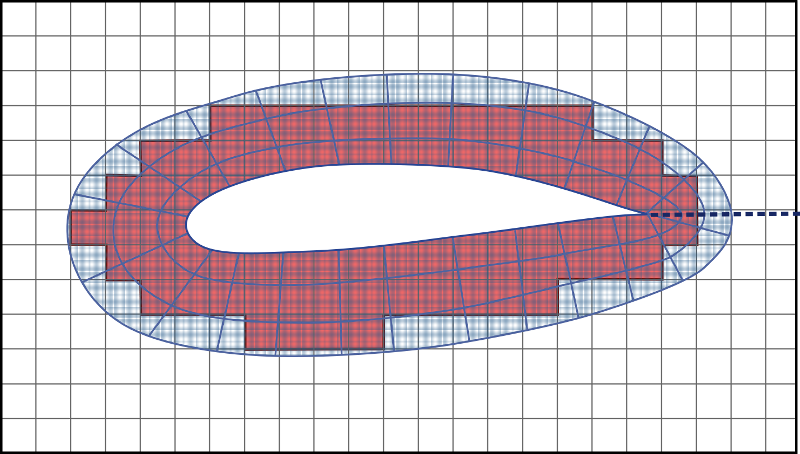}
  \label{fig:unblank-a}
}
\subfloat[Time $t + \Delta t$ with new blanking $\mathcal{I}_{t+\Delta t}$] 
{
  \includegraphics[width=.5\textwidth]{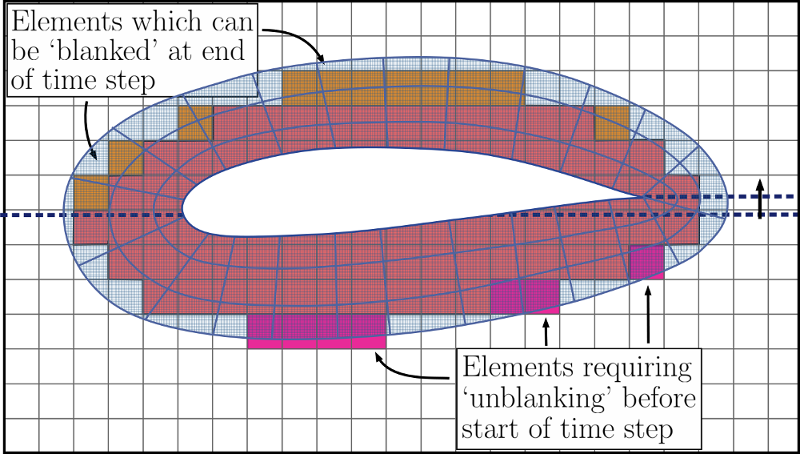}
  \label{fig:unblank-b}
}
\caption{Example of unblanking requirement.  From (a) to (b), several elements have changed status from hole to field status and require data in order to continue, while others have changed from field to hole but must remain until the end of the time step.}
\label{fig:unblank}
\end{figure}

When extended to moving and/or deforming grids, the AB method requires one additional high-level procedure not required by traditional methods built around finite-volume or finite-difference solvers, or high-order methods using volume interpolation and halos of fringe cells.  The issue which arises from grids in relative motion is exemplified in Figure \ref{fig:unblank}.  In Figure \ref{fig:unblank-a}, the light red cells show where elements have been cut (`blanked') from the background grid due to the foreground airfoil grid.  Some time later in Figure \ref{fig:unblank-b}, the airfoil has moved upwards, and the new hole cut from the background grid contains a different set of elements.  The new elements which were previously blanked must be located and filled with solution data (`unblanked') beforehand so that the simulation can progress without using invalid data.

This is not an issue in traditional solver methods due to their method of interpolation.  Since reconstructing the viscous fluxes in a finite-difference or finite-volume solver requires data from neighbors and neighbors of neighbors, when a hole is cut out of the grid, a `halo' of fringe points must be created to complete the numerical stencils for all nodes near the hole.  Thus, as the grids move, any single node never transitions from `hole' to `field' status in a single time step; first, it must pass through the halo of fringe points and receive interpolated data for nearby field nodes to use before it transitions to a field node itself.

Since we are primarily interested in explicit multi-stage Runge--Kutta time integration methods, the identification and processing of elements to unblank must be performed before starting each time step, and the blanking status held constant for all stages of the time step.  Therefore, for moving grids, the overall overset grid connectivity algorithm for each time step is laid out in Algorithm \ref{alg:unblank}.  This assumes that the movement of the grid during any time step is relatively small (i.e., less than the size of the smallest element); this is a reasonable assumption given that, for explicit time stepping, the Courant–Friedrichs–Lewy (CFL) constraints limit the time step to a point where information should essentially not travel farther than the width of the smallest cell during a single time step.  Elements need to be unblanked relatively infrequently in practice due to this, particularly if the unblanking is occurring near an overset boundary where elements are typically much larger than near-wall elements.  Since the point connectivity and data interpolation for element unblanking occurs only once when it is needed (as opposed to at every sub-stage of the RK time step), the additional overhead incurred when elements require unblanking is negligible to the overall simulation.

\begin{algorithm}
\caption{\label{alg:unblank} Artificial Boundary unblank procedure}
\begin{algorithmic}[1]
\State  Store current blanking status from time $t-\Delta t$ as $\mathcal{I}_{t-}$
\State  Calculate (or estimate) the position of the grid at time $t+\Delta t$
\State  Perform the hole cutting procedure to compute $\mathcal{I}_{t+}$
\State  Move the grid back to the current time $t$
\State  Perform the hole cutting procedure to compute $\mathcal{I}_t$
\State  For body grids cut by solid wall surfaces:
  \begin{enumerate}
  \item  Any cell which becomes a hole cell (lies inside the wall) in $\mathcal{I}_{t+}$ can be set to hole in $\mathcal{I}_{t}$
  \item  Any hole cell in $\mathcal{I}_{t-}$ which is no longer a hole cell in $\mathcal{I}_{t}$ and $\mathcal{I}_{t+}$ must be unblanked
  \end{enumerate}

\State  For background grids cut by overset boundary surfaces:
  \begin{enumerate}  
  \item  Any hole cell in $\mathcal{I}_{t-}$ which is no longer a hole cell in $\mathcal{I}_{t}$ or $\mathcal{I}_{t+}$ must be unblanked
  \end{enumerate}

\State  For all elements which must be unblanked:
  \begin{enumerate}
  \item  Gather all internal degrees of freedom (solution point locations) of the unblank elements
  \item  Perform the point connectivity process on these points, rather than the artificial boundary points
  \item  With both the grid and solution at time $t$, interpolate the solution to these points using the normal interpolation routines
  \end{enumerate}
\State  For each stage $i$ of the multi-stage Runge-Kutta time step:
  \begin{enumerate}
  \item  Move the grid to its position at time $t + c_i \Delta t$
  \item  Perform the normal artificial boundary point connectivity with blanking 
  $\mathcal{I}_{t}$
  \item  Perform data interpolation to the AB points
  \end{enumerate}  
\end{algorithmic}
\end{algorithm}

The distinction between off-body and near-body grids in the hole cutting and unblanking is to ensure a valid connectivity solution at all times.  When using overset boundaries to perform the hole cutting, to ensure we always have overlap between the grids, we must not blank potential hole cells until we have ascertained we no longer need them; conversely, we must unblank cells previously marked as holes before we need them to perform interpolation to the overset boundary.  When cutting with solid wall boundaries, we must ensure that no field cells end up inside the solid boundary at any point, so future hole cells must be removed at the beginning of the step.  Conversely, we cannot unblank a hole cell until we are sure it is no longer inside a solid boundary, so cells are left as holes until determined otherwise at the start of a time step.

\section{Parallel Overset Connectivity}
\label{S:pconn}

As opposed to the implicit hole-cutting now commonly used among finite volume-based overset solvers running on CPU-based hardware, such as the PUNDIT domain connectivity library used by HELIOS, the combination of high-order curved grids and modern highly-parallel hardware such as GPUs offers an opportunity for a simple, efficient hole-cutting procedure.

The approach we use is a direct cut method similar to that of both Galbraith \cite{Galbraith13} and Noack \cite{noack07}, but adapted for curved elements and GPUs.  Without loss of generality we will assume here that the system will consist of two types of grids: geometry-containing `near-body' (NB) grids, which will have an inner solid wall boundary surface and (typically) an outer artificial boundary \cite{Galbraith13, Galbraith14, crabill16}; and background  or `off-body' (OB) grids, which contain no geometry and are used to fill the overall simulation domain.  Off-body grids are still allowed to have pre-defined overset/artificial boundary surfaces; such a grid could be used as `patch' to locally refine an area in a manner similar to nested Cartesian grids.  Background grids will be cut by predefined overset surfaces, while body grids will only be cut by solid wall boundaries.  This approach can lead to larger amounts of overlap than implicit hole cutting in the case of multiple body grids. However, since a typical body grid created for an overset system will only extend a short distance from the geometry this is not expected to be an issue in practice.  Furthermore, compared to an implicit hole-cutting approach, the amount of data which must be exchanged is still significantly less, as only wall and overset boundary nodes need to be communicated, not the volume elements.

\subsection{Existing Direct Cut Methods}
\label{S:dcold}

\begin{figure}[t]
\centering
\subfloat[]{\includegraphics[width=.33\textwidth]{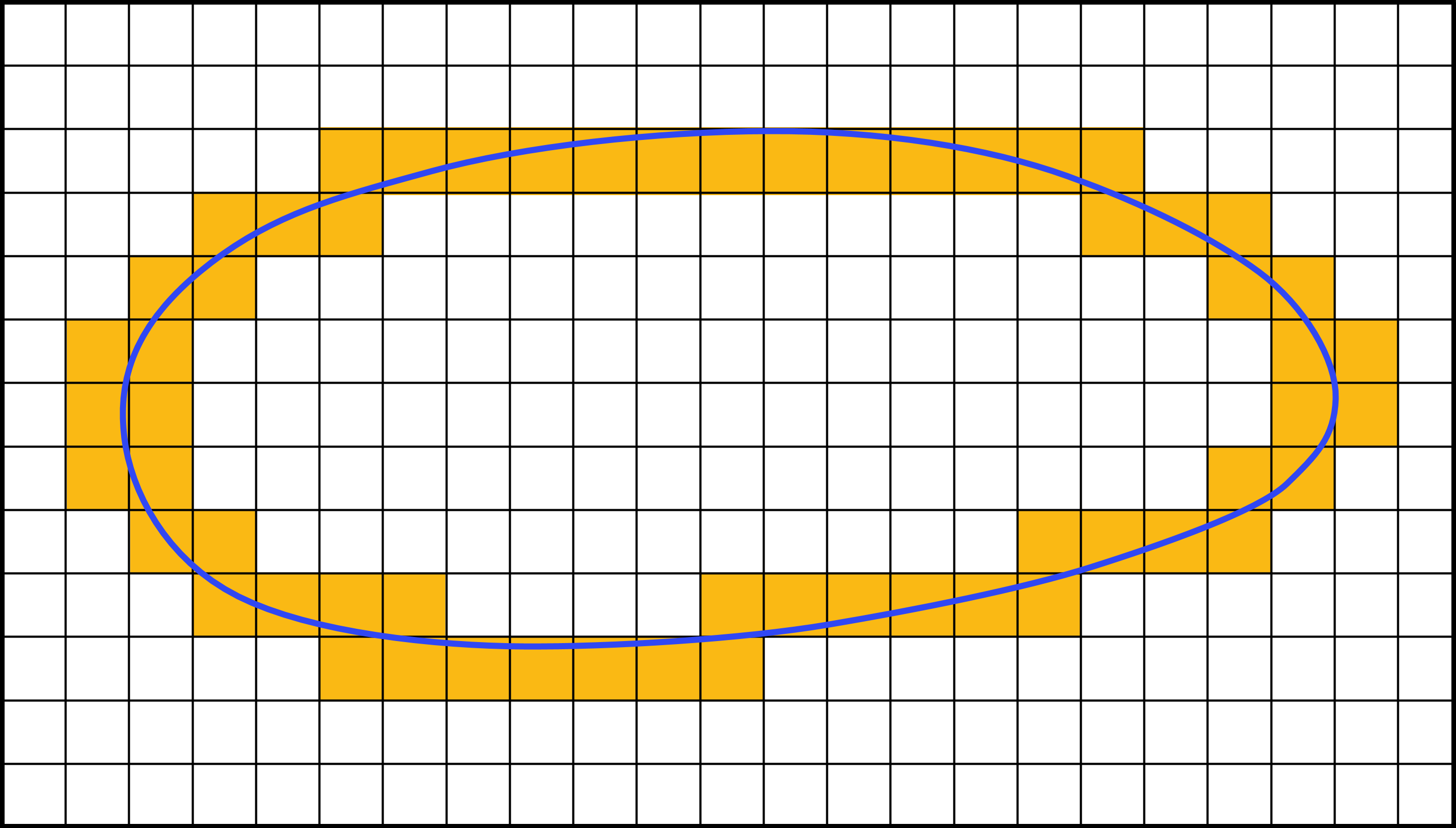} \label{fig:dcfill-a}}
\subfloat[]{\includegraphics[width=.33\textwidth]{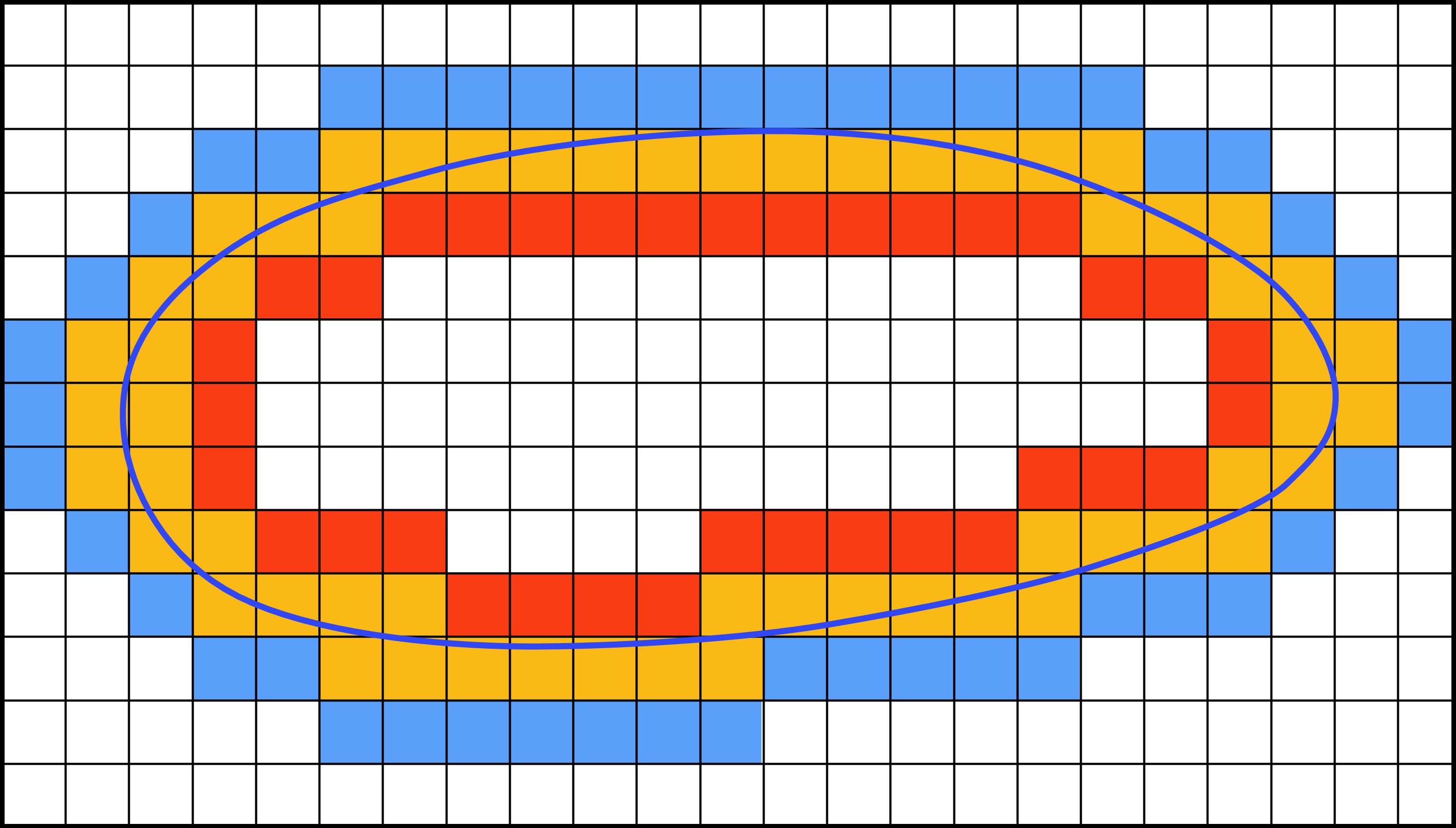} \label{fig:dcfill-b}} 
\subfloat[]{\includegraphics[width=.33\textwidth]{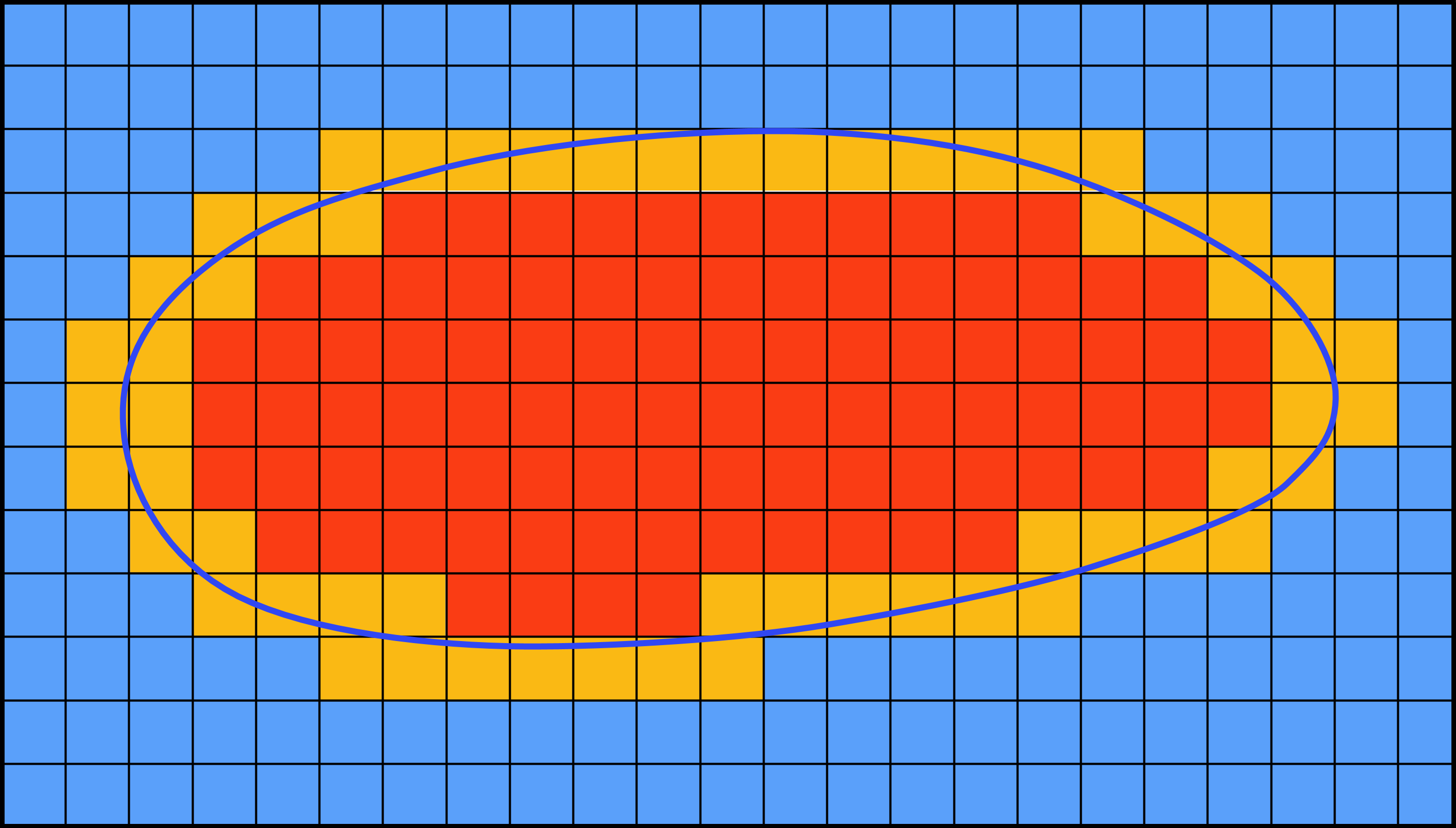} \label{fig:dcfill-c}}
\caption{Example of the Direct Cut process, based upon a paint-fill approach. (a) Locating all elements which are `cut' by the boundary. (b) Flagging all neighboring elements as being to the inside or the outside of the boundary. (c) `Paint-filling' the remainder of the grid by marching out from elements with known flags.}
\label{fig:dc-fill}
\end{figure}

Existing direct cut methods are built around a 3-stage process shown in Figure \ref{fig:dc-fill}.  First, geometric search algorithms and intersection checks are used to find the elements that are cut by a cutting surface (Figure \ref{fig:dcfill-a}).  Second, the neighbors of all the cut elements are found, and the normal direction of the nearest cutting face is used to determine which side of the surface it lies on and assign an appropriate blanking status (hole or field) (Figure \ref{fig:dcfill-b}).  Third, starting from the elements with a known status, a `paint-fill' procedure is used to assign a status to all remaining elements in the grid (Figure \ref{fig:dcfill-c}).

\begin{figure}
\centering
\subfloat[]{\includegraphics[width=.5\textwidth]{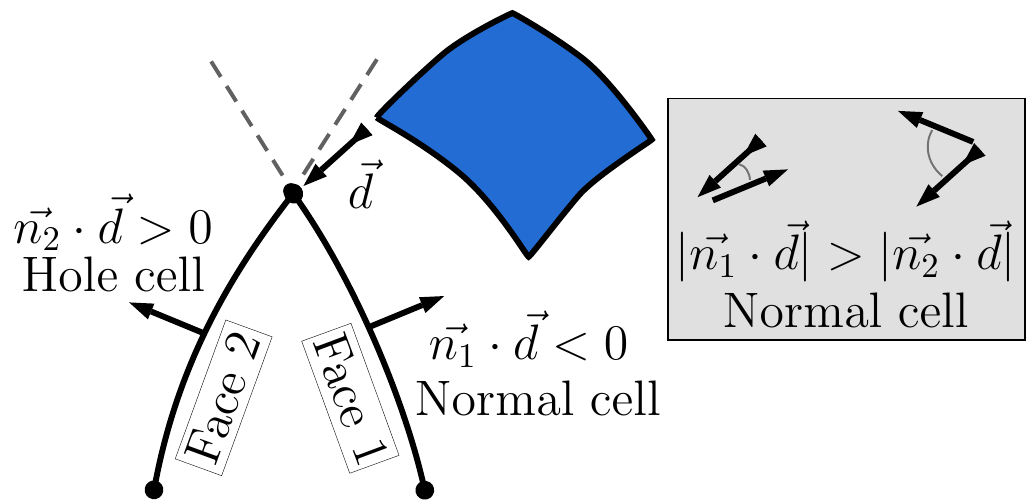}}
\hspace{6pt}
\subfloat[]{\includegraphics[width=.4\textwidth]{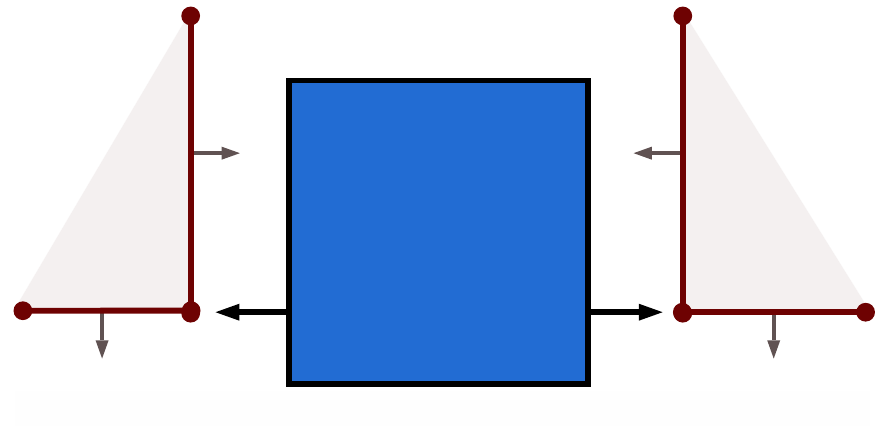}}
\caption{(a) Example of 2 equidistant faces to an element providing opposite cutting results. Taking the combination with the larger `visibility' (dot product magnitude) produces the correct result. (b) Example where taking the average of the face normals fails.}
\label{fig:multinorm}
\end{figure}

Galbraith's Direct Hole Cut method was the first application of a direct cut approach to grids with curved elements.  In his method, first, a KD-Tree is built from the cutting surface faces and used to locate all elements whose bounding box intersects that of a cutting face. To ensure success of the paint-fill process, all neighbors of these elements are also added.  Each element in the list then computes its distance to the cutting boundary using the Nelder--Mead minimization algorithm; this is done for each element-face pair by finding the shortest vector from a point inside the element to a point on the face.  The minimization variables are the reference coordinates of a point within the element and on the face, with constraints to enforce that points remain in the element and face, respectively.  The isoparametric mappings of the element and face---which can be of arbitrary order---are used to map the reference positions to physical positions and compute the vector.  Finding a vector of length 0 means that the element/face pair intersect; otherwise the dot product between the normalized minimal vector and the normalized surface normal vector on the cutting face is used to determine a signed distance and hence blanking status.  In the case of an element being approximately equidistant from multiple faces, extra steps are required.  If one of the vector/normal combinations has a significantly larger dot product magnitude, that dot product is used to determine the blanking status.  If the dot product magnitudes are approximately equal in magnitude but opposite in sign, then the normal vectors are averaged to provide a more accurate surface normal vector for the dot product.  If the dot products are of similar magnitude and produce the same result however, averaging should not be used.  As the case in Figure \ref{fig:multinorm} demonstrates, several normals can easily cancel out when taking the average if the element lies within a concave region of the cutting boundary.

Even in 2D, however, the use of nonlinear optimization to find the minimum distance between a curved element/face pair is unreliable and expensive; Galbraith gives several examples in which an incorrect local minimum can be found, and proposes a method of avoiding this issue. In 3D, the situation is even worse; simple tests on curved elements indicate that the process regularly fails to converge and, when it does converge, often locates a local, as opposed to global, minimum.  The paint-fill procedure is also problematic for large-scale problems on parallel distributed systems, as the procedure is inherently serial and would require multiple points of synchronization and communication between the processes comprising each grid.  Given these issues, it is clear that a more direct non-iterative element/face distance algorithm is needed.  Taking inspiration from computer graphics, our new algorithm is based upon a triangular representation of the geometry, and relies on fast box-box, line-line, and triangle-triangle distance calculations with an upper bound on computation required.  Such calculations require a relatively high number of floating point operations (FLOPs) per byte of memory used, and hence are well suited to modern hardware, characterized by an abundance of compute capability and limited memory bandwidth.  Most of these calculations are also independent and can be performed in parallel.  Combined with a hierarchical series of approximate distance calculations to reduce the number of detailed distance calculations required, the algorithm runs quickly and efficiently on modern GPU hardware.

\subsection{Parallel Direct Cut Algorithm}
\label{S:pdc}

In our approach, we turn the standard direct cut algorithm inside-out: rather than searching for all elements which intersect a cutting boundary, we instead have all elements calculate the (signed) minimum distance to the boundary. This eliminates the need for a paint-fill process by processing all elements in the grid in parallel to directly determine whether each element is inside, outside, or intersecting with the boundary.

The algorithm consists of a series of comparisons and distance calculations between elements and a cutting boundary, starting with large number of extremely fast but rough comparisons, and then moving on to a much smaller number of costly but far more accurate distance calculations.  Here we will assume that each partition contains a unique subset of a single grid, and that an arbitrary number of grids may be present, each with an arbitrary number of overset or solid wall boundary conditions.  Further, it will be assumed that any grid which contains a solid wall boundary will be classified as a \textit{near-body} (NB) grid, while all other grids will be classified as \textit{off-body} (OB) grids.  Lastly, although we will discuss the algorithm in the context of hexahedral elements, it is equally valid for all of the typical element types used for CFD simulations.  The algorithm described herein has been implemented into TIOGA.

During preprocessing, the axis-aligned bounding box (AABB) of each partition of each grid is stored.  Cartesian auxiliary maps of each potential cutting surface (both overset and solid boundaries) are constructed (on a per-grid basis) and exchanged with all other grids.  The auxiliary maps (hole maps) are constructed using a paint-fill process similar to that previously shown in Figure \ref{fig:dc-fill}, with the simplification that the hole map is a structured axis-aligned Cartesian grid of cubes of length $ds$.  The paint-fill procedure starts at the boundaries of the Cartesian grid, which are tagged with a value of `1' and constructed so as to guarantee they lie outside the surface boundary. From the `1' values at the boundary the interior is filled in until the surface boundary is reached, defined as any Cartesian cell with a value of `2'.  Upon completion, all cubes outside the surface are tagged with `1', all cubes intersecting the surface boundary are tagged with `2', and all cubes within the boundary are tagged with `0'.  The `0' and `1' values are then flipped, so that a value of `0' means outside the surface, `1' means entire within the surface, and `2' means at or near the surface.  Since finding the cube containing a point is a trivial procedure, this gives a fast and simple way to determine the approximate blanking status of elements by checking either their centroid or corner node positions.

At the start of the algorithm, each partition gathers all of its potential cutting surfaces, separated into overset and wall surfaces, along with an AABB for each.  Each partition shares its local cutting surface bounding boxes with all partitions belonging to other grids; which surface depends on the grid type of the receiver.  For near-body grids, only solid wall boundaries will be used; for off-body grids, only overset boundaries will be used.  A communication map is then constructed for each partition based upon overlap between its overall bounding box and cutting-surface bounding boxes received from other processes.  Process $A$ will receive data from process $B$ if the bounding box of $A$ intersects with the global bounding box of the cutting surface to which $B$ is part of, and if the bounding box of $A$ is near the local cutting surface bounding box of process $B$.  Here, `near' is taken to be a small percentage, typically ${\sim}3-5\%$, of the largest dimension of the global bounding box of the cutting surface.  The exact value to use depends upon the maximum error expected in the Cartesian auxiliary map of the surface.  A typical hole map will use 60--80 cubes along its longest dimension, meaning the maximum error in this discrete representation of the geometry is expected to be around $\sqrt{3 \cdot (1/60)^2} \approx 3\%$ of the longest dimension.  This tolerance ensures that any errors in status assignment caused by the hole map can be corrected with more accurate distance checks later in the algorithm.

\begin{figure}
\centering
\includegraphics[width=.55\textwidth]{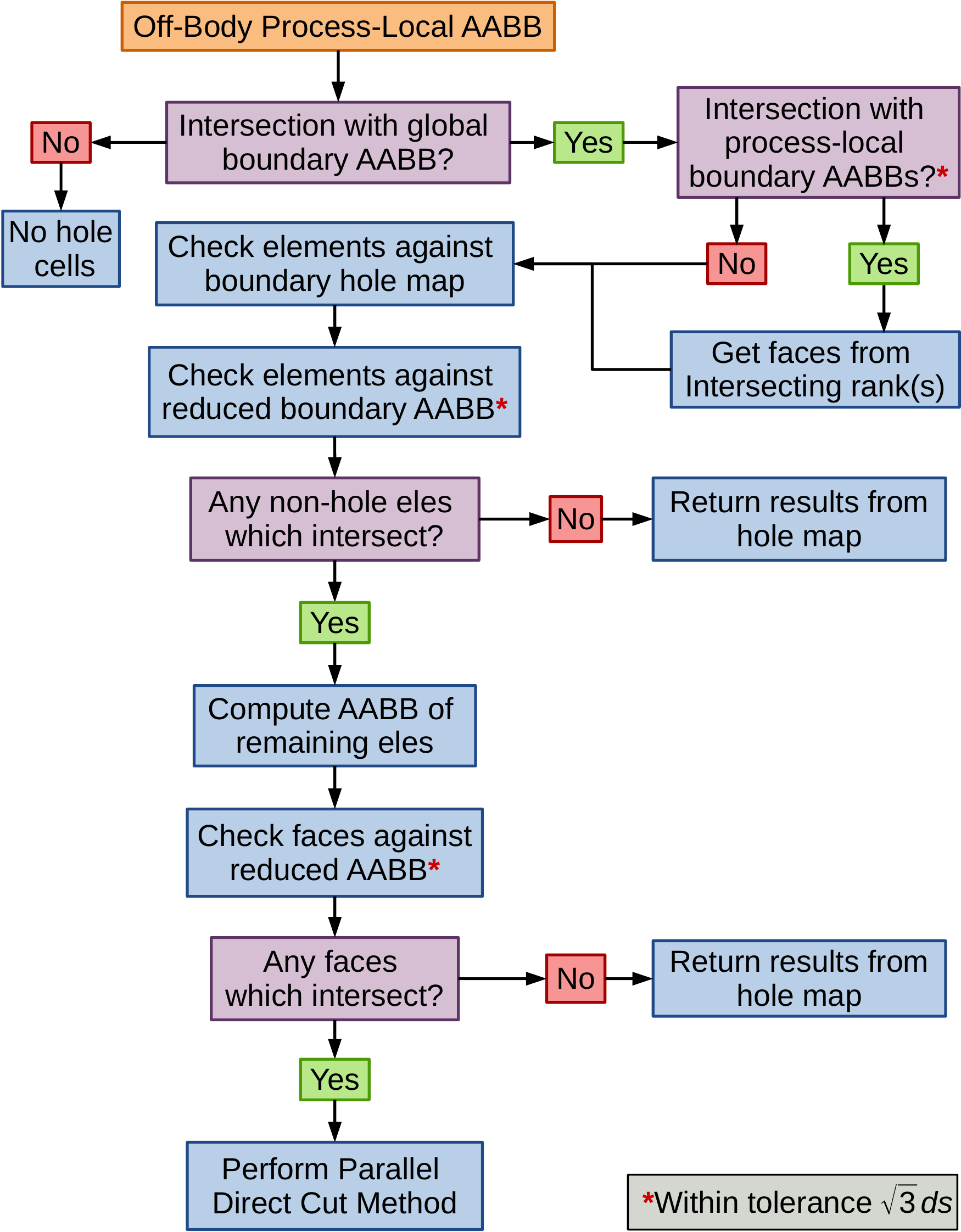} \\
\caption{High-level flowchart of the preprocessing leading into the parallel direct cut algorithm, including initial element / face filtering operations.}
\label{fig:flowchart}
\end{figure}

\begin{figure}
\centering
\includegraphics[width=.65\linewidth]{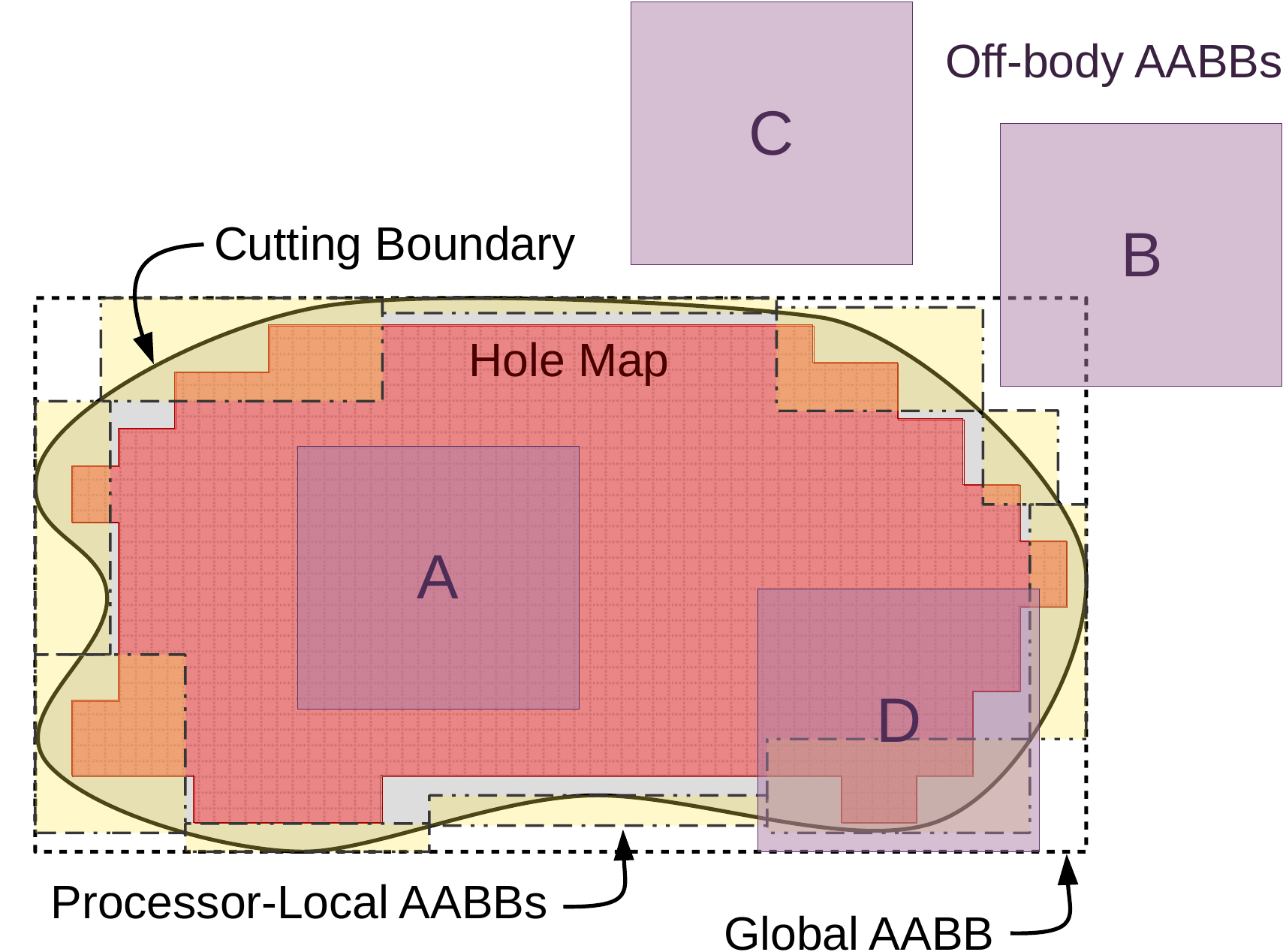}
\hspace{12pt}
\includegraphics[width=.3\textwidth]{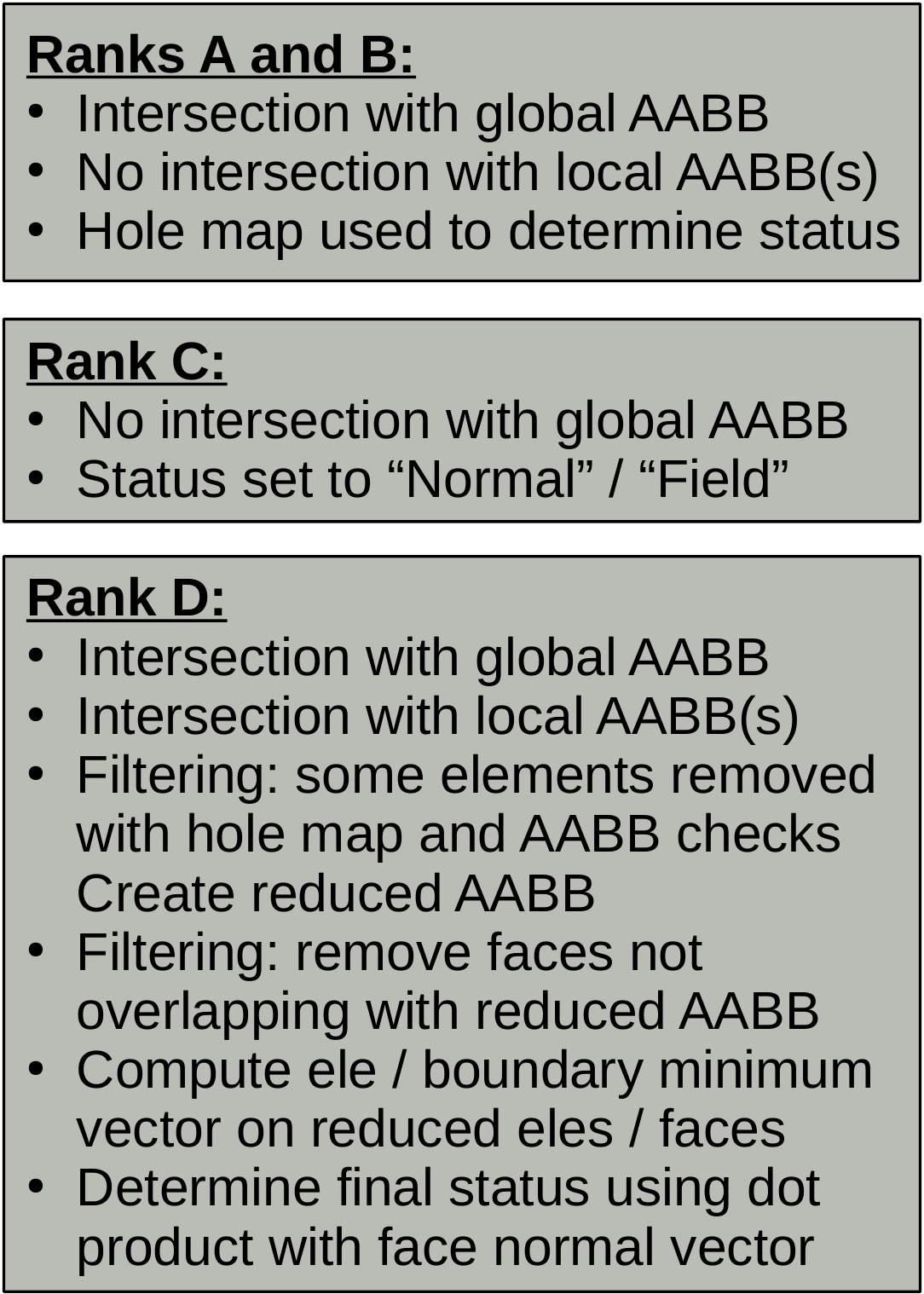}
\caption{Procedure followed for a single process in a variety of situations.}
\label{fig:flowchart-2}
\end{figure}

Finally, the nodes comprising each partition's cutting boundaries are sent out based upon this map; each partition receives the nodes into buffers organized by grid (i.e., all nodes from all partitions belonging to grid $g$ which have sent nodes to partition $p$ will be stored contiguously on rank $p$).  Once all partitions have received cutting surface data from all partitions which may contain relevant cutting faces, they begin the Parallel Direct Cut algorithm.  A high-level description of the preprocessing leading into the beginning of the algorithm is shown in Figure \ref{fig:flowchart}, with the route followed by a few specific cases shown in Figure \ref{fig:flowchart-2}.

\begin{figure}
\centering
\subfloat[]{\includegraphics[width=.3\textwidth]{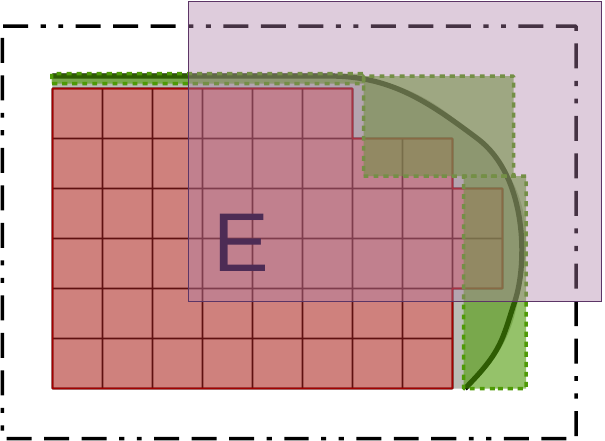}} 
\hspace{6pt}
\subfloat[]{\includegraphics[width=.3\textwidth]{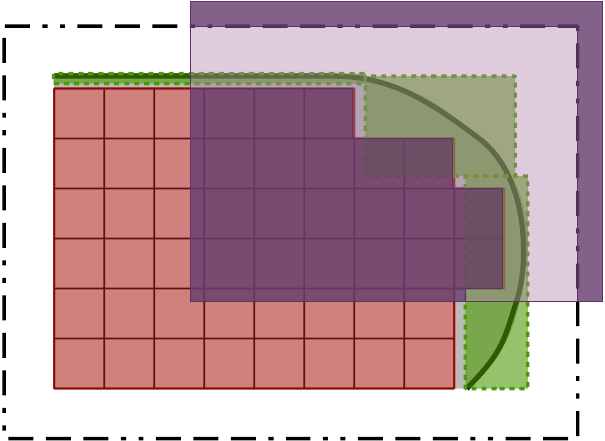}}
\hspace{6pt}
\subfloat[]{\includegraphics[width=.3\textwidth]{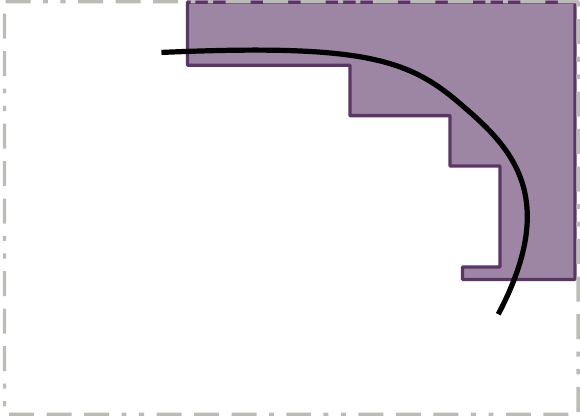}}
\caption{Example of the element/face filtering process for some process $E$.  (a) Hole map of surface, process-local surface bounding boxes from the four overlapping partitions, and reduced bounding box to compare elements against, with tolerance added.  (b) The elements which can be `filtered out' due to lying either within the hole map or outside the reduced bounding box. (c) The final set of filtered elements and faces to perform distance calculations on.}
\label{fig:dc-filter}
\end{figure}

First, the number of elements and faces to be compared in later steps can be reduced by using two quick `filter' functions.  Any elements which the hole map determines to be inside the boundary can immediately be tagged as such and excluded from further calculation.  Elements near the boundary according to the hole map (tagged with `2') and elements within $\sqrt{3} \cdot ds$ of the local cutting surface bounding box should remain. All remaining elements---now guaranteed to be outside the cutting surface ---may be tagged as normal cells and excluded from further calculation.  While creating this reduced list of elements, we also compute the bounding box of the set of filtered elements.  Assuming some elements remain, a similar filtering operation is also performed on the cutting surface faces given to each process.  Faces with $\sqrt{3} \cdot ds$ of the bounding box of the reduced/filtered set of elements are kept, and all others are discarded.  Figure \ref{fig:dc-filter} gives a schematic of the element and face filtering operations.

If no elements have passed the filtering stage, the results are copied back and the algorithm terminates.  Otherwise, some additional preprocessing is performed on the remaining elements and faces to compute several quantities for use in the later stages of the algorithm: 

\begin{enumerate}
\item Oriented bounding boxes (OBB) based upon corner nodes are computed for each element.  The OBB axes are chosen to be the eigenvectors of the covariance matrix of the nodes.
\item Approximate centroid locations based upon corner nodes are computed for each element.
\item Approximate centroid locations based upon corner nodes are computed for each face.
\end{enumerate}

\noindent
Using just the corner nodes extracts the linear component of the element or face, ignoring edge/surface/volume nodes if it is not linear.  We remark here that single-precision floating point values may generally be used in this part of the algorithm so as to reduce memory overhead.

\begin{figure}
\centering
\includegraphics[width=.65\textwidth]{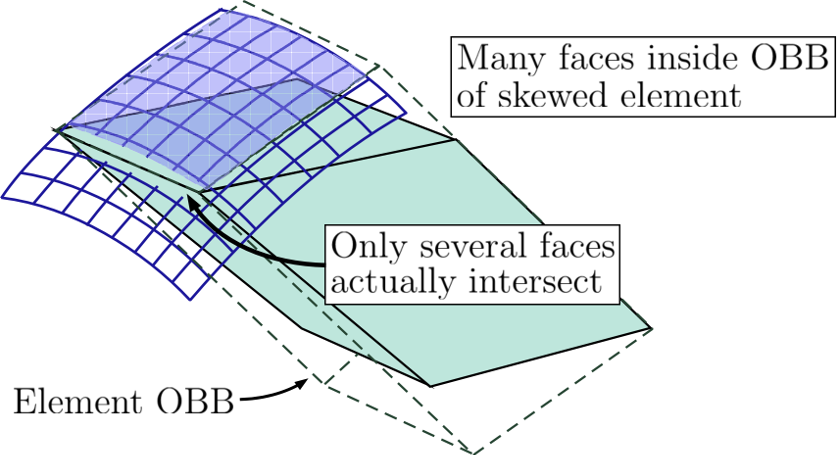}
\caption{Example showing case where relying on only bounding-box distance checks may fail, even for oriented bounding boxes.  Here, a large number of (small) cutting faces all intersect the (large, skewed) element's OBB, but only a small fraction actually intersect the element and could potentially be missed.  Further ranking the faces by their centroid-centroid distance improves the situation.}
\label{fig:dc-filt-0}
\end{figure}

Next, an approximate distance is computed from each filtered element to each filtered face.  The corner node positions of each face corner node are rotated into the OBB's axes and used to construct an AABB in these axes, then the distance between the two bounding boxes is computed.  This can lead to a large number of cutting faces with a distance of 0 to the element, as shown in Figure \ref{fig:dc-filt-0}; therefore, a small fraction of the centroid-centroid distance is added to the bounding-box distance to get an updated approximate distance suitable for sorting with fewer duplicate values.  The approximate distance is then stored to memory.  Only a small fraction of the centroid distance is used since there could be a large number of faces near an element which do not intersect its bounding box; this occurs with high aspect ratio and/or skewed elements, and we wish to ensure that faces whose bounding box intersects the element are prioritized over those that do not.

A separate kernel sorts the results for each element to keep only the closest $N_F^1$ faces per element, where $N_F^1$ is a constant known at compile time so that it can be used for sizing static arrays; a value of 16 is typically used.  The relatively large number of `approximate closest' elements retained at this stage is to ensure that, in the case of highly skewed elements and/or large differences in element sizes between grids, the chances of the actual closest element not being in the reduced list is quite small.  An example of where this issue may occur is shown in Figure \ref{fig:dc-filt-0}, where many faces are given nearly the same approximate distance to the element, but only a few actually intersect the element.  If it is known that very large differences in size between grids exist in the region of overlap, the value of $N_F^1$ can be increased as necessary.

As a second pass, for each element an improved approximate distance to each of its $N_F^1$ nearest faces is computed based upon linear representations of the elements and faces (only the corner nodes of each element and face are used).  The bilinear quadrilateral faces of each hexahedron are represented as two linear triangles, as is each quadrilateral face of the cutting surface.  We can then use a modified version of the M\"oller triangle-triangle intersection algorithm \cite{Moller_tri} to compute the shortest vector between each sub-triangle of the element and each sub-triangle of the boundary face.  To cut the amount of work in half, we first determine the nearest corner point of the element to the face's centroid, and only the 3 faces of the element sharing that corner point are used in the comparisons. These distances and nearest corners are sorted again to be left with only the nearest $N_F^2$ faces to each element, where $N_F^2$ is another, smaller constant integer, usually taken to be 4.  Again, for increased robustness when dealing with large differences in element sizes between grids, large aspect ratios, or highly curved elements, the value of $N_F^2$ can be increased.

\begin{figure}
\centering
\subfloat[]{\includegraphics[width=.33\textwidth]{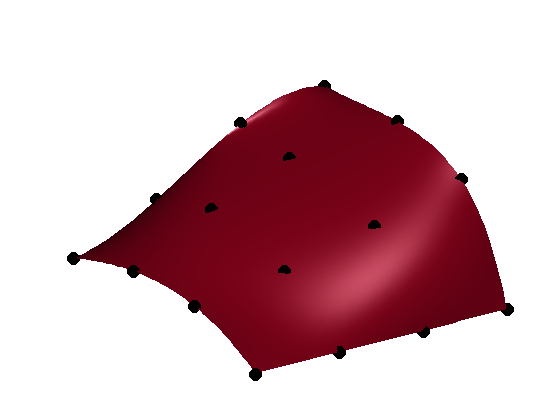}}
\subfloat[]{\includegraphics[width=.33\textwidth]{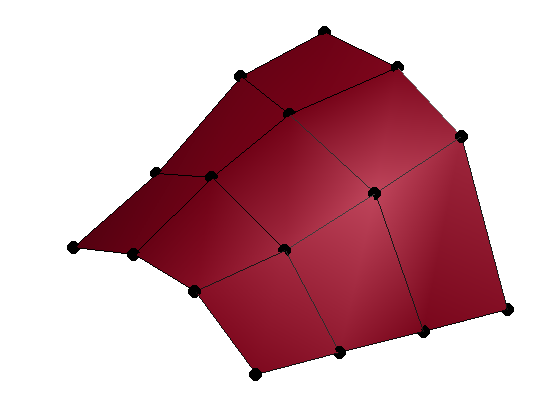}}
\subfloat[]{\includegraphics[width=.33\textwidth]{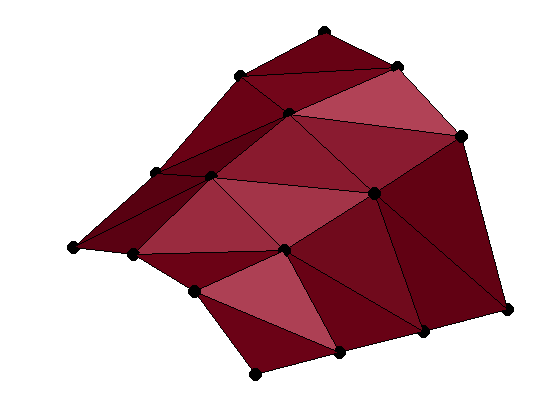}}
\caption{Example of splitting a high-order quadrilateral into linear triangles. (a) The exact tensor-product Lagrange polynomial representation; (b) Treating the tensor-product grid as a grid of linear quadrilaterals; (c) Splitting linear quads into triangles.}
\label{fig:quadsplit}
\end{figure}

Finally, for the most accurate element-boundary distance calculations, each high-order element face and boundary face is represented as grid of linear quadrilateral faces, which are once again split into two linear triangles, as shown in Figure \ref{fig:quadsplit}.  A similar splitting could be performed for element types with high-order triangular faces. Similarly to the previous step, we compute the shortest vector between each triangle-triangle combination and store the results, again using the previously determined near-corner point to cut the number of element/face distance calculations in half.  One final sort is performed to find the final nearest face for each element and the corresponding vector, and a dot product with the face's outward-oriented normal vector is used to determine which side of the face the element lies on.  In the case of several faces having approximately the same distance to an element, the corresponding normals are averaged into a new vector which is used in the dot-product calculation instead.  

The final blanking status of each cell can then be saved to memory and copied back; a separate function then takes the union of the hole cells created from each of the cutting boundaries to create the final blanking result.  The artificial boundary faces are then determined to be any face which has a hole cell on one side and a normal/field cell on the other.  MPI partition data is also needed to check blanking status between multiple processes of the same grid.  In this case, the process ID and local face ID of the opposite side of each MPI face is used to communicate the status of each MPI face.

The geometry linearization does incur some error in the representation of surface and element faces; however, any regions of tight curvature on a geometry of interest are likely to be more refined to capture the flow physics in that region, so there is a practical limit on the curvature of any single element face.  Additionally, out of necessity, the outer boundary of any near-body grid will be far smoother than the solid geometry inside; this outer overset boundary provides a much `cleaner' cutting surface.  The outer boundary could even be linearized to reduce the work required here, since the shape of the outer boundary is essentially arbitrary.  We also include a small relative tolerance based on element size when determining if two triangles intersect.  As such, errors arising from this linearization are expected to be negligible in practice.

Detailed pseudocode for each step of the above algorithm can be found in Appendix \ref{A:code}; for the full code listings see the TIOGA GitHub repository \cite{tioga}.

\subsection{Donor/Receptor Point Connectivity on GPUs}

Accelerating just the hole cutting process is not enough to speed up moving-grid overset simulations with an accelerated solver, since calculation of the donor/receptor point connectivity is also a work-intensive procedure.  At the completion of the Parallel Direct Cut method, each rank has a list of artificial boundary faces, each with some number of flux points (fringe points) for which donor elements must be found.  This requires a fast geometric search procedure to locate potential donor elements, followed by a point-containment check for curved elements to determine whether the point lies within the element, and if so, the exact position within the element.  A typical method for geometric searching within a mesh is to use an ADT.  As found by Soni et al. \cite{soni12}, construction of the ADT is a serial recursive process not well suited for accelerators.  However, if all grid motions are constrained to be rigid-body translations and rotations---something which is sufficient for a large number of calculations---then the ADT can be generated once for the initial grid configuration and accessed as the grids move through use of a rotation matrix and linear offset.  The ADT can thus be constructed once during preprocessing on the CPU and copied onto the GPU.

A recursive tree-traversal function is typically used to search through ADTs or other binary tree structures.  However, current programming paradigms for accelerators eschew recursion.  Furthermore, searching through a tree can also lead to divergent behavior, between threads/vector lanes.   In the worst case this can effectively serialize a substantial portion of the computation.

\begin{figure}
\centering
\includegraphics[width=.85\textwidth]{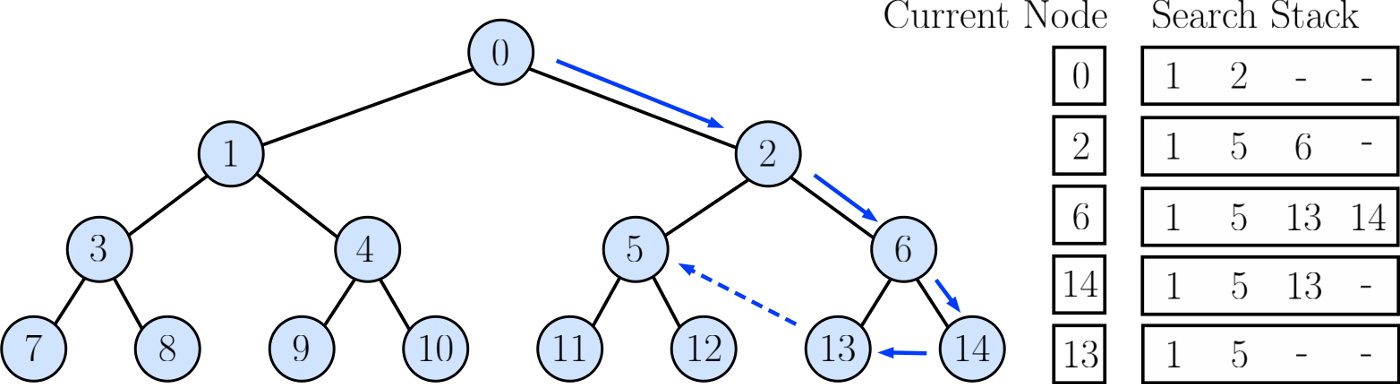}
\caption{Example stack-based ADT search with `worst case' of each node pushing both children onto the stack (i.e., a full tree traversal); the stack size must be equal to the height of the tree plus one (in this case, 4).}
\label{fig:adt-stack}
\end{figure}

The solution to these issues is twofold.  To avoid the need for a recursive function, a simple stack based approach can be employed wherein tree nodes are pushed and popped to a LIFO (last in first out) stack.  Since memory bandwidth is the primary bottleneck on modern hardware, the stack is a fixed-size local array; and since a LIFO stack leads to a depth-first search, the size of the array is equal to the number of tree levels that can be searched.  In the worst case of a complete tree traversal, a stack of size $N$ allows traversing a balanced tree with $2^{N}-1$ nodes (See Figure \ref{fig:adt-stack}; note that by construction the ADT is a balanced tree).  This allows a relatively small stack to be used to search large trees; for example, a stack of size 20 allows for searching a tree containing some $2^{20}-1 = 1\,048\,575$ nodes.  Second, to mitigate the issue of divergence, instead of performing the point-containment check at a tree node when a potential donor is found, the potential donor element is appended to a separate array.  Only once the search has finished are the work-intensive containment checks performed.

The point-containment check used herein is to perform Newton iterations on the isoparametric mapping of the element to locate the reference coordinates from the physical position.  Given the Lagrange shape basis for an element, the mapping from the reference element with domain $-1 \leq \xi,\eta,\zeta \leq 1$ to its physical location is:
\begin{equation} \label{eq:reftophys}
\mathbf{x}(\xi,\eta,\zeta) = \sum_{i=0}^{K} \sum_{j=0}^{K} \sum_{k=0}^{K} \mathbf{x}_{i,j,k} l_i(\xi) l_j(\eta) l_k(\zeta) 
\end{equation}
where $\mathbf{x}_{i,j,k}$ is the vector of physical coordinates for node $i,j,k$ in the tensor-product Lagrange basis defining the hexahedron's shape, and $l_i$ is the $i^{th}$ mode of the $K^{th}$-order Lagrange basis polynomial.  This can also be written in vector form as:
\begin{equation}
\mathbf{x}(\mathbf{\mathcal{X}}) = \mathbf{X} \mathbf{N}(\mathbf{\mathcal{X}})
\end{equation}
where $\mathbf{X}$ is the $3\times (K+1)^3$ matrix of node positions and $N(\mathbf{\mathcal{X}})$ is the vector comprising the $K^3$ tensor-product Lagrange bases evaluated at $\mathbf{\mathcal{X}}$.  

Given a physical position $\mathbf{x}^p$, we wish to determine the reference location $\mathbf{\mathcal{X}}^p$ which maps to $\mathbf{x}^p$.  Starting with an initial guess $\mathbf{\mathcal{X}}^0$, the function we wish to minimize is given by
\begin{equation}
\Delta(\mathbf{\mathcal{X}}^n) = \mathbf{x}^p - \mathbf{x}(\mathbf{\mathcal{X}}^n)
\end{equation}
We wish to update our guess $\mathbf{\mathcal{X}}^n$ by performing a Newton step (descending along the direction of maximum error in reference space):
\begin{equation} \label{eq:ref-update}
\mathbf{\mathcal{X}}^{n+1} = \mathbf{\mathcal{X}}^n + \frac{\partial \mathbf{\mathcal{X}}}{\partial \mathbf{x}}(\mathbf{\mathcal{X}}^n) \Delta(\mathbf{\mathcal{X}}^n) = \mathbf{\mathcal{X}}^n + \mathcal{J}^{-1}(\mathbf{\mathcal{X}}^n) \Delta(\mathbf{\mathcal{X}}^n)
\end{equation}
where $\mathcal{J}^{-1}(\mathbf{\mathcal{X}}^n)$ is matrix inverse of the Jacobian of the isoparametric mapping $\mathcal{J}$ evaluated at $\mathbf{\mathcal{X}}^n$:
\begin{equation} \label{eq:Jac}
\mathcal{J} = \frac{\partial(x,y,z)}{\partial(\xi,\eta,\zeta)} = 
\frac{\partial \mathbf{x}}{\partial \mathbf{\mathcal{X}}}
\end{equation}
with the individual terms of Equation \ref{eq:Jac} calculated as derivatives of Equation \ref{eq:reftophys}.

Since Newton's method generally converges quadratically, Equation \ref{eq:ref-update} usually converges in a small number of iterations. A maximum upper limit of 10 is used to handle highly skewed elements where the mapping is less well-behaved.  Additionally, since we limit $\mathbf{\mathcal{X}}^{n+1}$ to lie within a small tolerance of the range [-1,1], it is not possible for the algorithm to converge when the physical point $\mathbf{x}^p$ lies outside the element; instead it will find the nearest point on the boundary of the element.  This typically happens within a few iterations, at which point the method can converge no further and the error remains constant.

\subsection{Connectivity Robustness}
\label{S:testing}

\begin{figure}
\centering
\subfloat[Inner grid; iso, front, side]{\shortstack{
  \includegraphics[width=.45\textwidth]{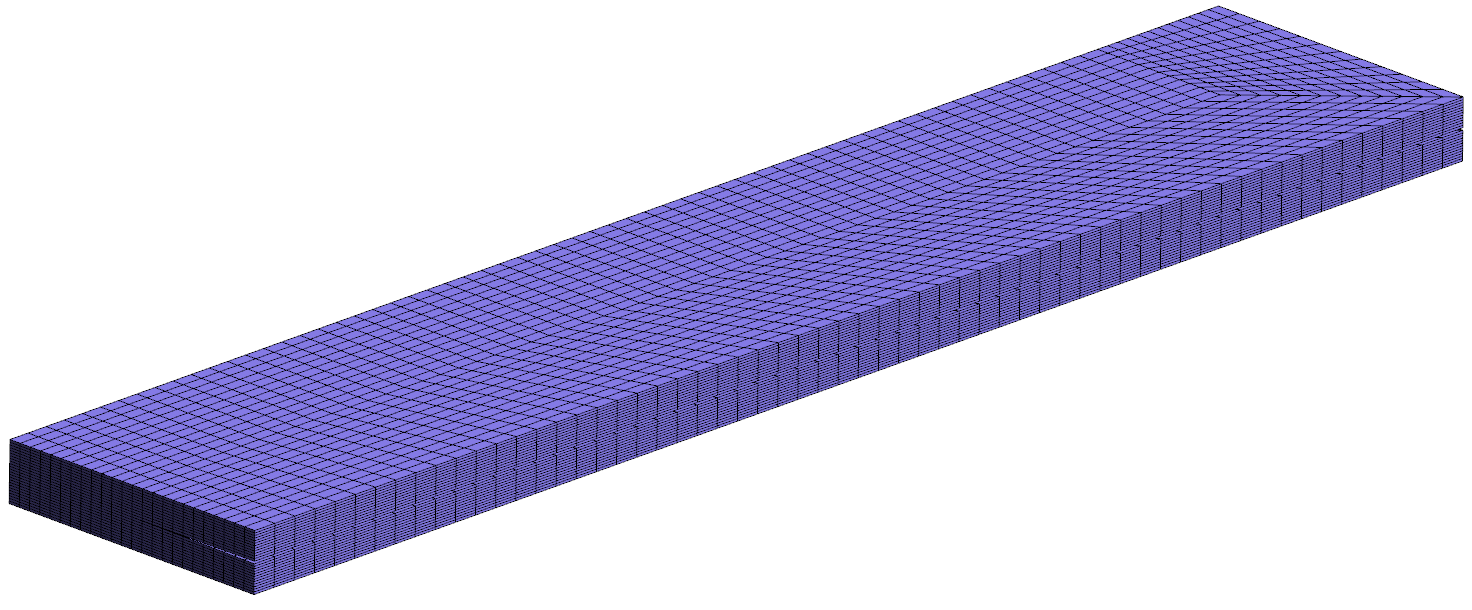}
  \\
  \includegraphics[width=.22\textwidth]{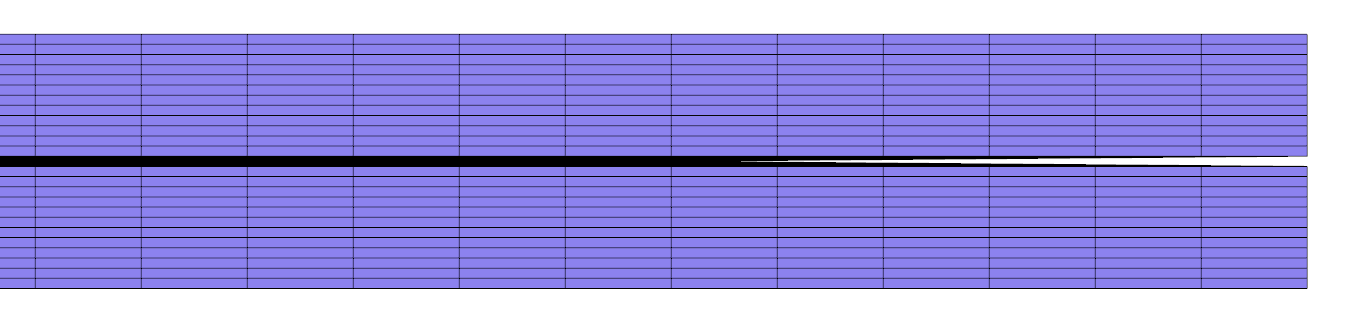}
  \includegraphics[width=.22\textwidth]{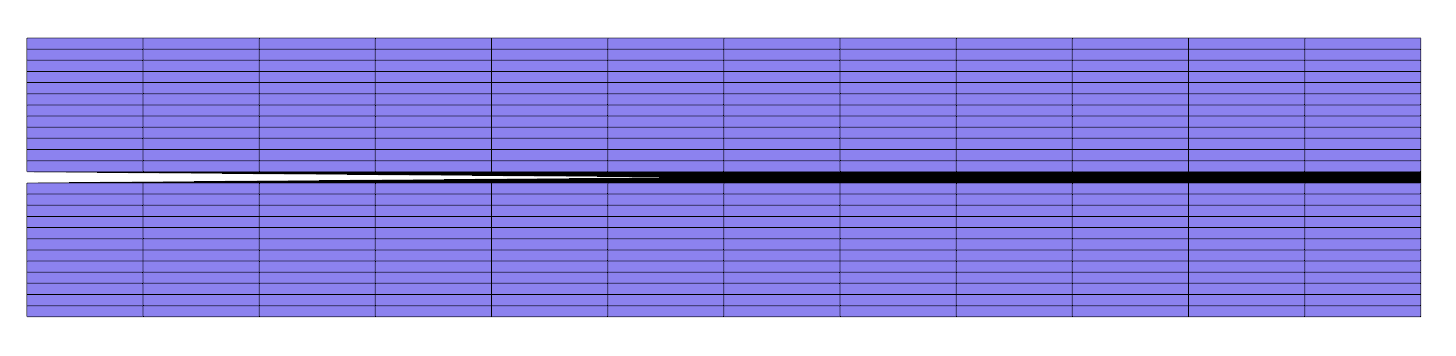}
}}
\hspace{6pt}
\subfloat[Outer grid]{\includegraphics[width=.4\textwidth]{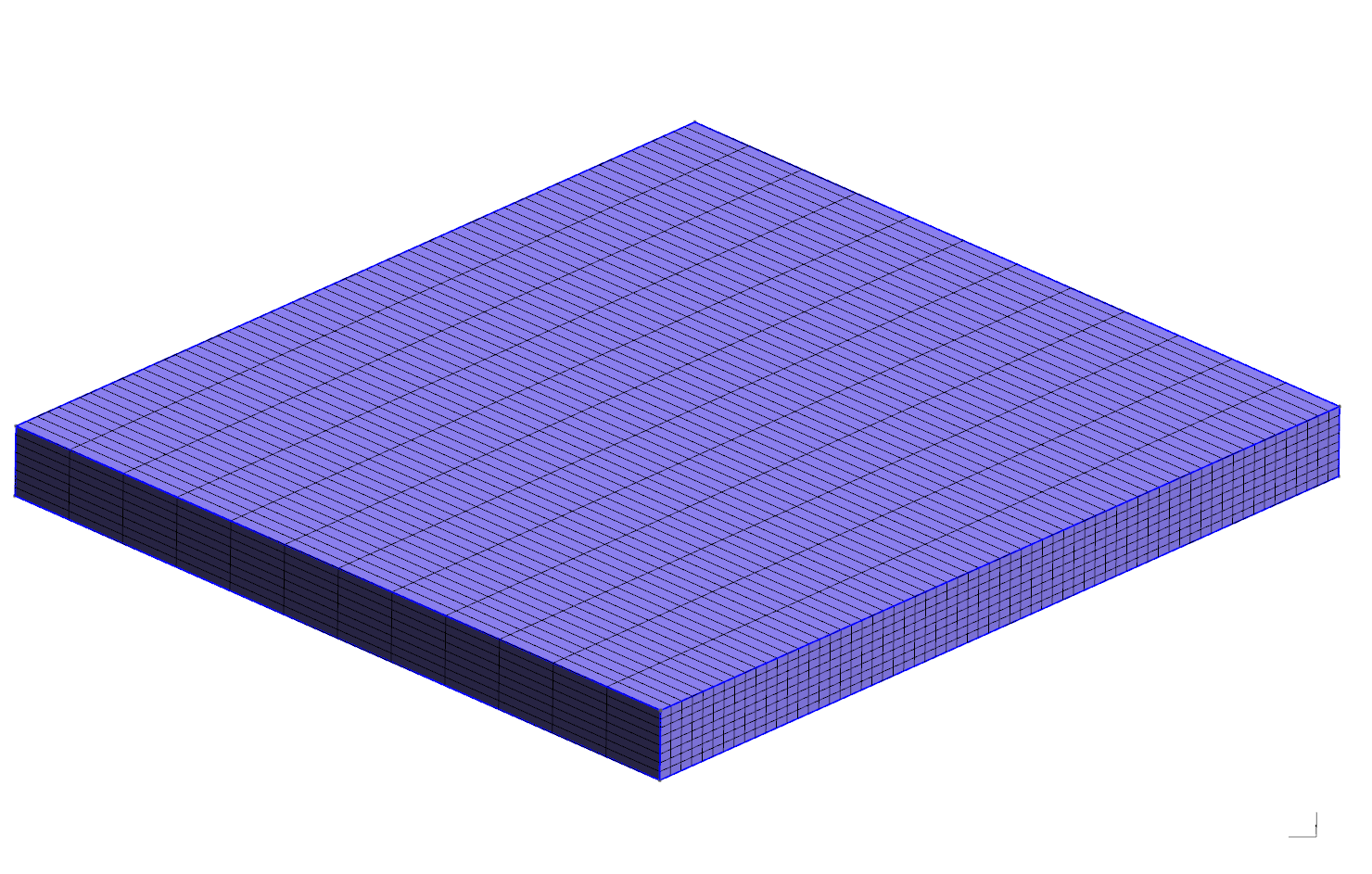}}
\caption{Geometry used to explore edge cases of the parallel direct cut method.}
\label{fig:dc-test}
\end{figure}

To verify that our proposed connectivity method will be robust when applied to a wide range of geometries, a variety of grid configurations have been tested.  First, a 2-grid test case has been developed to explore a number of `edge cases' where the method is most likely to fail.  The basic grid system used for these tests is shown in Figure \ref{fig:dc-test}, and features an inner grid with high aspect ratios and a groove on two sides, and a Cartesian background grid which can be varied in density in each direction, as well as rotated to apply skew between the elements of the two grids.  As discussed in Section \ref{S:pdc}, a large difference in element sizes between grids can lead to issues in the initial low-accuracy distance approximations if a large number of faces appear to be close to an element when in fact other faces are actually closer. This is tested through relative rotation of the two grids as well as assigning widely different element densities between them. The Parallel Direct Cut method produced a valid hole cutting result for all combinations of rotation, aspect ratio, element density, and groove sizes.

\begin{figure}
\centering
\subfloat[Grid configuration]{\includegraphics[width=.3\textwidth]{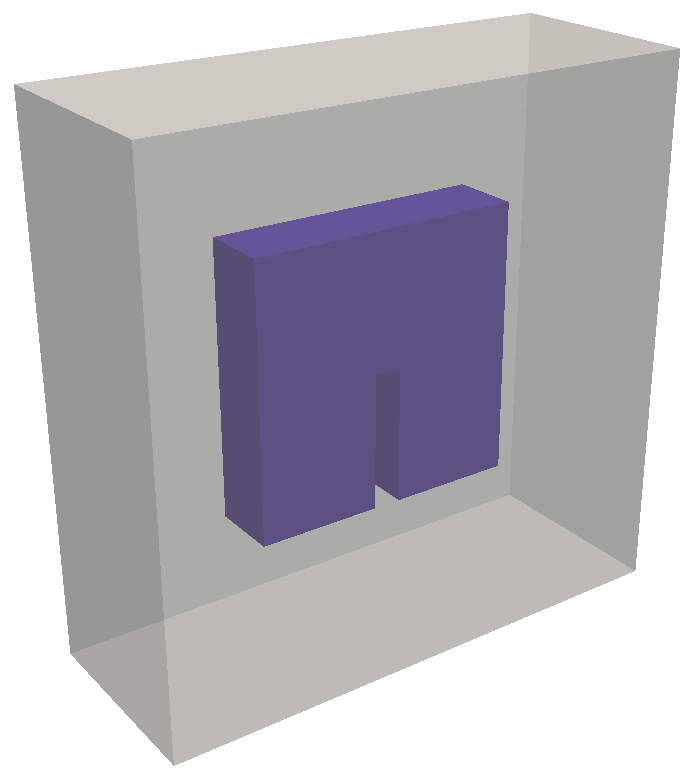}}
\subfloat[Front view of grids]{\includegraphics[width=.33\textwidth]{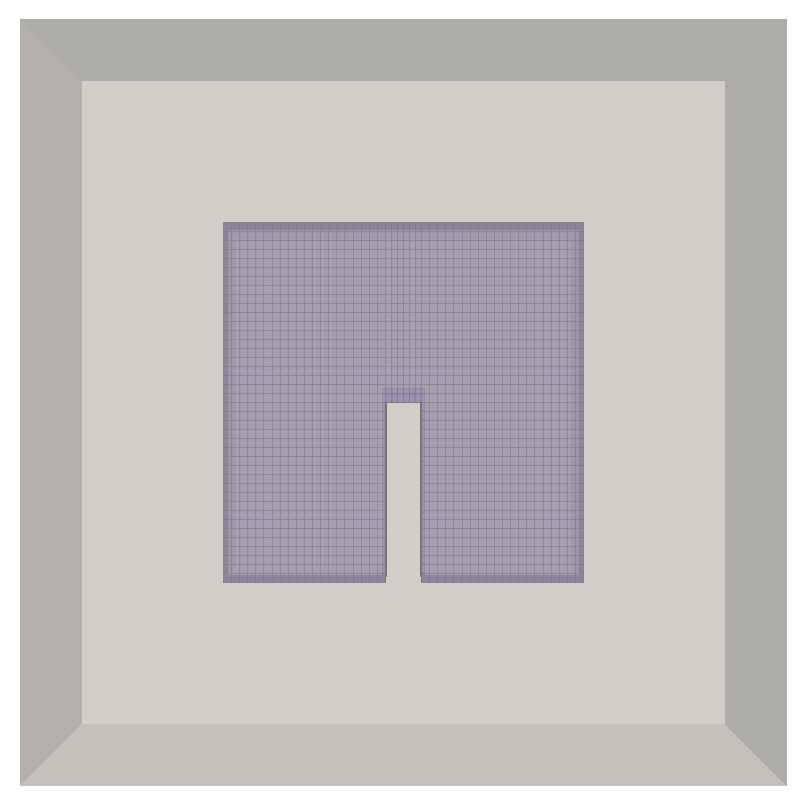}}
\subfloat[Final hole cut]{\includegraphics[width=.33\textwidth]{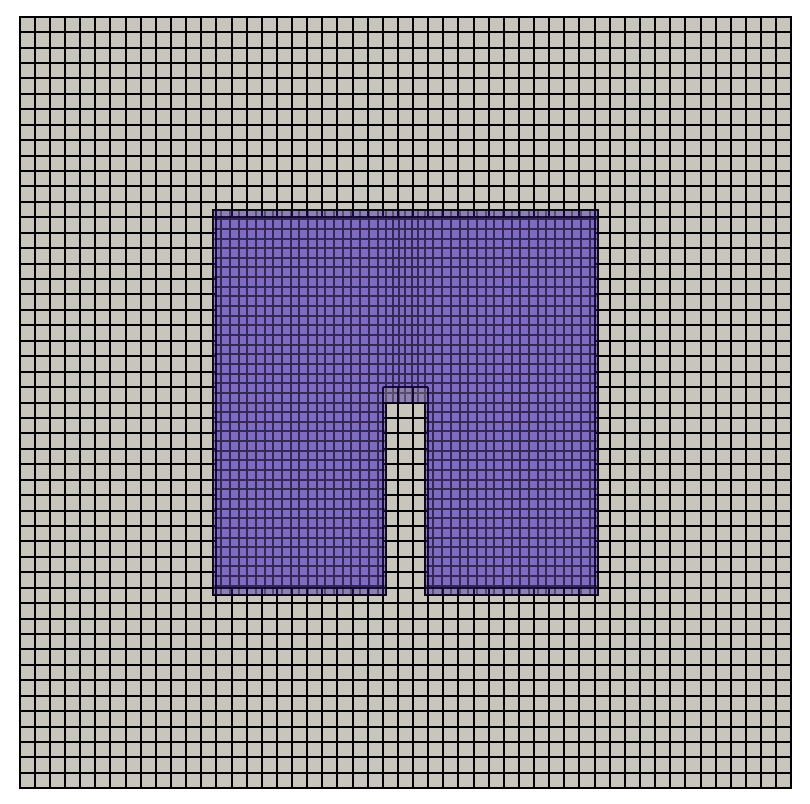}}
\caption{Slot geometry.}
\label{fig:slothole}
\end{figure}

A similar test case is shown in Figure \ref{fig:slothole}, designed specifically to test the final status assignment algorithm.  The inner grid has a straight-sided slot through it, and several elements of the background Cartesian grid lie directly centered in the slot, with some additionally equidistant from faces around the corners of the slot.  As discussed in the previous section, with equidistant faces on both sides of these elements giving the same blanking status, the face normals must not be averaged as described in Figure \ref{fig:multinorm}; instead, the face normal/minimum vector combination yielding the largest magnitude dot product should be used.

\begin{figure}
\centering
\includegraphics[width=.7\textwidth]{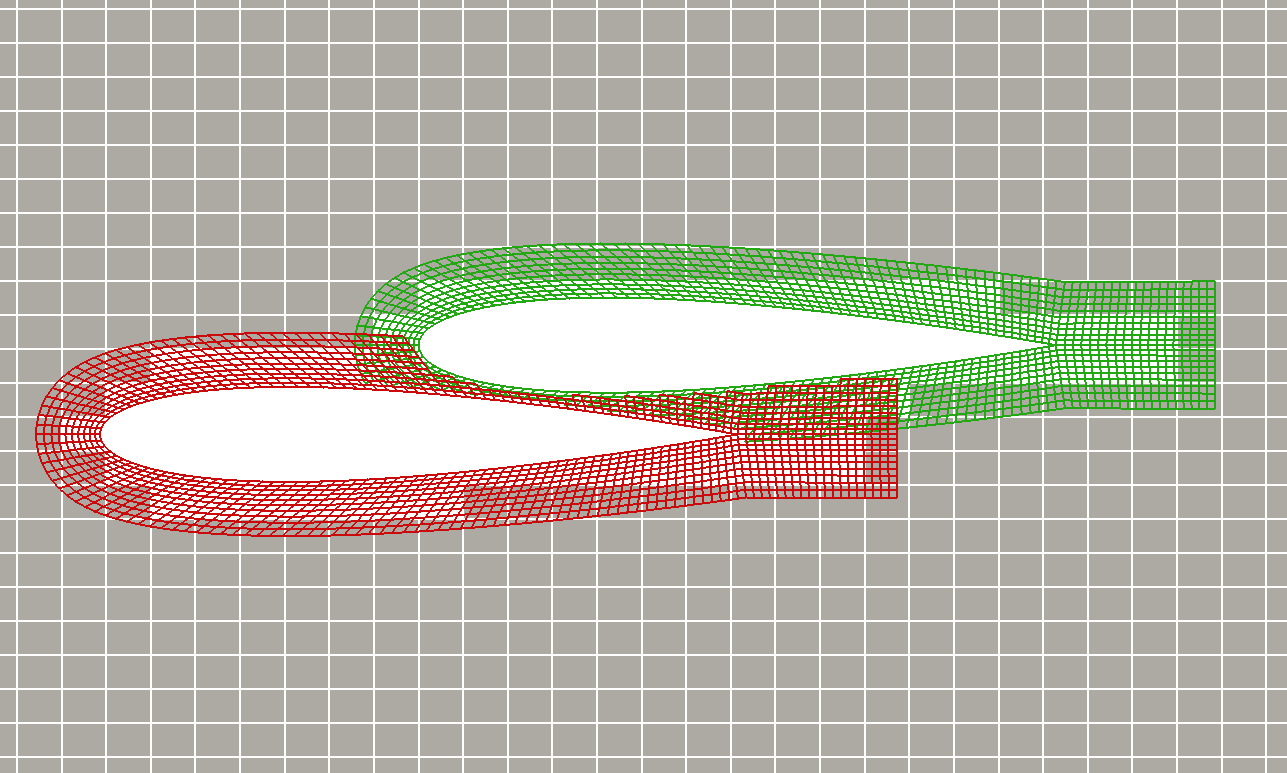}
\caption{Final grid configuration for two NACA airfoils in close proximity.}
\label{fig:doublenaca}
\end{figure}

The next test case consists of two NACA0012 airfoils in close proximity, similar to the type of configuration encountered in high-lift systems with multi-element airfoils or airfoils with slats and flaps.  The airfoils are created as cubically curved quad grids and extruded a small distance in the out-of-plane direction.  The hole cut solution when combined with a simple, coarse Cartesian background grid is shown in Figure \ref{fig:doublenaca}.

\begin{figure}
\centering
\subfloat[TSTO geometry]{\includegraphics[width=.35\textwidth]{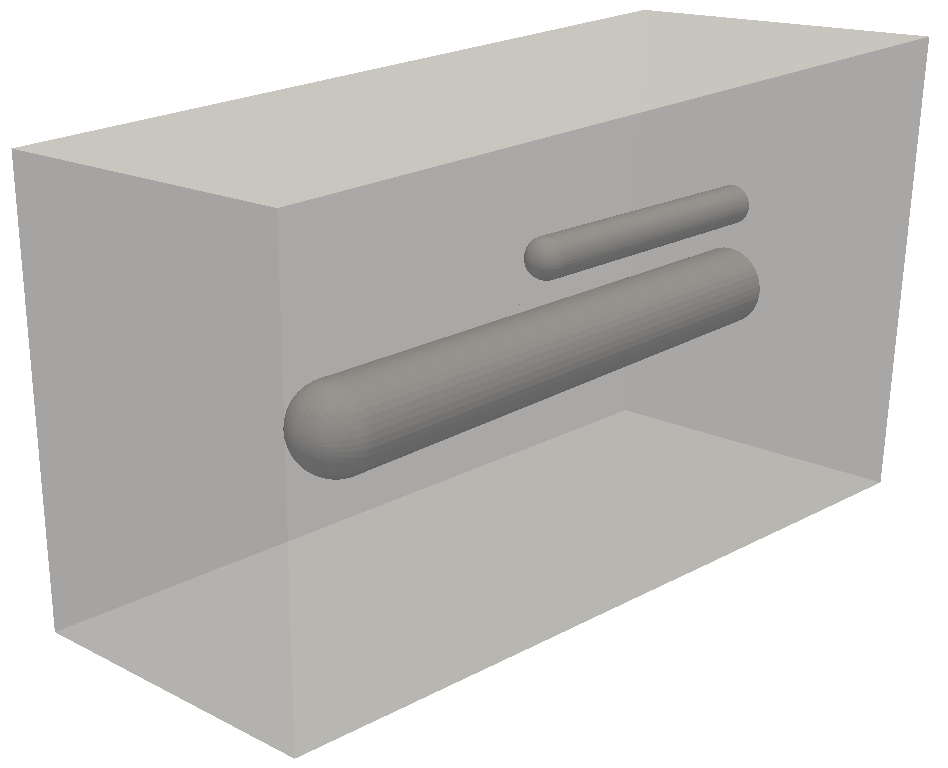}}
\subfloat[Front-view slice]{\includegraphics[width=.22\textwidth]{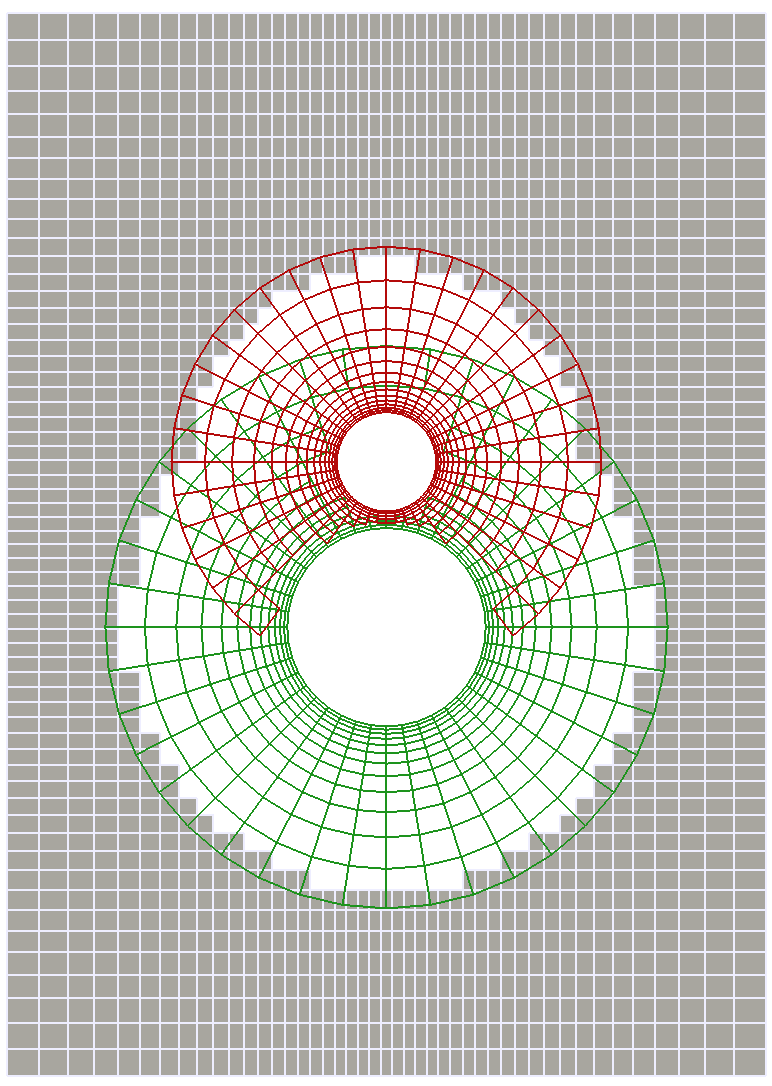}}
\subfloat[Side-view slice]{\includegraphics[width=.37\textwidth]{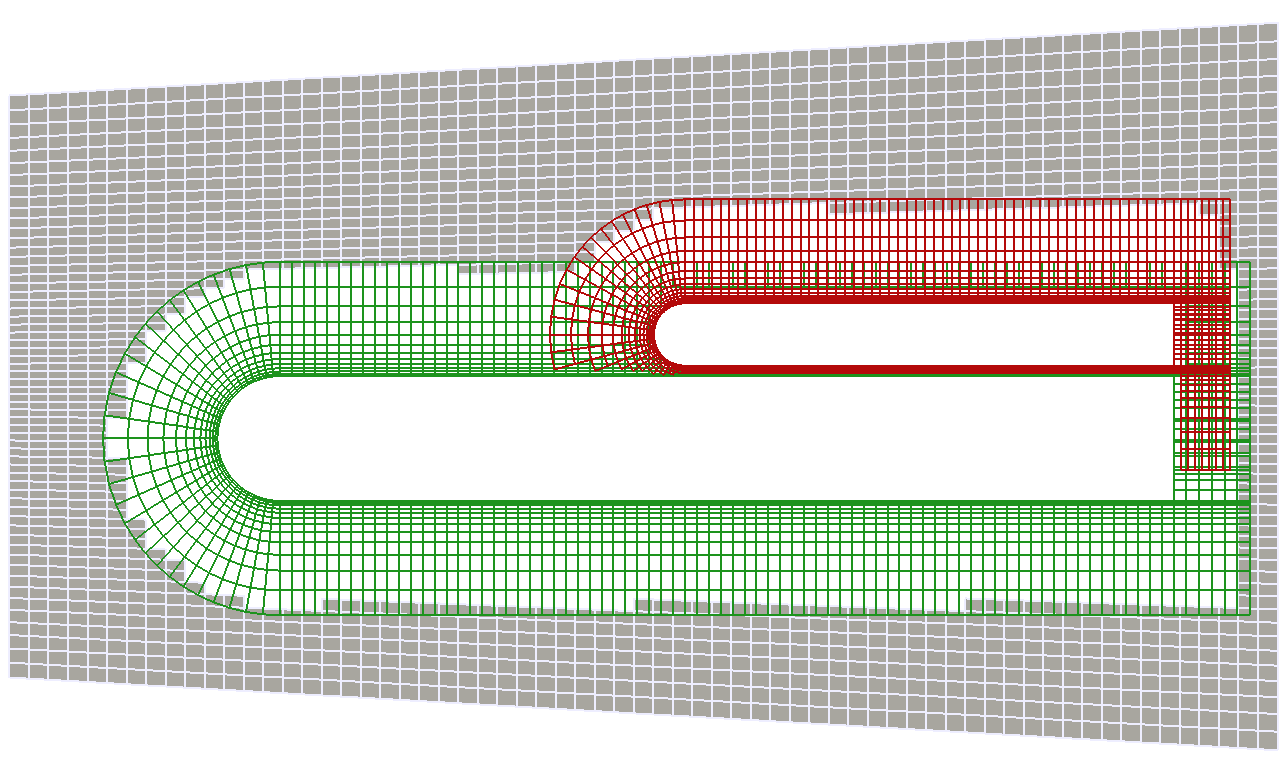}}
\caption{Geometry for the TSTO test case: two hemispherical cylinders and simple background grid.  The slices in (b) in (c) show the cylinders positioned as close as possible while generating a valid grid configuration.}
\label{fig:tsto-geo}
\end{figure}

\begin{figure}
\centering
\subfloat[Final hole cut in background grid]{\includegraphics[width=.45\textwidth]{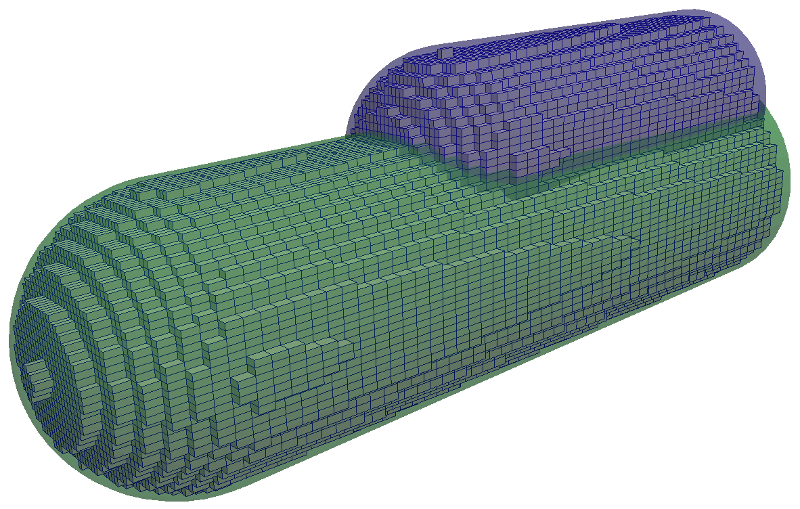}}
\subfloat[Modified boundaries of body grids]{\includegraphics[width=.45\textwidth]{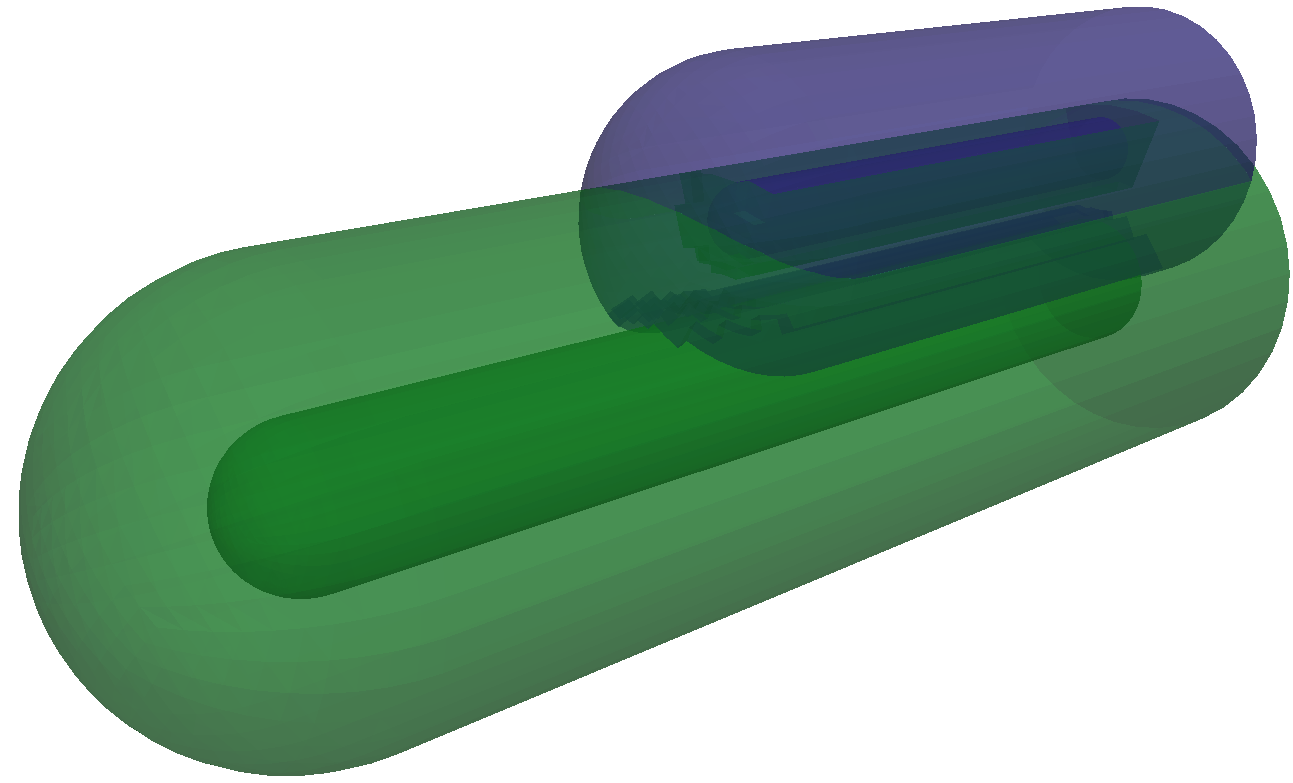}} \\
\subfloat[Hole cut in lower body grid]{\includegraphics[width=.45\textwidth]{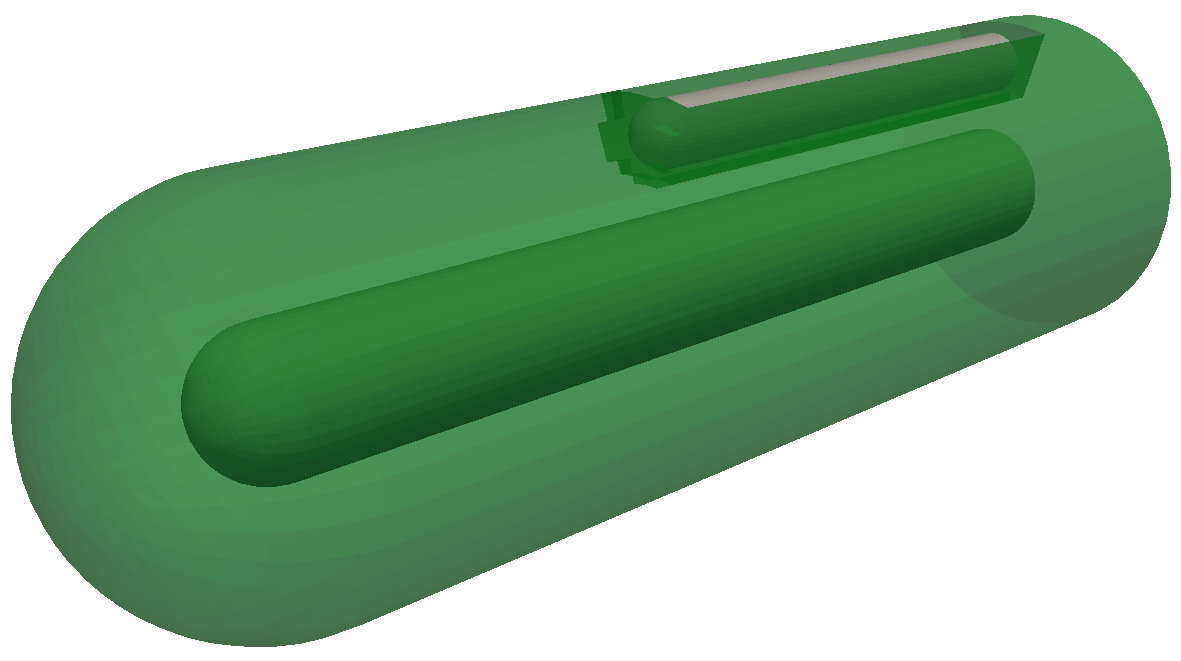}}
\subfloat[Hole cut in upper body grid]{\includegraphics[width=.45\textwidth]{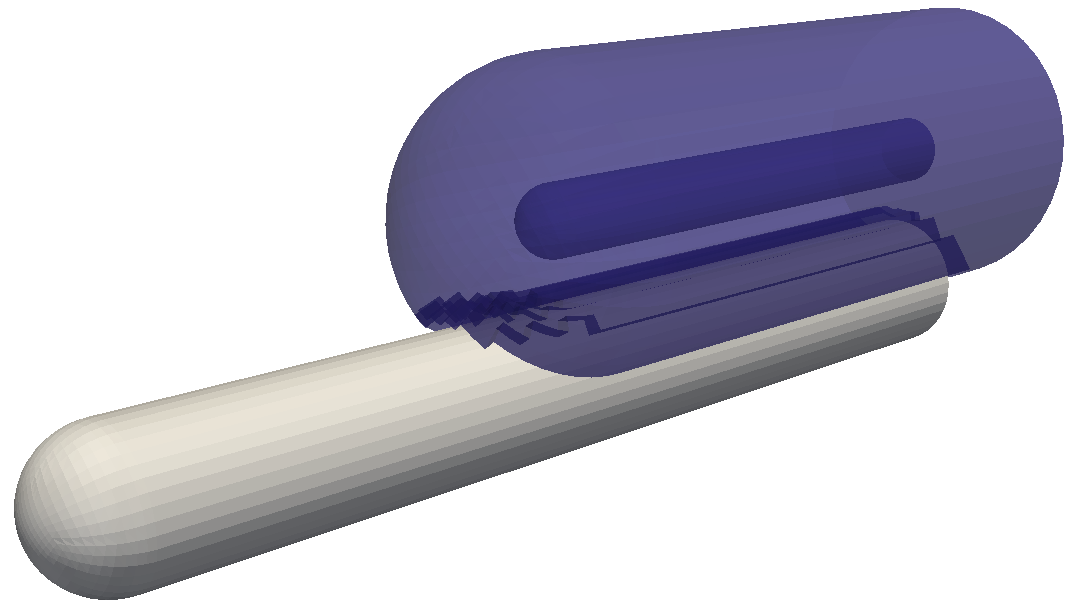}}
\caption{Hole cutting solution for the TSTO case.}
\label{fig:tsto-holes}
\end{figure}

The last geometric test case is a 2-body `Two Stage To Orbit' (TSTO) test case initially created by Yamamoto et al. \cite{yamamoto02}.  We add in a simple background grid to fill out a larger domain.  The test case consists of two hemispherical cylinders of different size in close proximity, and serves to test the ability of the method to handle multiple bodies in close proximity, as well as its ability to handle curved grids.  The body grids use cubically curved hexahedra, while the background grid uses linear hexahedra.  The geometry used is shown in Figure \ref{fig:tsto-geo} along with front-- and side--view slices through the final grid configuration.  3D views of the final hole cutting configuration are shown in Figure \ref{fig:tsto-holes}.  Note that more overlap than strictly necessary is left between the two body grids due to the use of the wall surfaces to perform the cutting between them.

\section{Numerical Experiments}
\label{S:results}

All test cases and results described herein are run on the XStream computing cluster, which utilizes NVIDIA Tesla K80 GPUs.  Each K80 board appears as two logical devices to the host system, and it is these individual devices we will be referring to when discussing the number of GPUs.  All grids are partitioned using METIS \cite{metis}.  We define the number of degrees of freedom (DOF) in any simulation as the total number of solution points multiplied by the number of field variables, which in the case of the three dimensional Navier--Stokes equation is five. In the following section we describe the Taylor--Green and golf ball test cases, and discuss our results in relation to previous studies.  All test cases described in this section were performed with explicit time stepping using a 5-stage, 4th-order, low-storage Runge--Kutta time-integration scheme (RK4(3)5[2R+]C) \cite{kennedy00}.  The time step size is adjusted throughout each simulation by a PI-type controller \cite{butcherODE} with the error estimates based upon the embedded 4th order/3rd order pair.

\subsection{Taylor--Green Vortex}
\label{S:TGV}

\subsubsection{Background}

The first 3D Navier--Stokes test case we shall consider is simulation of the Taylor--Green vortex decay.  It is particularly attractive as a test case due to the lack of boundary conditions, simple initial condition, and availability of spectral DNS data from van Rees et al \cite{vanrees11}.  The annual High-Order Workshop has used it several times as a `difficult' test case, and many groups have submitted results from a variety of high-order methods for comparison \cite{how2,how4}.

The initial conditions for the Taylor--Green vortex test case are specified as
\begin{equation}
\begin{split}
u &= \hphantom{-}U_0 \sin(\frac{x}{L}) \cos(\frac{y}{L}) \cos(\frac{z}{L}), \\
v &= -U_0 \sin(x/L) \cos(y/L) \cos(z/L), \\
w &= 0, \\
p &= P_0 + \frac{\rho_0 U_0^2}{16} ( \cos(\frac{2x}{L}) + \cos(\frac{2y}{L}))(\cos(\frac{2z}{L}) + 2), \\
\rho &= \frac{p}{RT_0},
\end{split}
\end{equation}
where $T_0$ and $U_0$ are constants chosen such that the Mach number based on $U_0$ is $0.1$.  The domain is a periodic cube with extents $-\pi L \leq x,y,z \leq \pi L$ and the viscosity is chosen such the the Reynolds number based upon $U_0$ and $L$ is $1\,600$.  The test case follows the breakdown of the initial condition up to $t = 20t_c$, where the convective time $t_c = L/U_0$.

\subsubsection{Overset Domain Setup}

A total of three simulations were performed (i) a reference run on a single grid, (ii) a run with a translating overset inner grid, and (iii) a run with a rotating overset inner grid.  For the two overset test cases, the inner grid was sized to $40\%$ of the background grid and initially centered at the origin, and number of elements along each direction of the inner grid was taken to be slightly more than $40\%$ of the number used for the background grid.  For the translating overset grid case, the inner grid rigidly translates in a figure-8 pattern defined by the equation
\begin{equation}
\begin{split}
x_0 &= x_0 + A_x \sin(2\pi f_x t), \\
y_0 &= y_0 + A_y (1 - \cos(2\pi f_y t)), \\
z_0 &= z_0 + A_z \sin(2\pi f_z t), \\
\end{split}
\end{equation}
where $A_x = A_z = 6\pi/10$, $A_y = 3\pi/10$, $f_x = f_z = 10 \pi / 169$, and $f_y = 20 \pi / 169$.  The rotating test case kept the inner grid fixed at the origin, and simply applied an angular velocity vector to the center of the inner grid defined by $\omega_x = \omega_y = 10 \pi/169$, and $\omega_z = 10 \pi/338$.  In all cases the simulation was run to a physical time of $169$s using the previously described explicit adaptive time-integration scheme.

\subsubsection{Results}

We are interested in the time histories of both the total kinetic energy and enstrophy integrated over the domain:
\begin{equation}
E_k = \frac{1}{\rho_0 U_0^2 |\Omega|} \int_{\Omega} \rho \frac{\mathbf{v \cdot v}}{2} d \mathbf{x},
\end{equation}
\begin{equation}
\varepsilon = \frac{1}{\rho_0 U_0^2 |\Omega|} \int_{\Omega} \rho \frac{\boldsymbol{\omega} \cdot \boldsymbol{\omega}}{2} d \mathbf{x},
\end{equation}
where $|\Omega| = (2\pi)^3$ is the volume of the domain, $\mathbf{v}$ is the velocity vector, and $\boldsymbol{\omega}$ the vorticity.  The enstrophy is of interest as it relates to the kinetic energy dissipation rate $\epsilon = \frac{dE_k}{dt}$; in incompressible flows they are related as $\epsilon = 2 \frac{\mu}{\rho_0} \varepsilon$.

\begin{figure}
\centering
\subfloat[Inner grid colored by velocity.]{
  \includegraphics[width=.5\textwidth]{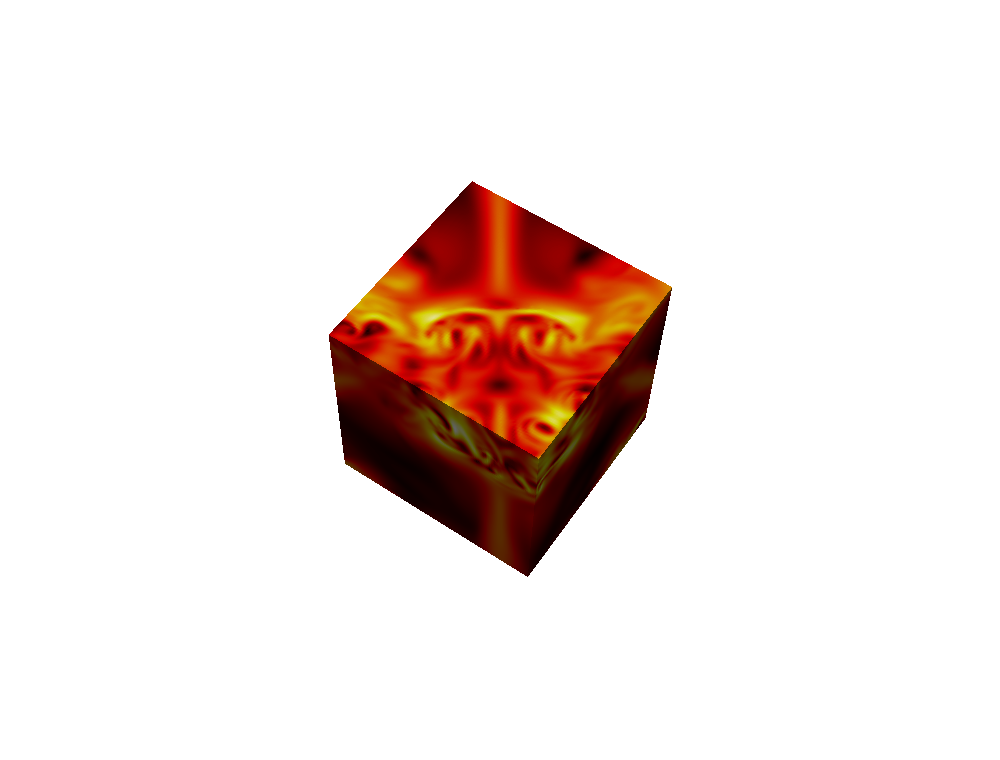}
  \label{fig:tg-vis.a}
}
\subfloat[Background grid with artificial boundaries.]{
  \includegraphics[width=.5\textwidth]{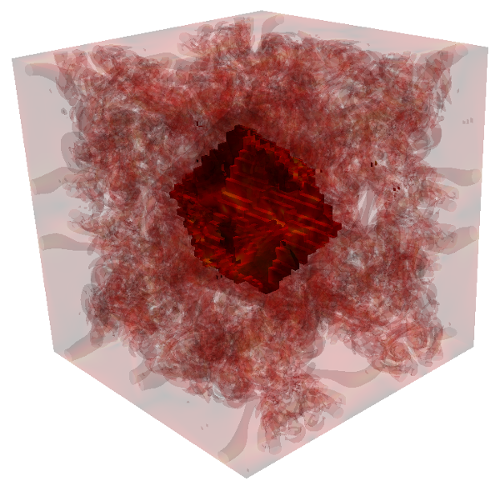}
  \label{fig:tg-vis.b}
}
\caption{Sample visualizations of Q-criterion isosurfaces, along with inner-grid and outer-grid artificial boundary surfaces, colored by velocity magnitude.}
\label{fig:tg-vis}
\end{figure}

\begin{figure}
\centering
\subfloat[]{
  \includegraphics[width=.5\textwidth]{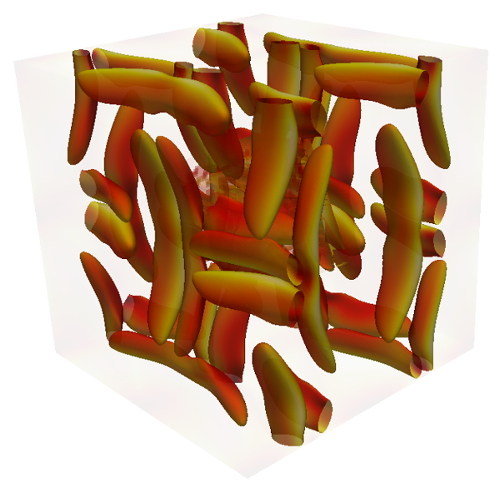}
}
\subfloat[]{
  \includegraphics[width=.5\textwidth]{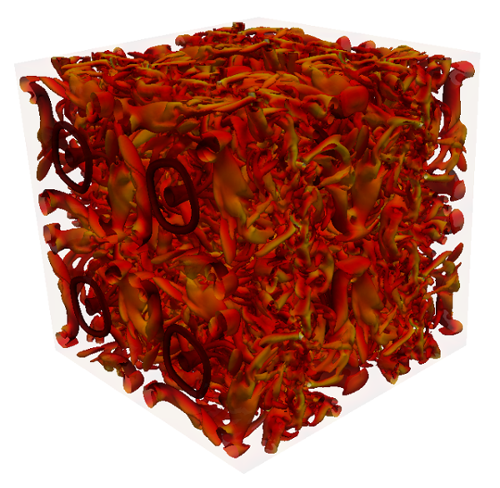}
}
\caption{Visualizations of Q-criterion at two different times. (a) Near $t=0$, (b) at $t \approx 14s$}
\label{fig:tg-vis-2}
\end{figure}

Figure \ref{fig:tg-vis} shows the positions of the grids at one instant of the rotating overset test case; \ref{fig:tg-vis.b} shows the background grid with faint isosurfaces of Q-criterion along with the inner artificial boundary surfaces which have been created through the hole cutting process by the inner grid in the orientation shown in \ref{fig:tg-vis.a}.  Figure \ref{fig:tg-vis-2} show isosurfaces of Q-criterion at two time instants, highlighting the progression of the test case starting from large-scale structures which break down into many small-scale structures which the method is capable of tracking and preserving.

\begin{figure}
\centering
\includegraphics[width=.85\textwidth]{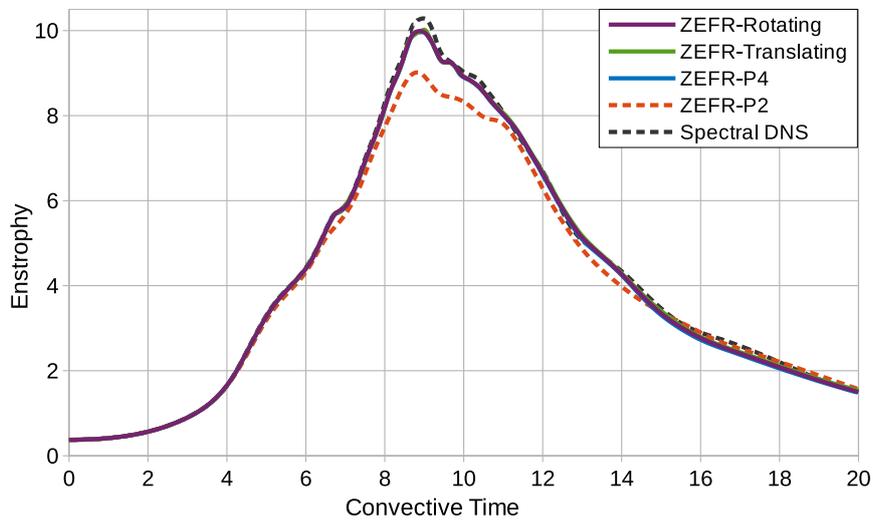}
\caption{Comparison of enstrophy histories between a reference `exact' solution, reference $3$rd order solution, and our current 5th order solutions on a single grid and on moving overset grids, all ZEFR solutions using $5 \cdot 256^3$ DOF.}
\label{fig:tg-enst}
\end{figure}

\begin{figure}
\centering
\includegraphics[width=.7\textwidth]{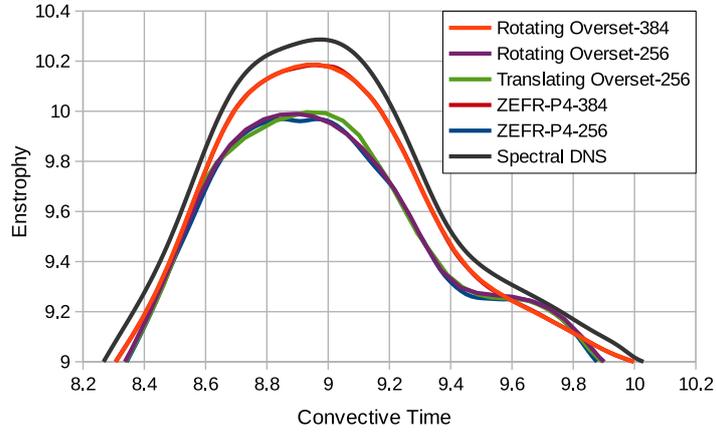}
\caption{Closeup of peak enstrophy for all 5th order $5 \cdot 256^3$ DOF ZEFR test cases and the `exact' solution, along with results from refining the ZEFR cases to $5 \cdot 384^3$ DOF.}
\label{fig:tg-enst-peak}
\end{figure}

\begin{figure}
\centering
\includegraphics[width=.85\textwidth]{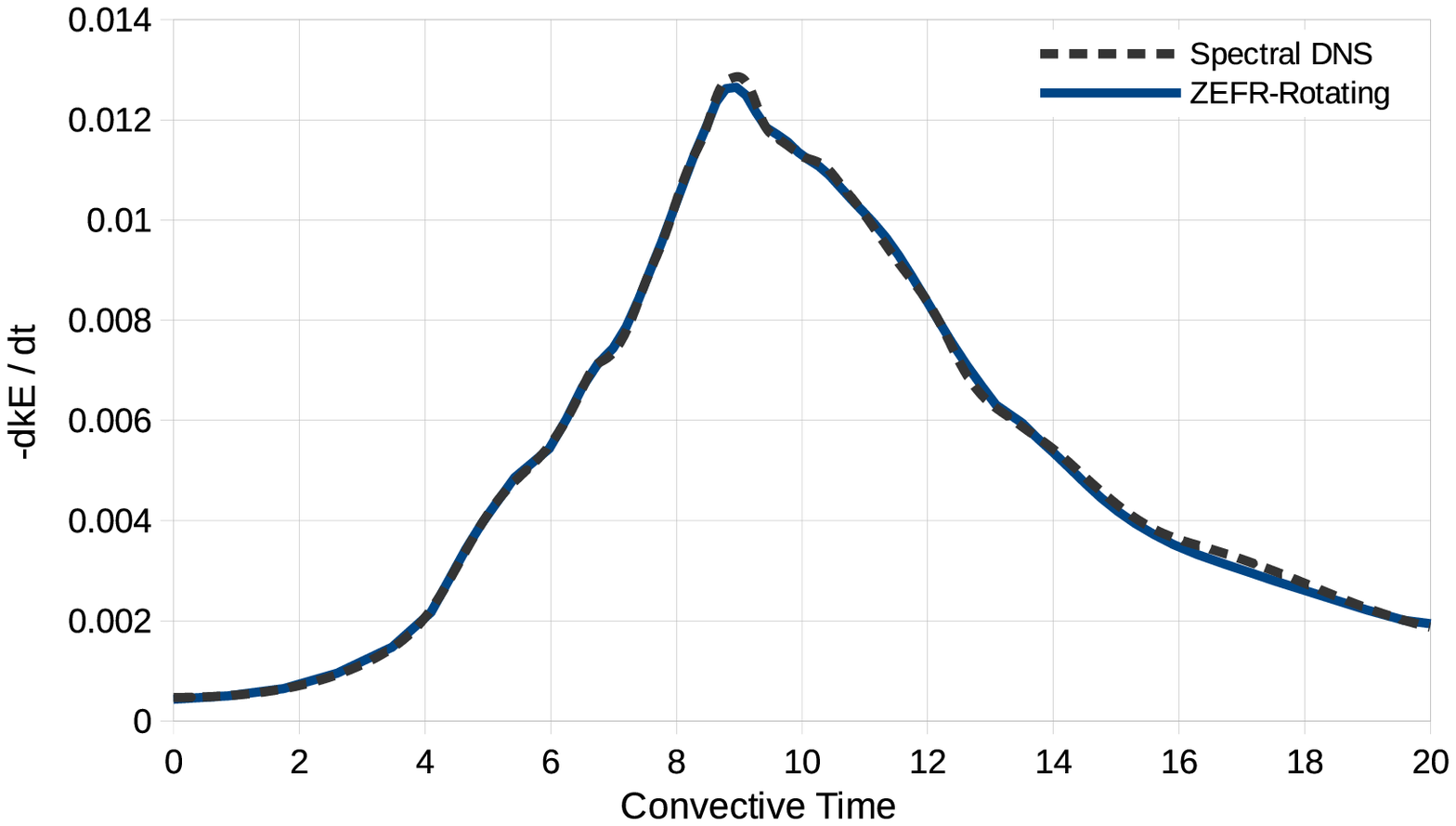}
\caption{Comparison of kinetic energy dissipation rate histories between the reference `exact' solution and the present 5th order $5 \cdot 256^3$ DOF solution on a rotating overset grid system.}
\label{fig:tg-disp}
\end{figure}

Figure \ref{fig:tg-enst} shows the time history of enstrophy for our overset-grid runs, as well as two other solutions for comparison: a spectral DNS solution with $5 \cdot 512^3$ degrees of freedom considered as an `exact' solution \cite{pyfrstar}, and a 3rd order single-grid ZEFR solution.  All of the results shown besides the spectral DNS results use a total of approximately $5 \cdot 256^3$ degrees of freedom.  The moving overset grid results were post-processed by effectively `unblanking' all holes cut in the background grid, which was then processed identically to the single grid case. The results from all cases run with ZEFR are indistinguishable, and agree quite well against the reference solution; Figure \ref{fig:tg-enst-peak} shows a closeup view of the peak enstrophy where this can be seen more easily.  A grid refinement to $5 \cdot 384^3$ DOF is also shown, with the results following more closely to the exact solution as expected.  Figure \ref{fig:tg-disp} provides a similar comparison of the kinetic energy dissipation rate with the $5 \cdot 256^3$ DOF rotating overset case vs. the `exact' solution.  Although the surface interpolation utilized in the artificial boundary approach is not fully conservative, these results show that in practice the error in conservation of quantities such as kinetic energy is negligible.  This agrees with the results of Galbraith et al. \cite{Galbraith14} and Crabill et al. \cite{crabill16}, who showed that conservation error decreases at the same order as solution error, even for moving grids.

\subsection{Golf Ball}
\label{S:golf}

As a test of our method applied to external flow problems, the second 3D Navier--Stokes test case we shall consider is flow over a golf ball.  We first consider a static golf ball, for which numerous computational and experimental studies exist for comparison.  Second, we consider a spinning golf ball to more fully test our method for moving grids and complex fluid dynamics.

The flow physics behind the phenomenon of drag reduction of dimpled spheres---golf balls---has been investigated since at least the 1970s \cite{bearman76,mehta85}.  Early studies relied primarily on wind tunnel experiments, typically collecting force data, and occasionally also performing flow visualizations with, for example, oil streaks. It is only recently, with the advent of large-scale LES and DNS simulations of modest Reynolds numbers, that the accurate, predictive CFD simulation of a golf ball has allowed deeper insight into the effects of dimples.  

The typical goal of a golf ball design is to maximize the range it can be driven in a straight line.  This primarily leads to the desire to reduce its drag as much as possible, with a secondary goal of minimizing variation in side forces to maintain straight--line flight.  Putting backspin on the ball will produce a lift force via the Magnus effect, which extends the flight time and distance of the ball.  Drag reduction is mostly due to the dimples, which are sized to create a series of separation bubbles that will lead to early transition in the unstable shear layer above the bubbles.  The exact size, depth, and arrangement of dimples all contribute to the final aerodynamic characteristics of a golf ball under various conditions.

The rotation of a golf ball is typically defined with the non-dimensional spin parameter $V/U$, where $V$ is the equatorial velocity of the ball and $U$ is the ball's speed of flight.  The typical range of rotation speeds for a realistic golf ball range from $2\,000$ to $4\,000$ rpm (33 to 66 $s^{-1}$), giving spin parameters in the range of $0.1$ to $0.2$ \cite{bearman76, mehta85, aoki10}.  In contrast to smooth spheres, which exhibit a negative Magnus effect near the critical Re \cite{bearman76,muto12}, the forces generated by a spinning golf ball increase monotonically with the spin parameter.

\subsubsection{Previous Studies}

The first noteworthy experimental investigation of the aerodynamics of golf balls under a variety of flow conditions is by Bearman and Harvey in 1976 \cite{bearman76}.  Their study used wind tunnel testing of scaled golf ball models to compare the characteristics of round vs. hexagonal dimples, with a smooth sphere used to assess the validity of their experimental setup.  The hexagonally dimpled ball also had far fewer dimples than the ``conventional" ball ($240$ vs. $330$ or $336$).  They found that the hexagonally dimpled ball had a lower drag coefficient ($C_D$) and higher lift coefficient ($C_L$) over most of the Re and spin rate range of interest, hypothesizing that the hexagonal dimples led to more discrete vortices due to the straight edges of the dimples.  The effect of the dimple edge radius was not studied.  For both dimple types however, they showed that the dimples serve to reduce the critical Re at which a drag reduction occurs, and that the drag coefficient remains nearly constant for a large range of Re after this point.  

Another detailed wind tunnel study was more recently performed by Choi et al. \cite{choi06}.  They studied both fully-dimpled and half-dimpled spheres without rotation.  In comparison to Bearman and Harvey, they used only round dimples with a much smaller depth ($k/d = 4 \cdot 10^{-3}$ for Choi et al. vs. $k/d = 9 \cdot 10^{-3}$ for Bearman and Harvey, where $k$ is the dimple depth and $d$ is the sphere diameter) and also with a larger number of dimples ($392$ vs. approximately $330$).  Their results showed a slightly higher critical $Re$ (${\sim}80\,000$ vs. ${\sim}50\,000$) with a slightly lower $C_D$ (${\sim}0.21$ vs. ${\sim}0.25$) afterwards, with a much more noticeable rise in $C_D$ after the initial drop.  Velocity data collected with a hot-wire anemometer was used to confirm that the turbulence generated by the free shear layer over the dimples led to an increase in momentum near the surface of the golf ball after reattachment, and that the separation angle remained at a constant $110^o$ after the critical $Re$.

Further studies have been performed using a combination of RANS \cite{ting02,ting03}, LES \cite{aoki10, li15}, DNS \cite{smith10}, and wind tunnel experiments \cite{aoki10,chowdhury16}.  Li et al. proposed a link between small-scale vortices created at the golf ball dimples and a reduction in side-force variations at supercritical Reynolds numbers.  Also, both Ting \cite{ting03} and Chowdhury et al. \cite{chowdhury16} have shown a positive correlation between dimple depth the supercritical drag coefficient, and a negative correlation between dimple depth and critical Reynolds number. Aoki et al. performed wind tunnel experiments on static and spinning golf balls and showed a positive correlation between lift force and spin rate (though smaller in magnitude than the negative lift force generated by a similar smooth sphere), and a slight correlation between drag and spin rate as well.

\subsubsection{Golf Ball Geometry}
\label{S:golf-geo}

\begin{figure}
\subfloat[Side View.]{
  \includegraphics[width=.33\textwidth]{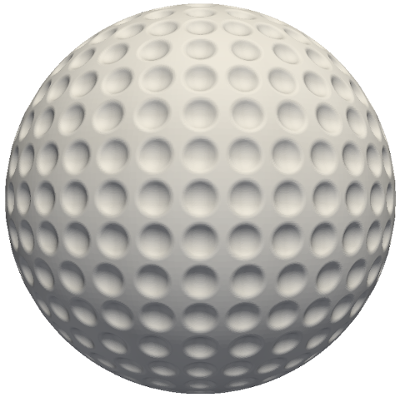}
  \label{fig:golf-side}
}
\subfloat[Top View.]{
  \includegraphics[width=.33\textwidth]{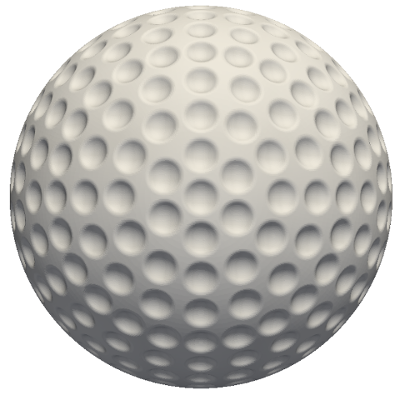}
  \label{fig:golf-top}
}
\subfloat[Closeup of mesh in a dimple.]{
  \includegraphics[width=.33\textwidth]{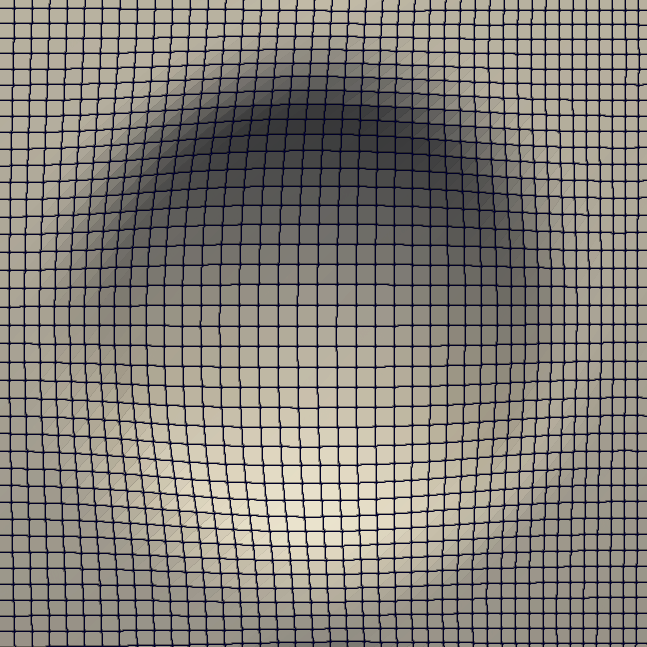}
  \label{fig:dimple-mesh}
}
\caption{Golf ball surface}
\label{fig:golf-surf}
\end{figure}

In this study, the golf ball surface geometry was created as a parameterized CAD model with 19 rows of circular dimples (9 rows per hemisphere + 34 dimples around centerline), for a total of $388$ dimples, as shown in Figure \ref{fig:golf-surf}.  The golf ball diameter is $42.7$mm, the dimple depth is $6.41 \cdot 10^{-4}$m ($k/D = .015$), and the dimple diameter is a constant $2.99$mm ($c/D = 7.0 \cdot 10^{-2}$).  The dimple edges are filleted with a radius of $0.75$mm.  The surface was exported in the STL format and used within the multiblock structured mesh generator GridPro \cite{GridPro} to create a spherical grid with a boundary layer.  The surface of the golf ball was divided into $24$ roughly square regions, each with a resolution of $144 \times 144$ quadrilaterals, with $60$ layers in the radial direction, for a total of $29\,859\,840$ linear hexahedra, or $1\,105\,920$ cubically curved hexahedra after agglomeration.  The first cell height was chosen to be at an estimated $y^+$ value of 6.667 ($3.4 \cdot 10^{-5}$m), the first $18$ layers were held to a constant thickness, and the remaining $42$ layers were allowed to grow out to a final outer diameter of $31.82$mm.  The first cell height was chosen so that once the first three layers of cells near the wall were agglomerated into cubically-curved hexahedra and run with 4th order tensor-product solution polynomials), the first solution point inside the element would lie at a $y^+$ of approximately 1. The mesh was output in the CGNS structured multiblock format and imported into HOPR (High-Order Pre-Processor) \cite{HOPR}, a utility which can agglomerate the cells of a structured mesh into high-order curved  hexahedra.  The new high-order mesh, in an HDF5-based HOPR-specific format, was then converted using PyFR into the PyFR mesh format \cite{pyfr}, which ZEFR has the capability to read.

This pseudo-structured golf ball grid was then combined with a mostly Cartesian background grid created in Gmsh \cite{gmsh} to fill the desired extents of the full computational domain.  The box has a width and height of $0.6832$m (16 times the golf ball diameter $D$), and length $1.0248$m ($-12D$ to $12D$).  A refined region was created in the area to be occupied by the golf ball, with a refined wake region stretching out to the rear of the domain for a total of $715\,750$ linear hexahedra elements.

\subsubsection{Static Golf Ball}
\label{S:golf-static}

\begin{figure}
\centering
\includegraphics[width=.85\textwidth]{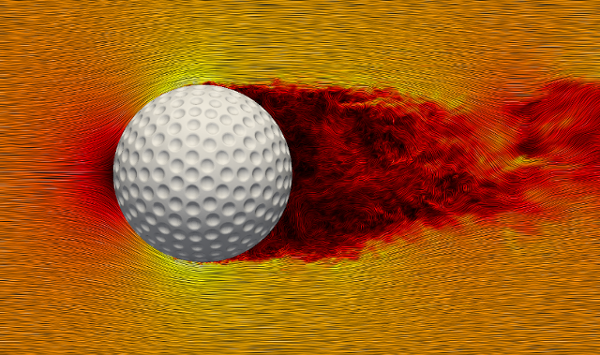}
\caption{View of approximate streamlines and velocity magnitude field through the $y=0$ plane (golf ball centerline).}
\label{fig:vel-static-1}
\end{figure}

\begin{figure}
\centering
\subfloat[Time history of force coefficients]{
\includegraphics[width=.6\textwidth]{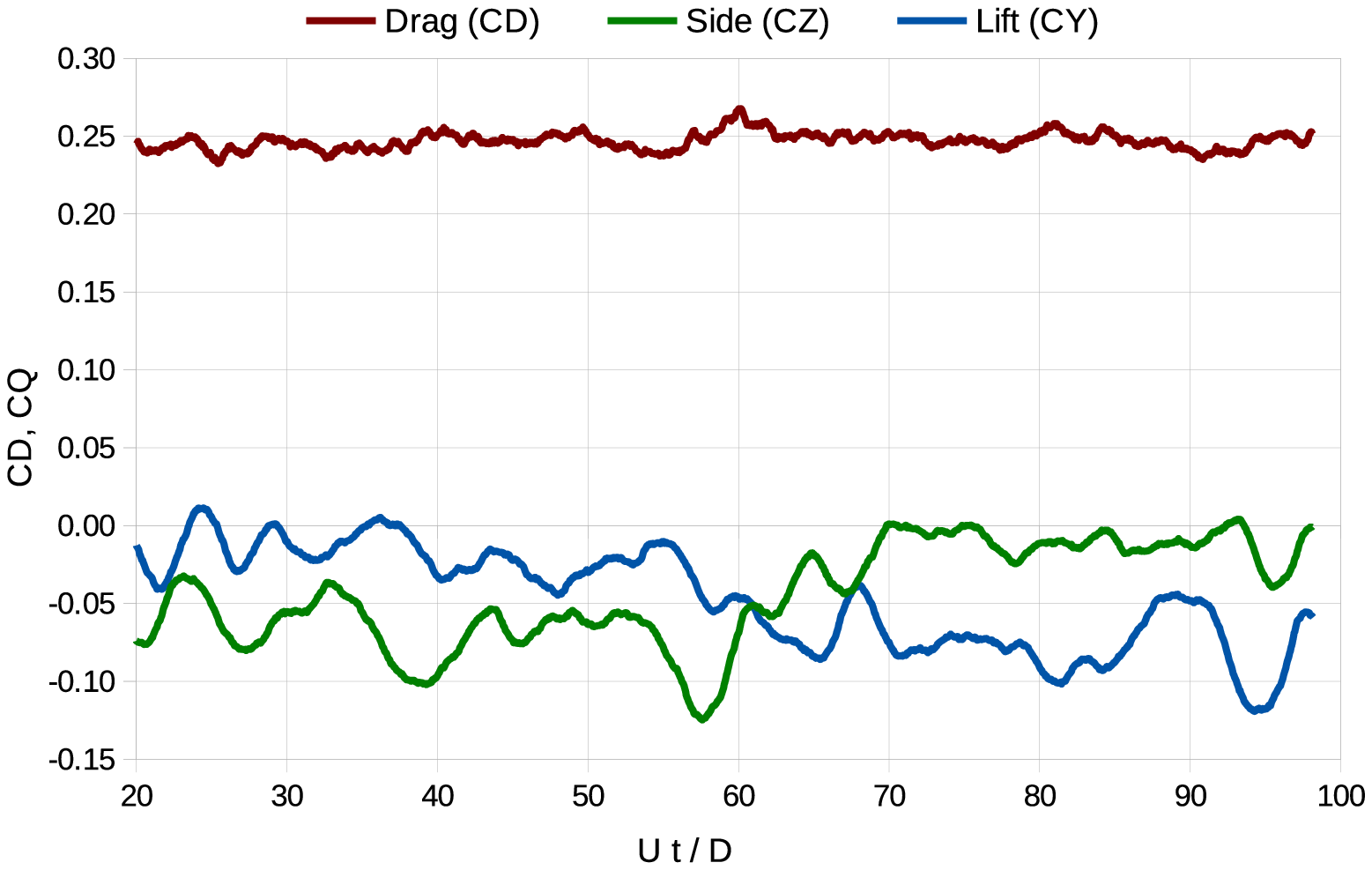}
\label{sfig:force-static-hist}
}
\subfloat[Side forces polar]{
\includegraphics[width=.35\textwidth]{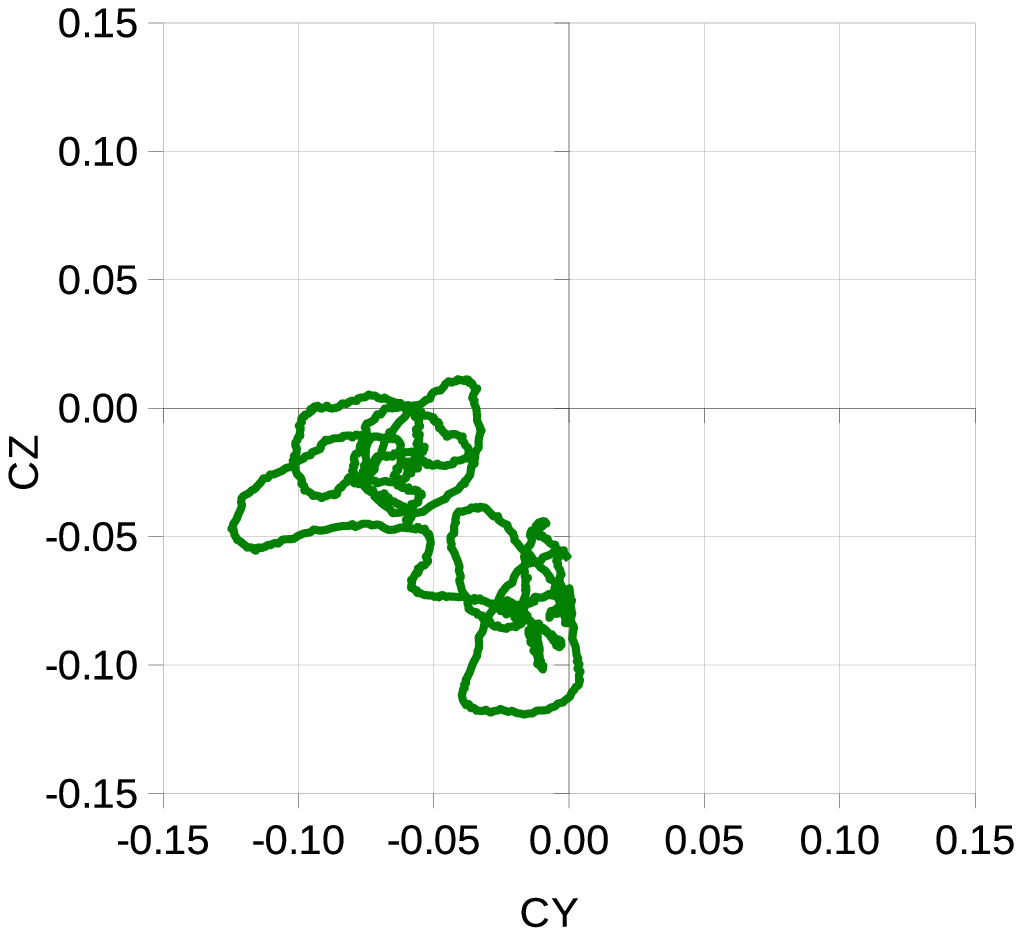}
\label{sfig:force-static-polar}
}
\caption{Force coefficients for the static golf ball simulation.}
\label{fig:forces-static}
\end{figure}

The simulation was advanced in time using the same adaptive RK54[2R+] scheme as before.  The flow is along the $x$-axis, with a Reynolds number of $150\,000$ based upon the golf ball diameter of $0.0427$m, and a Mach number of $0.2$.  The full physical freestream conditions used (scaled such that the freestream velocity is 1) are shown in Table \ref{tab:golf-fs}.  An instantaneous view of velocity contours and approximate streamlines in the mid plane of the ball are shown in Figure \ref{fig:vel-static-1}.  The time histories of the drag and both side forces are shown in Figure \ref{sfig:force-static-hist}, and a polar plot of the two side forces $C_Y$ and $C_Z$ are shown in Figure \ref{sfig:force-static-polar}.

\begin{table}
\centering
\caption{\label{tab:golf-fs}Simulation conditions for all golf ball test cases.}\vskip12pt
\begin{tabular}{r   l | r  l}
\toprule
Reynolds number &  $150\,000$    & $\rho$ &  $1.0$kg/$\text{m}^3$ \\
Mach     &  $0.2$         & $V$    &  $1.0\,$m/s \\ 
Prandtl  &  $0.72$        & $P$    &  $17.85714286\,$Pa \\
$\gamma$ &  $1.4$        & $R$    &  $17.85714286\,$J/(Kg K) \\
$L$      &  $0.0427\,$m  & $T$    &  $1\,$K \\ \bottomrule
\end{tabular}

\end{table}

\begin{figure}
\centering
\subfloat[Time history of force coefficients]{
\includegraphics[width=.6\textwidth]{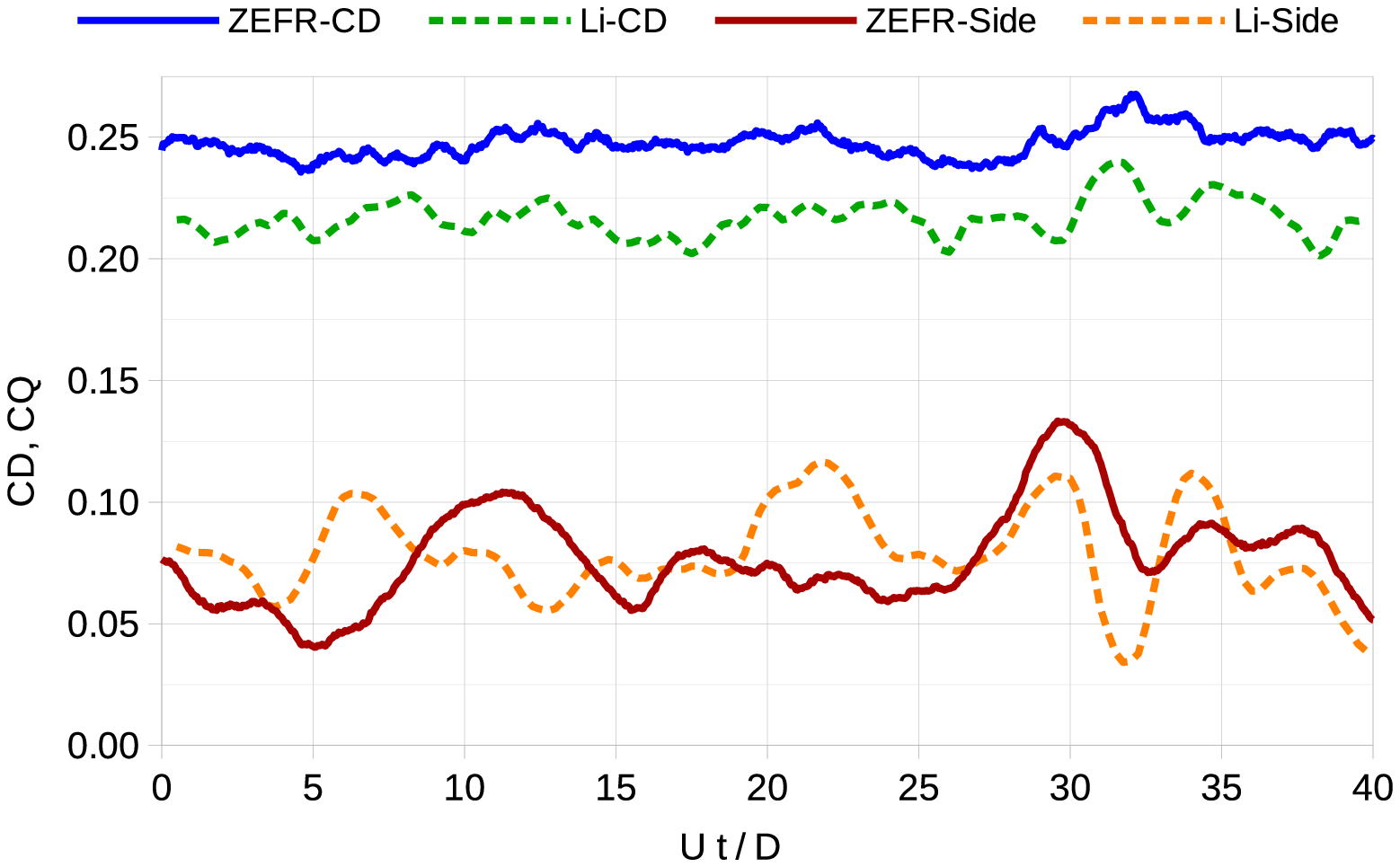}
\label{sfig:force-hist-compare}
}
\subfloat[Side forces polar]{
\includegraphics[width=.35\textwidth]{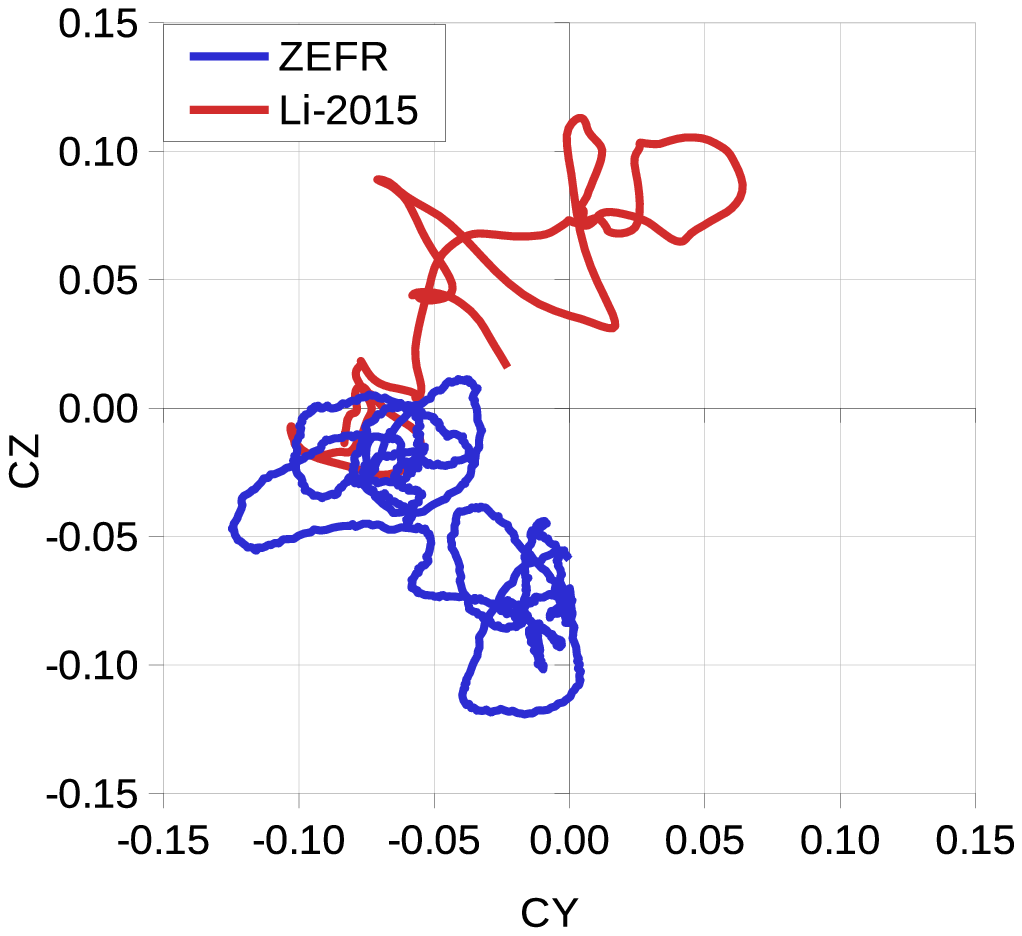}
\label{sfig:force-polar-compare}
}
\caption{Force coefficients for the static golf ball simulation, compared with those produced by Li et al. under similar conditions.}
\label{fig:force-compare}
\end{figure}

As a verification that our results are correct, our force coefficient histories are also plotted alongside those generated by Li et al. \cite{li15} for a very similar case in Figure \ref{fig:force-compare}.  The conditions for their study were $Re = 110\,000$ incompressible flow; a lower Reynolds number than that used here, but still corresponding to the transcritical regime where the drag coefficient should remain nearly constant.  A second-order finite-volume LES solver was utilized for their simulation, using implicit time-stepping.  Their golf ball had $392$ dimples with a dimensionless diameter $c/D = 9.0 \cdot 10^{-2}$ and depth $k/D = 0.5 \cdot 10^{-2}$.  An unstructured hexahedral grid was used with a total of approximately $1.45 \cdot 10^6$ elements in the domain, with overall extents $-13D \leq x \leq 13D$ and $-5.6D \leq y,z \leq 5.6D$.  

Since the dimples primarily change the drag, it would be expected that the side forces should be quite similar between the two cases. Indeed, that is the case as shown in Figure \ref{fig:force-compare}; the average and standard deviation of the side force histories are nearly identical.  The side-force polar plot shows this as well; the two studies show similar trajectories, simply offset by a rotation about the axis of the flow. The present study was run for much longer ($100$ passes vs. $40$), leading to a more visibly bimodal polar plot, but the trends remain the same.  The drag histories are also in agreement; the offset between the two is to be expected, as the dimple depth used here is far greater than the dimple depth used by Li et al.  Results from a variety of studies have shown a direct correlation between dimple depth and supercritical drag coefficient, along with an inverse correlation to the critical Reynolds number.

\begin{figure}
\centering
\subfloat[]{
\includegraphics[width=.3\textwidth]{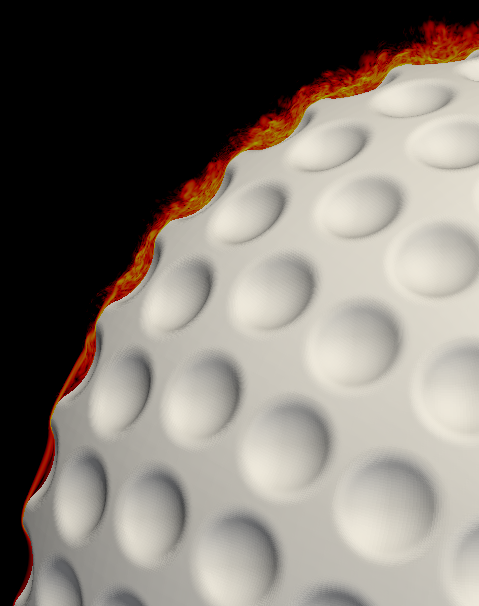}
}
\subfloat[]{
\includegraphics[width=.67\textwidth]{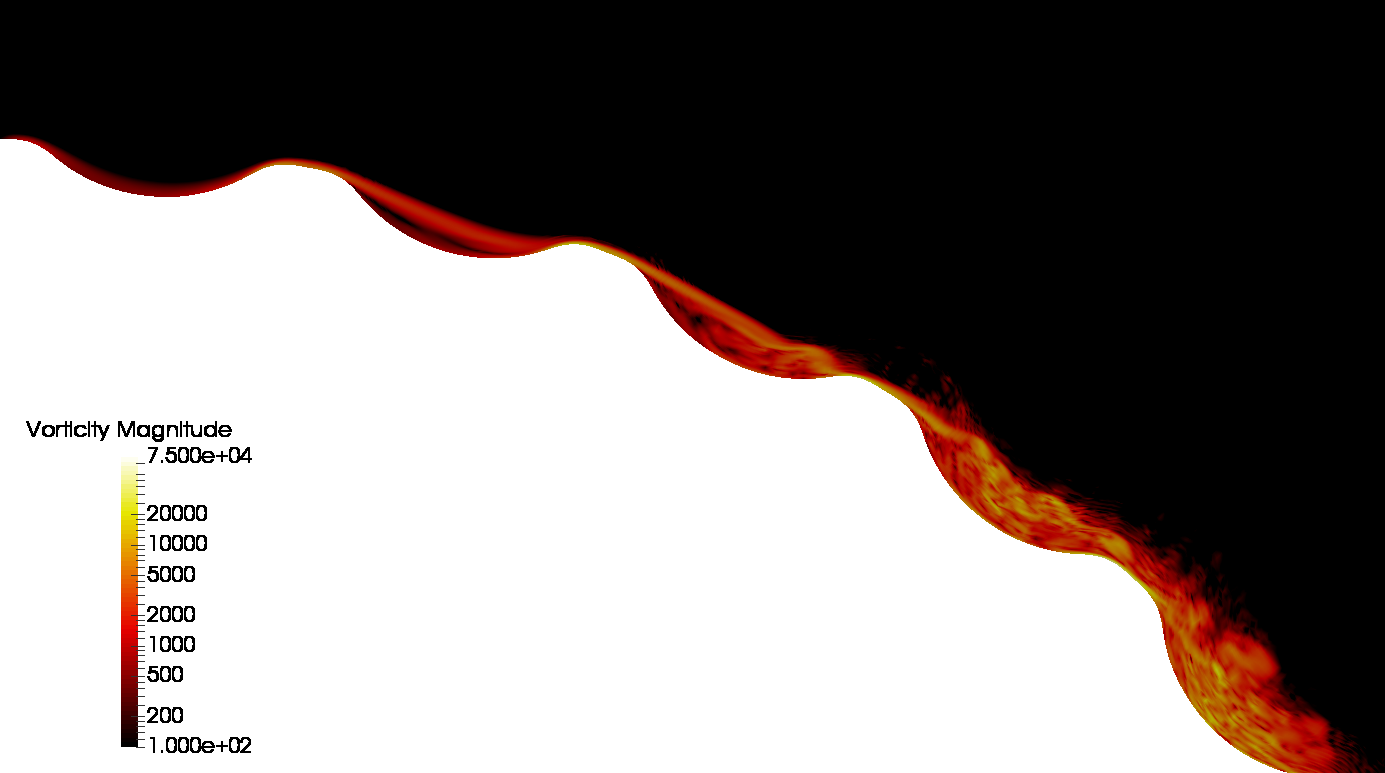}
}
\caption{Closeup view of log of vorticity magnitude through the $y=0$ plane showing boundary layer transition occurring in the shear layer above a dimple.  In (b), the stagnation point is at the top left of the image.}
\label{fig:vort-static-1}
\end{figure}

\begin{figure}
\centering
\subfloat[]{
\includegraphics[width=.5\textwidth]{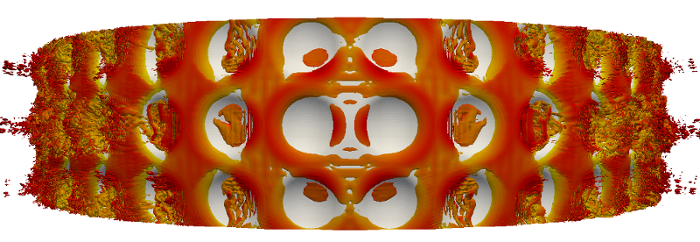}
} \\
\subfloat[$z > 0$]{
\includegraphics[width=.44\textwidth]{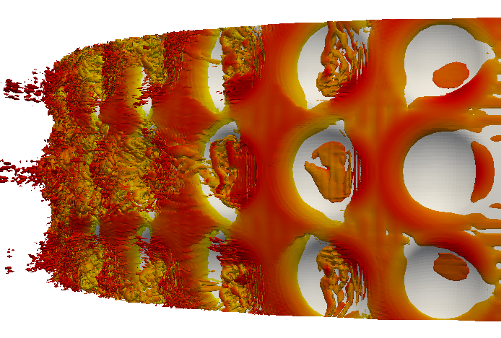}
\label{fig:q-vort-a}
}
\subfloat[Closeup on transition areas. ($z<0$)]{
\includegraphics[width=.53\textwidth]{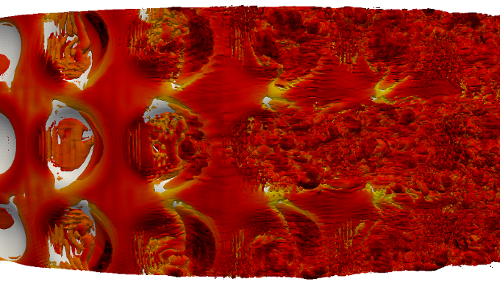}
\label{fig:q-vort-b}
}
\caption{Isosurfaces of Q-criterion colored by log of vorticity magnitude in a strip along the $y=0$ centerline of the golf ball.  Note that the location of transition is clearly visible as starting from the instabilities in the free shear layer over the dimples not far from the stagnation point.}
\label{fig:q-vort-1}
\end{figure}

\begin{figure}
\centering
\includegraphics[width=.45\textwidth]{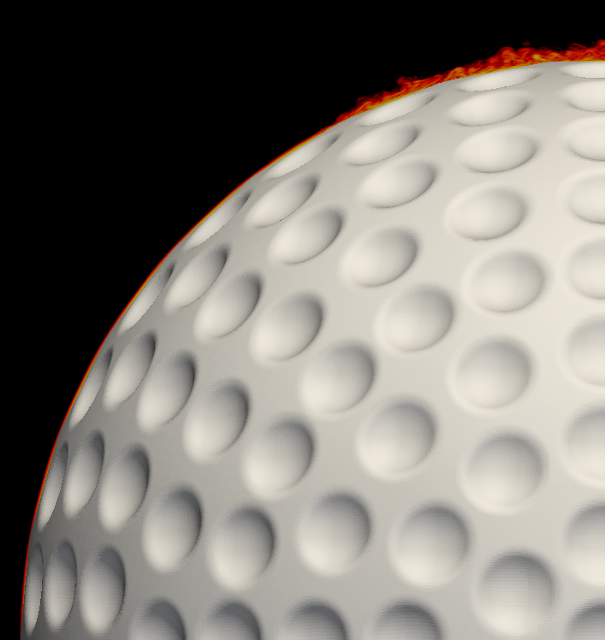}
\caption{Log of vorticity magnitude shown along the centerline between two rows of dimples close to $y=0$.  Note that the flow between the dimples remains laminar until the point of transition clearly visible after the 6th dimple from the stagnation point.}
\label{fig:cent-trans}
\end{figure}

In Figures \ref{fig:vort-static-1} and \ref{fig:q-vort-1}, we confirm the results of others in showing that the mechanism of transition is the growth of instabilities in the shear layer which forms over the recirculation regions inside the dimples not far from the stagnation point.  Figure \ref{fig:vort-static-1} shows contours of instantaneous vorticity magnitude through the $y=0$ plane (mid plane of the golf ball).  Separation and reattachment can be seen in dimples near the stagnation point, then near the top of the image, the shear layer breaks down and becomes turbulent.  Figure \ref{fig:q-vort-1} shows the same from a view above the centerline of the golf ball, with the three lines of dimples around $y=0$ shown looking down at the stagnation point.  Figure \ref{fig:q-vort-a} shows the strip for $z>0$ (the flow is roughly radially symmetric from the stagnation point).  The dimples near the stagnation point have well-defined reattachment areas; but when the flow reaches the next dimple, instabilities are visible in the shear layer over the dimple, clearly seen in Figure \ref{fig:q-vort-b}.  Over the 4th row of dimples (beginning at $31^\circ$ from the stagnation point), the shear layer has broken down and become fully turbulent.  The turbulent region then spreads out downstream from each dimple until the entire flow becomes turbulent by ${\sim}63^\circ$.  Figure \ref{fig:cent-trans} shows the point at which the smooth line between two rows of dimples becomes turbulent.

\subsubsection{Spinning Golf Ball}
\label{S:golf-spin}

In order to fully test the effectiveness of our moving-grid overset capability, we next move on to the case of a spinning golf ball.  We keep the golf ball fixed at the origin, but apply a constant rotation rate around the $z$-axis; to fall in line with other studies, we choose a non-dimensional spin rate $\Gamma = \omega r / U_{inf} = 0.15$.  All other physical flow parameters are left the same.  As discussed in Section \ref{S:motion}, we simplify our handling of moving grids through use of rotation matrices to map between the updated and original positions of the golf ball grid.  To handle arbitrary rigid-body motion, however, the full 6 DOF equations of rigid-body motion are integrated in time to keep track of the current translation and rotation of the inner golf ball grid (although the translation is not modified here, in the future the calculated surface forces on the golf ball could be integrated in time for a full 6 DOF simulation).

\begin{figure}
\centering
\subfloat[Time history of force coefficients for the spinning golf ball.]{
\includegraphics[width=.6\textwidth]{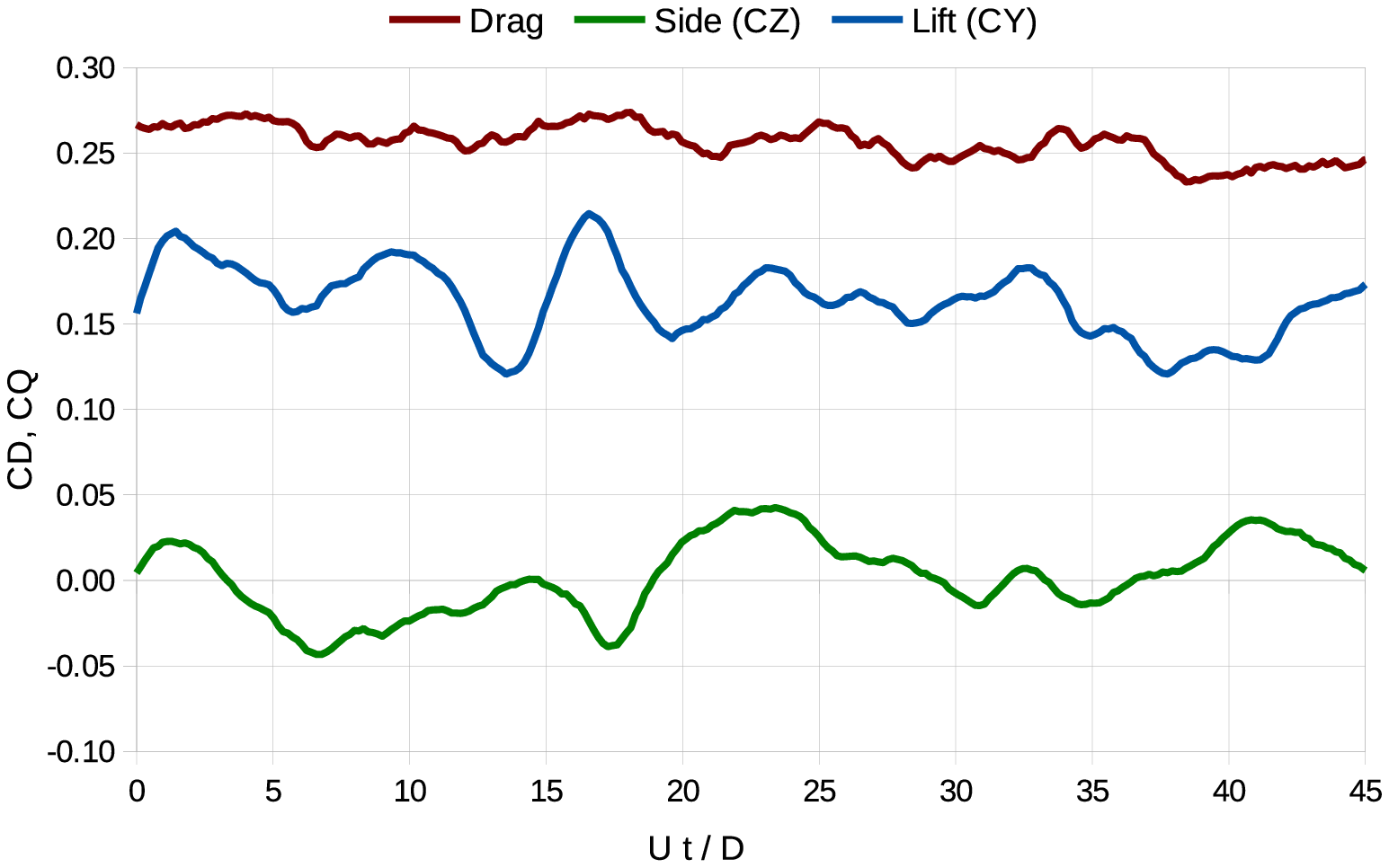}
\label{sfig:force-spin-hist}
}
\subfloat[Side forces polar.]{
\includegraphics[width=.35\textwidth]{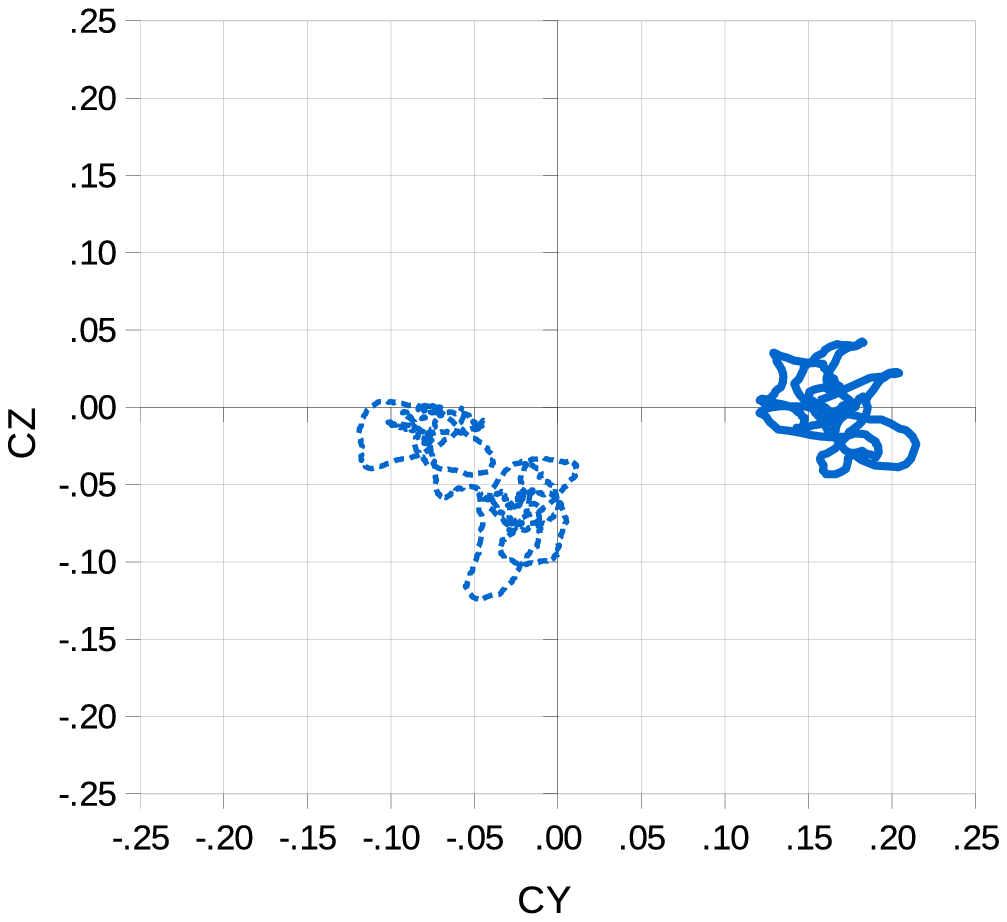}
\label{sfig:force-spin-polar}
}
\caption{Force coefficients for the spinning golf ball simulation.  The dashed curve in (b) is a re-plot of static golf ball forces.}
\label{fig:force-spin}
\end{figure}

\begin{figure}
\centering
\includegraphics[width=.8\textwidth]{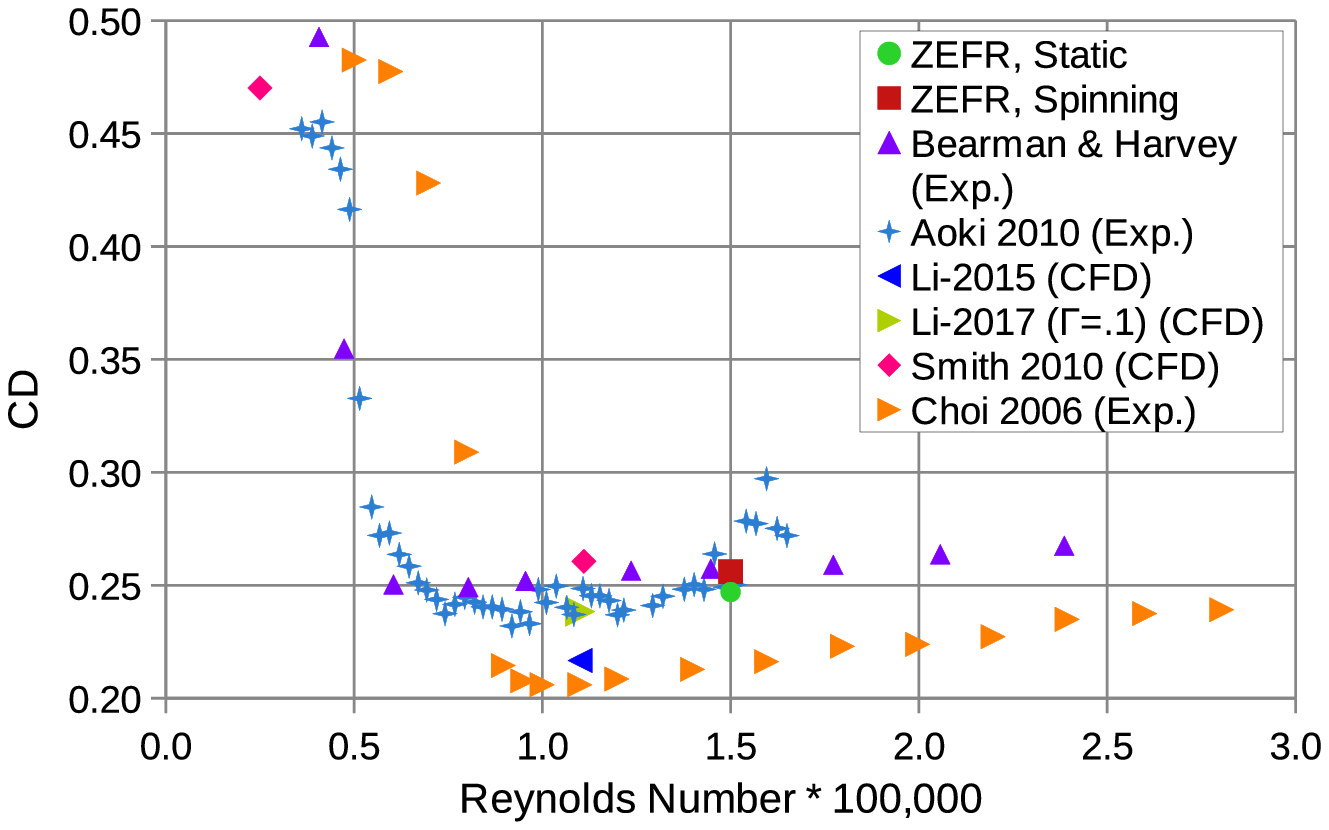}
\caption{Comparison of $C_D$ values from several experimental and computational studies alongside the present results produced with ZEFR.}
\label{fig:cd-study-compare}
\end{figure}

Our average $C_D$ values are compared against the results from a number of other studies, both experimental and computational, in Figure \ref{fig:cd-study-compare}.  As expected from previous literature, the spin induces a slightly higher drag coefficient than the static case but imparts a more regular variation in side forces upon the golf ball; the averages for all force coefficients (with standard deviations) are summarized in Table \ref{tab:forces} and the time history is shown in Figure \ref{fig:force-spin}.  While the out-of-plane side force ($C_Z$) hovers near zero, the lift ($C_L$ or $C_Y$) hovers around a value of $0.16$, with moderate variations in time.  However, looking at a polar plot of the side forces, the oscillations are far more constrained than in the static case, where the axially symmetric nature of the flow allows the wake to oscillate randomly with no preferred direction.  In addition to providing a sizable lift force, the spin has the effect of imposing some structure and a more preferred direction to the oscillations of the wake.

\begin{table}
\centering
\caption{Summary of average force coefficients for the static and spinning golf balls; $C_Q$ refers to the combined magnitude of the lift and side forces $C_Y$ and $C_Z$.  Present results compared to the similar study from Li et al.}\vskip12pt
\begin{tabular}{cccc}
\toprule
 \;    & Static                        & Spinning ($\Gamma = .15$) & Li 2015 (Static) \\ \midrule
$C_D$  & $\hphantom{-}0.2469 \pm 0.005$ & $0.256 \pm 0.010$ & $\hphantom{-}0.217 \pm 0.008$ \\ 
$C_Q$  & $\hphantom{-}0.076 \pm 0.020$  & $0.165 \pm 0.021$ & $\hphantom{-}0.079 \pm 0.019$ \\ 
$C_Y$  & $-0.047 \pm 0.032$             & $0.164 \pm 0.021$ & $           -0.029 \pm 0.045$ \\
$C_Z$  & $-0.044 \pm 0.032$             & $0.002 \pm 0.022$ & $\hphantom{-}0.046 \pm 0.040$ \\ \bottomrule
\end{tabular}
\label{tab:forces}
\end{table}

\begin{figure}
\centering
\includegraphics[width=.82\textwidth]{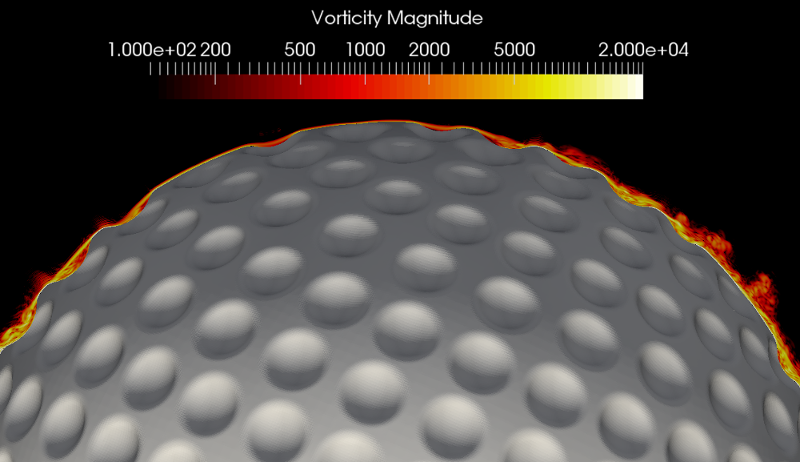}
\caption{Closeup view of $z=0$ centerplane showing log of vorticity magnitude.  The stagnation point is at the top of the image and the ball is spinning counter-clockwise.}
\label{fig:vort-trans-spin}
\end{figure}

\begin{figure}
\centering
\includegraphics[width=.9\textwidth]{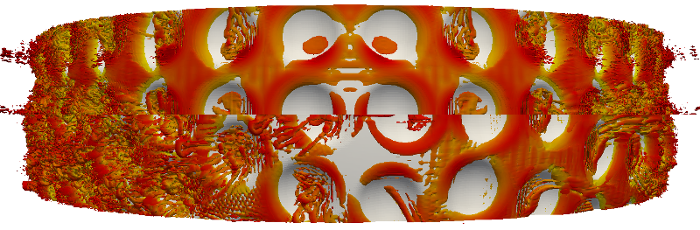}
\caption{Isosurfaces of Q-criterion colored by log of vorticity magnitude along the centerline of the golf ball. Top: Static golf ball, bottom: spinning golf ball at $T=49s$}
\label{fig:q-front-compare}
\end{figure}

While the overall flow features are quite similar from the static to the spinning golf ball, a few differences can clearly be seen.  Figure \ref{fig:vort-trans-spin} shows how the locations of transition shift; Figure \ref{fig:q-front-compare} likewise shows a direct comparison between flow features near the stagnation points of the static and spinning cases.  On the advancing side of the stagnation point, both figures show that the increased relative velocity over the surface of the golf ball lead to earlier transition; conversely, on the retreating side, the transition is slightly delayed relative to the static case.

These are believed to be the first high order simulations of static and spinning golf balls, and represent an advance in the state of the art in both scale-resolving CFD and overset grid calculations.  Many interesting applications are now within reach of high-fidelity simulation, including high-lift systems, turbomachinery, and a variety of multicopters and small-scale unmanned aerial vehicles (UAVs).

\section{Computational Performance}
\label{S:perf}

It is important to evaluate the computational efficiency of our method, as high accuracy is useless if the method is too expensive to produce a result in a reasonable amount of time.  Therefore, in this section, detailed profiling results and performance metrics will be presented on real test cases --- specifically, the preceding test cases of the Taylor--Green vortex and the static/spinning golf ball.

All cases in this section have been run on the Stanford Research Computing Facility's XStream supercomputer, a Cray CS-Storm GPU compute cluster.  XStream has a 1.0 petaflops total computing capability comprised of dense nodes with 8 x NVIDIA Tesla K80 GPUs (16 logical GPUs), 2 x 10-core Intel Xeon E5-2680 v2 CPUs, and 1 x FDR Infiniband card.  CUDA-aware MPI was enabled to allow the MPI distribution to operate directly on the memory resident on each GPU, and handle the CPU/GPU memory transfers automatically in the background, or even use direct GPU memory access with RDMA or PCI-e.

\subsection{Performance of the Taylor--Green Test Case}

To evaluate the performance of our new algorithm, the previous test cases have been profiled in detail and compared to equivalent non-overset cases.  Both static and moving overset grids are compared.  For the Taylor--Green case, the grids are partitioned such that the inner grid is run on one rank, and the background grid is run on three ranks. In order to ensure that the GPUs are being utilized efficiently, each rank contains ${\sim}46\,000$ elements; we size the background grid to have $52^3$ elements and the inner grid to have $36^3$ elements.  Each rank has approximately $14\,980\,000$ DOF for $p=3$ and $29\,260\,000$ DOF for $p=4$.

\begin{figure}
\centering
\includegraphics[width=.6\textwidth]{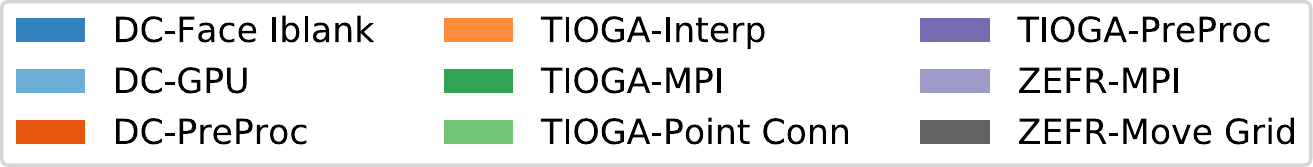} \\
\subfloat[$p=3$, Static]{\includegraphics[width=.5\textwidth]{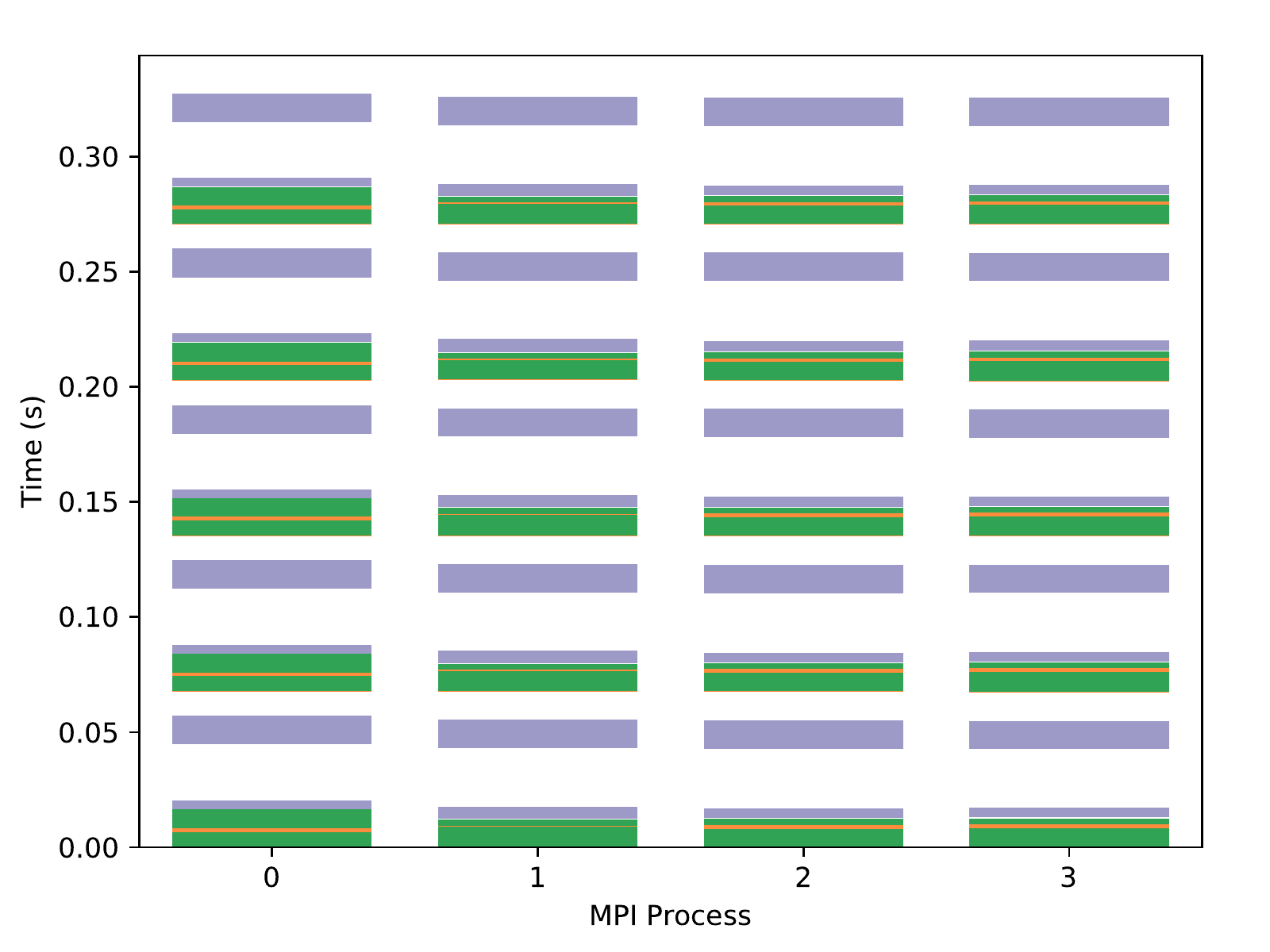}}
\subfloat[$p=4$, Static]{\includegraphics[width=.5\textwidth]{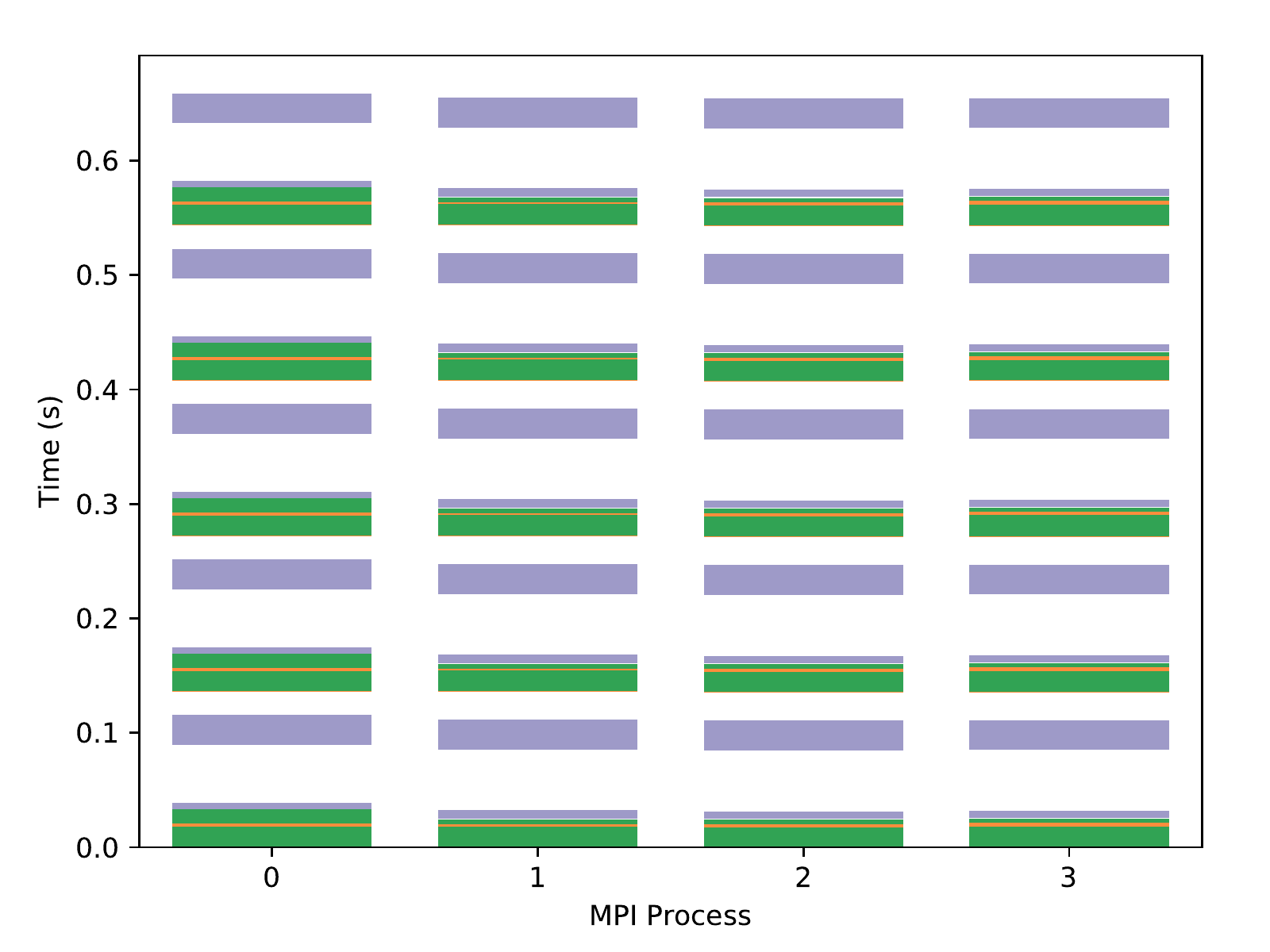}} \\
\subfloat[$p=3$, Moving]{\includegraphics[width=.5\textwidth]{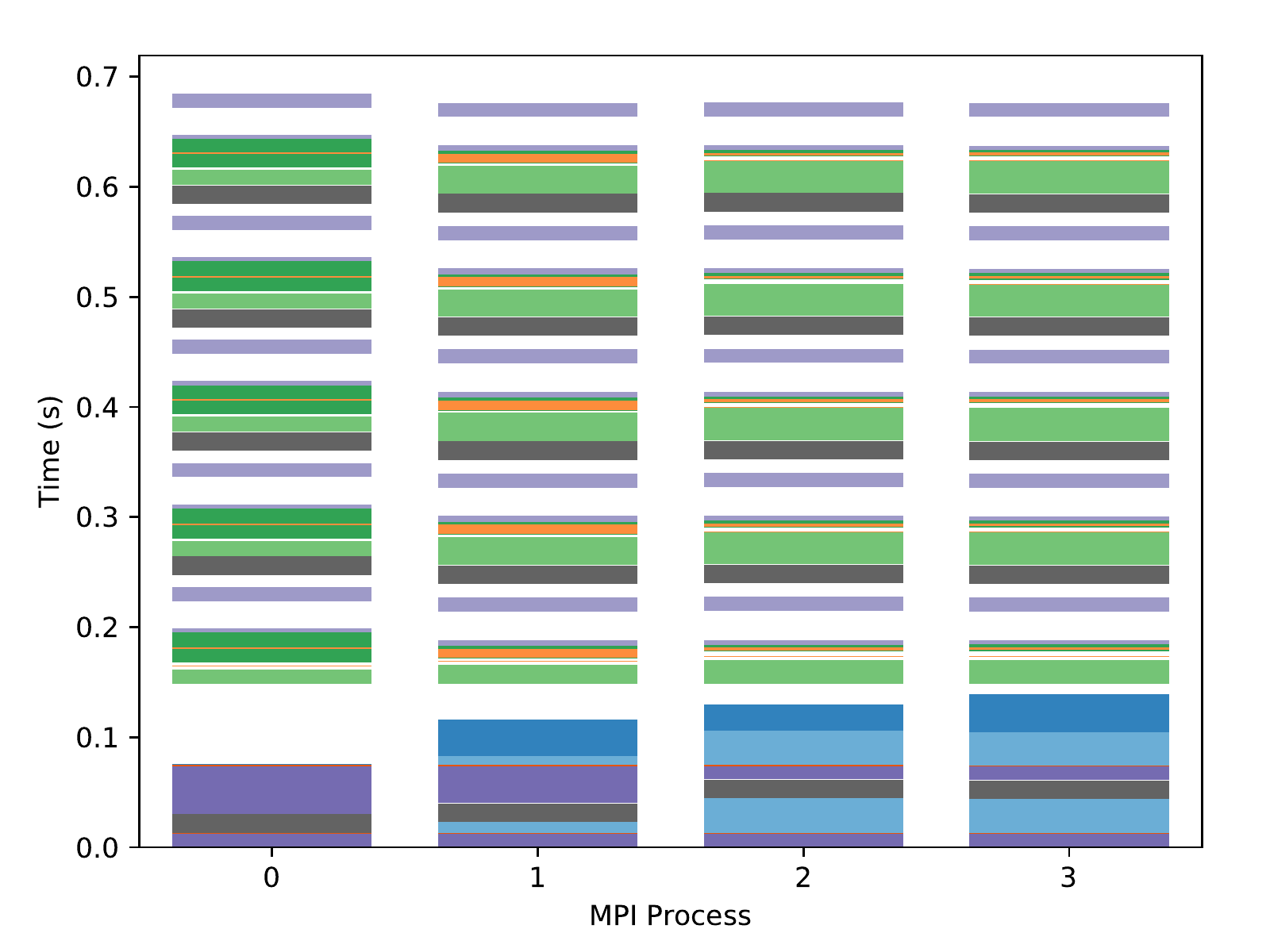}}
\subfloat[$p=4$, Moving]{\includegraphics[width=.5\textwidth]{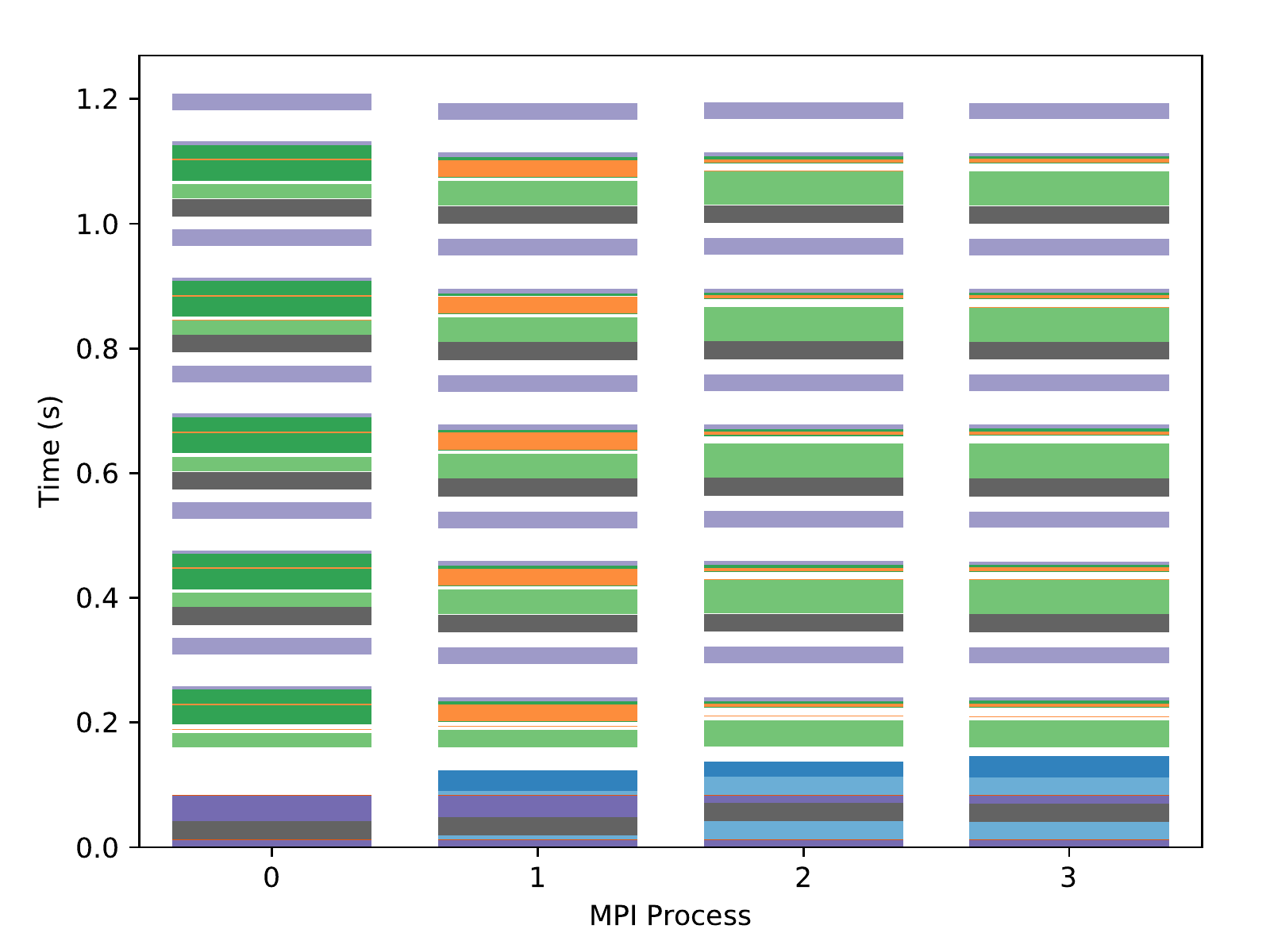}}
\caption{Timeline of overset grid work during one full five-stage explicit time step over all ranks of the Taylor Green test case.  Top: Static overset grids; Bottom: Moving overset grids.  The prefix `DC' refers to work specific to our Direct Cut method; `TG' refers to standard procedures used within TIOGA; and `ZEFR' refers to work performed within the solver.}
\label{fig:timeline-tg}
\end{figure}

Figure \ref{fig:timeline-tg} shows the timeline of overset-related work performed during one complete time step, consisting of five Runge--Kutta stages, for the static and moving overset Taylor--Green test cases for two polynomial orders.  Since only overset- and MPI-related functions are plotted, the white space in each timeline corresponds to work solely being done in the solver --- i.e., the underlying FR operations.  Note that non-blocking MPI sends/receives are used throughout ZEFR and TIOGA, so that much of the MPI communication time is overlapped with useful work in the solver.  For both $p=3$ and $p=4$, the unblanking procedure---moving the grid and performing the hole cutting twice to locate cells which must be added/removed from the grid---takes up the first ${\sim}0.15$ seconds of the time step, with the remaining time devoted to the residual calculation and Runge--Kutta stage updates.  This highlights the efficiency of high-order methods; if coarser grids with higher polynomial orders can be used, the amount of time spent on geometry-related operations relative to the numerical scheme is reduced. The time required to complete the unblank procedure is fairly evenly split between the MPI communication of search points and the construction of the hole maps (`TIOGA-PreProc'), the MPI communication required to ensure consistency in assigning a blanking status (normal/hole/artificial boundary) to all MPI interfaces (`DC-Face Iblank'), and the actual Parallel Direct Cut procedure (`DC-GPU').  Once within the main residual computation, the main sources of overhead are the moving of the grid and corresponding geometry-related updates with ZEFR (`ZEFR-Move Grid'), and the point connectivity process required to find updated donor elements for all fringe points (`TIOGA-Point Conn'), which includes communication of search points, searching the ADT, and computing interpolation weights.  On paper the TIOGA data interpolation and MPI communication (`TIOGA-MPI') appear to take a sizable portion of the wall time as well.  However, since these operations are overlapped with other useful work within the solver, the actual impact on the runtime is somewhat less than it appears.

\begin{figure}
\centering
\includegraphics[width=.55\textwidth]{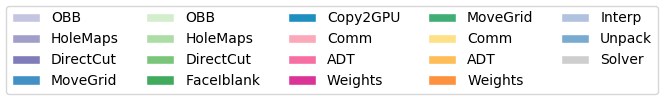} \\
\subfloat[$p=3$, Moving]{\includegraphics[height=.43\textwidth]{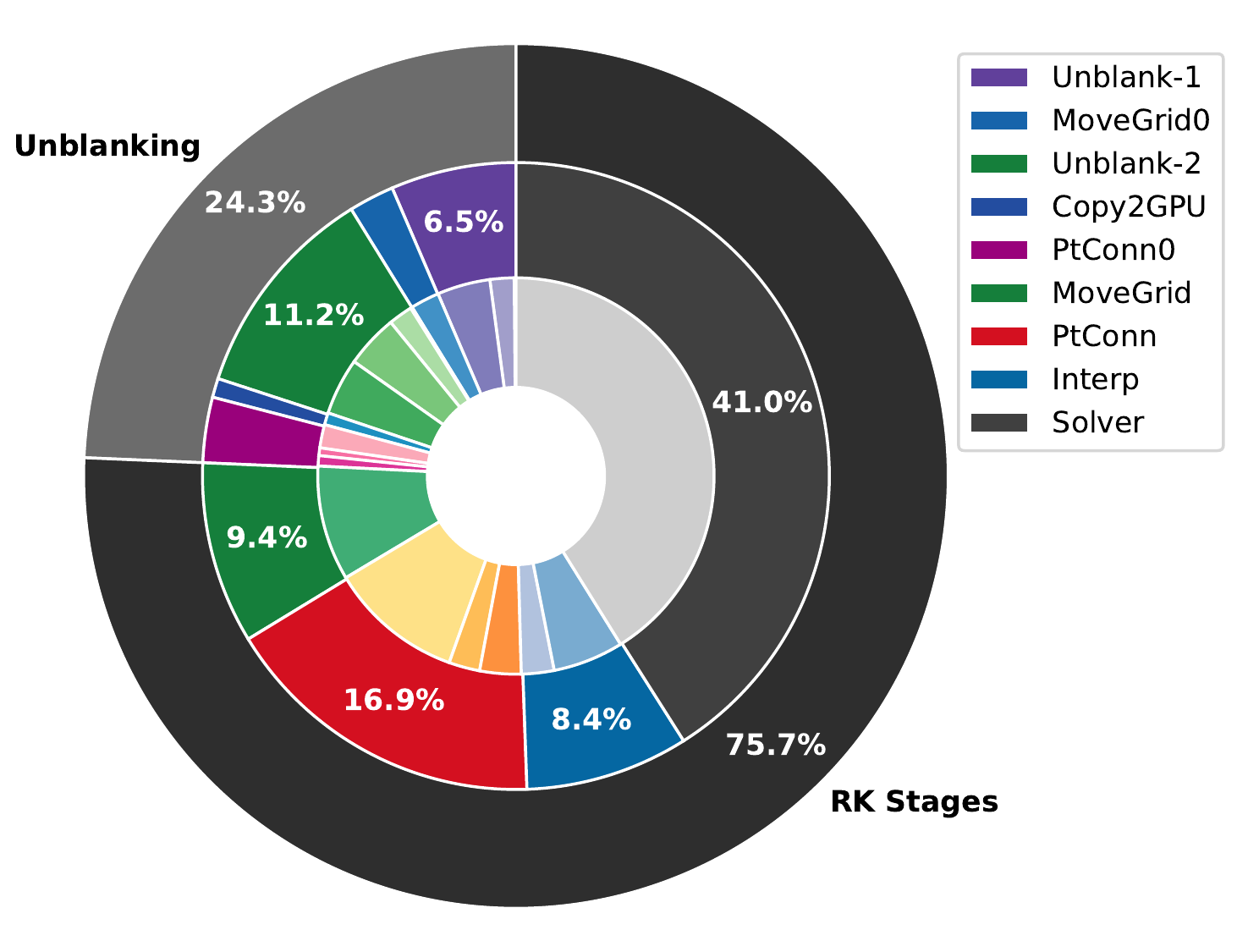}}
\subfloat[$p=4$, Moving]{\includegraphics[height=.43\textwidth]{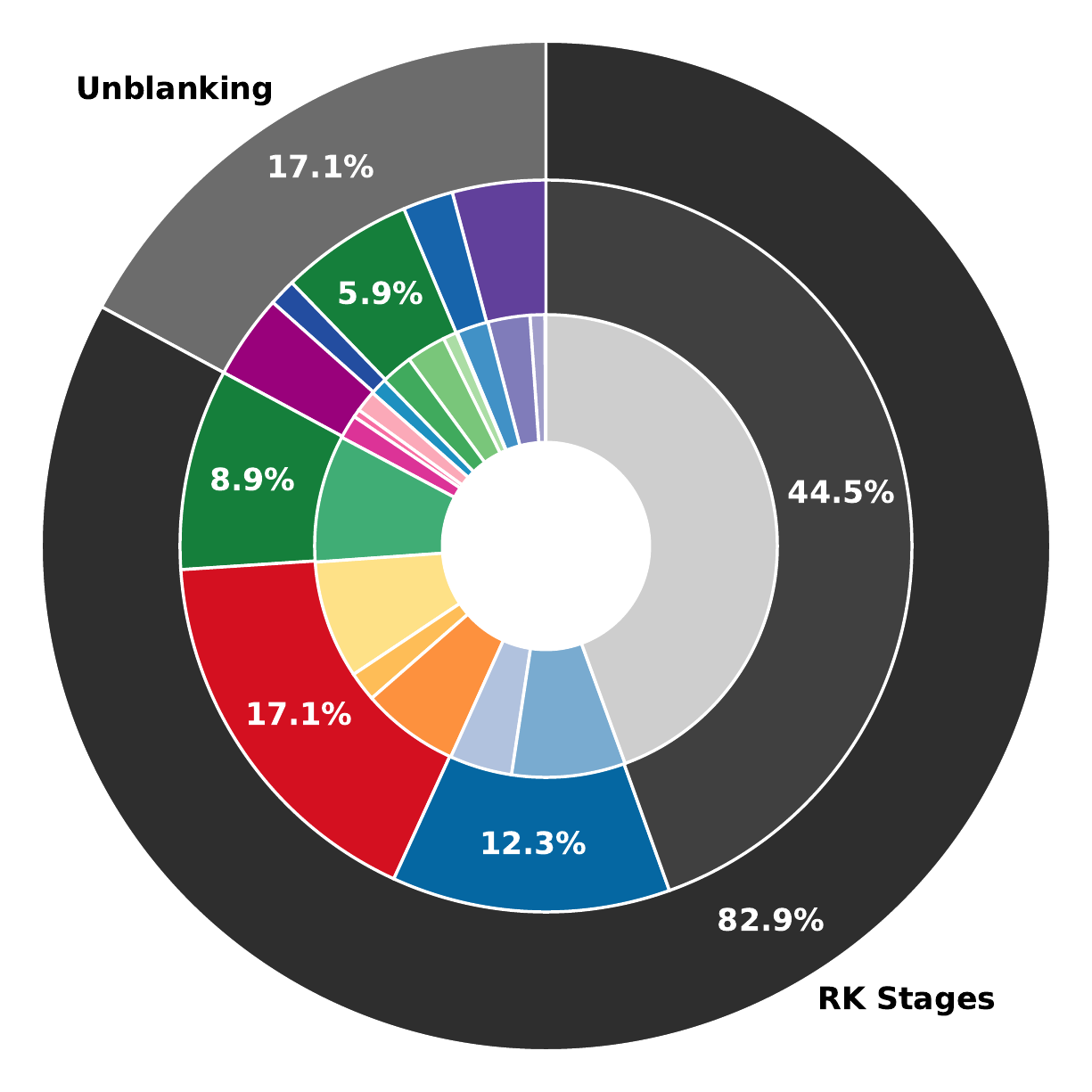}}
\caption{Breakdown of time spent on overset-related tasks for moving grids on GPUs (single node, 4-GPU case).  Middle layer: High-level overset tasks; Inner layer: Lower-level subtasks for each slice of the middle layer.}
\label{fig:pie-tg}
\end{figure}

This same data from the moving-grid cases is also plotted proportionally in Figure \ref{fig:pie-tg}.  Here, the timing data from the four MPI ranks have been averaged to get representative values for the overall simulation.  The outer ring of each donut plot shows the proportion of time spent on the unblanking procedure at the start of each time step vs. the Runge--Kutta stages.  The middle ring breaks up the overset work into its high-level components (with the legend in the middle), and the innermost ring breaks down each component further (with the legend at the top). Legend items appear in CCW order in the plots, starting at the top of each ring.  `Unblank-1' and `Unblank-2' refer to the two stages of the unblank procedure, where the hole cutting is performed for the time step's beginning and ending grid positions.  This includes updating each MPI rank's OBB and hole map (`OBB' and `HoleMaps'), in addition to performing the Parallel Direct Cut procedure (`DirectCut') and determining the final blanking status of all faces (`FaceIblank'). `PtConn' refers to the point connectivity process, where fringe points are sent to possible donor ranks (`Comm'), donors are found with an ADT search (`ADT'), and interpolation weights are calculated from the donor search results (`Weights').  The final operations displayed are `Copy2GPU', which is the host-to-device copy of the new iblank values for elements and faces, and `Interp', which involves both the actual interpolation kernel as well as the MPI buffer packing, sending/receiving, and unpacking time (`Unpack').

The advantages of using higher polynomial orders are quite clear here.  Since both cases use the same grid, the unblanking time is nearly identical between the two.  However, the unblanking procedure comprises 7\% less of the total time for the $p=4$ case due to the additional work done by the solver.  Within the RK stages, the update of the grid positions takes up a substantial amount of time.  Since this consists mostly of several matrix-matrix multiplications to update the rigid-body dynamics of the grids, there is not much that can be done to reduce this time; it is completely independent of the domain connectivity method.  The single most expensive operation (in terms of wall time) during the RK stages is the communication of fringe node positions between each rank and its potential donor ranks.  This is because the point connectivity operation within TIOGA is not overlapped with any other useful work; the inter-rank communication latency is not hidden.  It may be possible in the future to separate the point connectivity process into two stages, with useful work in between, but at present it would require invasive changes to both ZEFR and TIOGA to implement.

\subsection{Performance of the Golf Ball Test Case}

The large scale of the golf ball test case provides a very useful platform on which to assess the performance of ZEFR and the Parallel Direct Cut method at scale.   The golf ball grid uses cubically curved hexahedra, while the background grid is a structured box using linear hexahedra.  A total of $715\,750$ elements are used in the background grid, and $1\,105\,920$ in the golf ball grid.  The background and body grids are partitioned into 19 and 29 ranks respectively, giving an average of $37\,951$ elements per GPU.  The first case is run across 3 nodes of Stanford University's XStream cluster, each of which has 8 K80 boards (each with two logical GPUs), 2 Intel Xeon E5-2680 v2 CPUs, and 1 Infiniband card.

\begin{figure}
\centering
\includegraphics[width=.6\textwidth]{Legend-H.pdf} \\
\subfloat[$p=3$, Static]{\includegraphics[width=.5\textwidth]{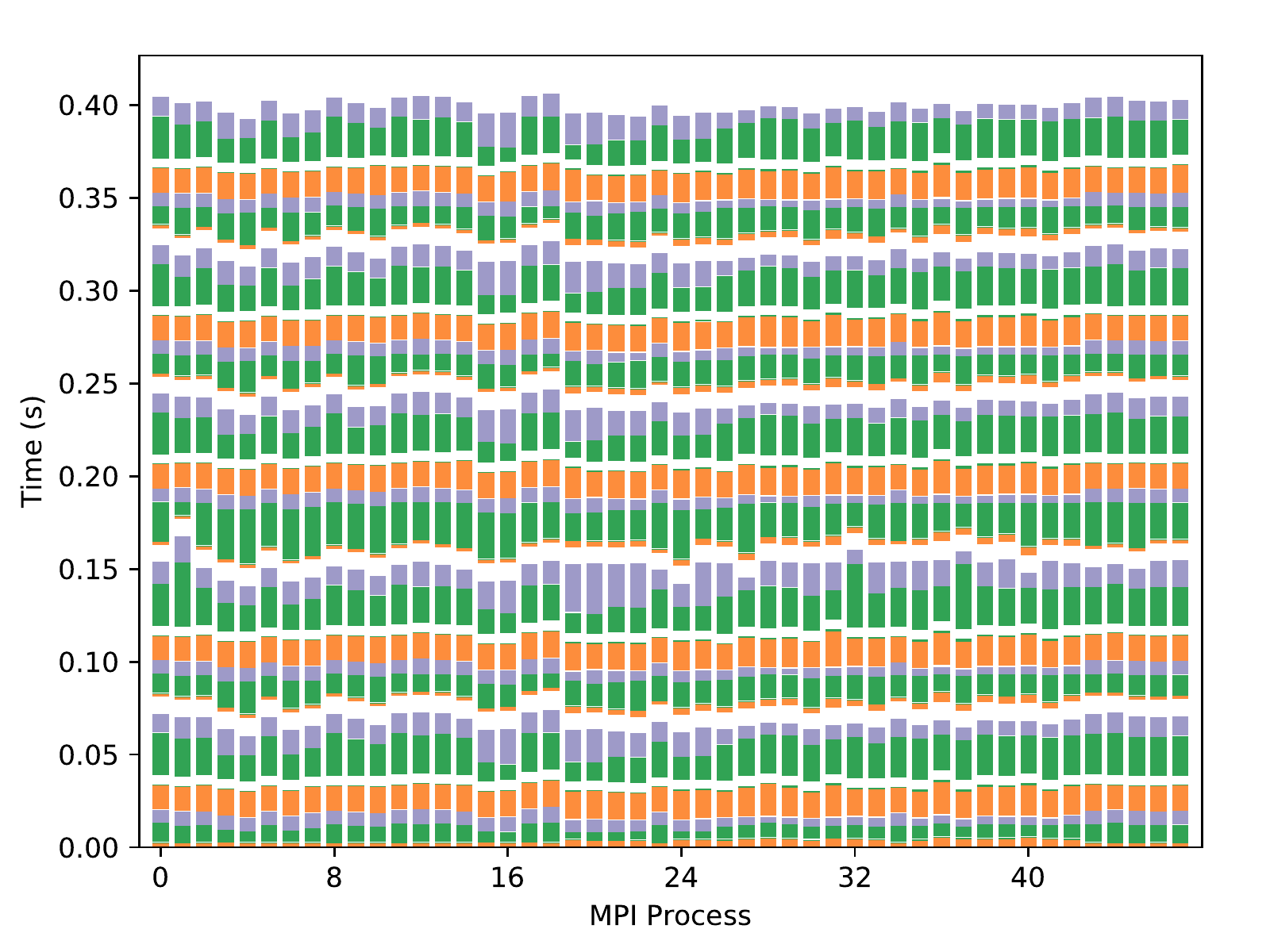}}
\subfloat[$p=4$, Static]{\includegraphics[width=.5\textwidth]{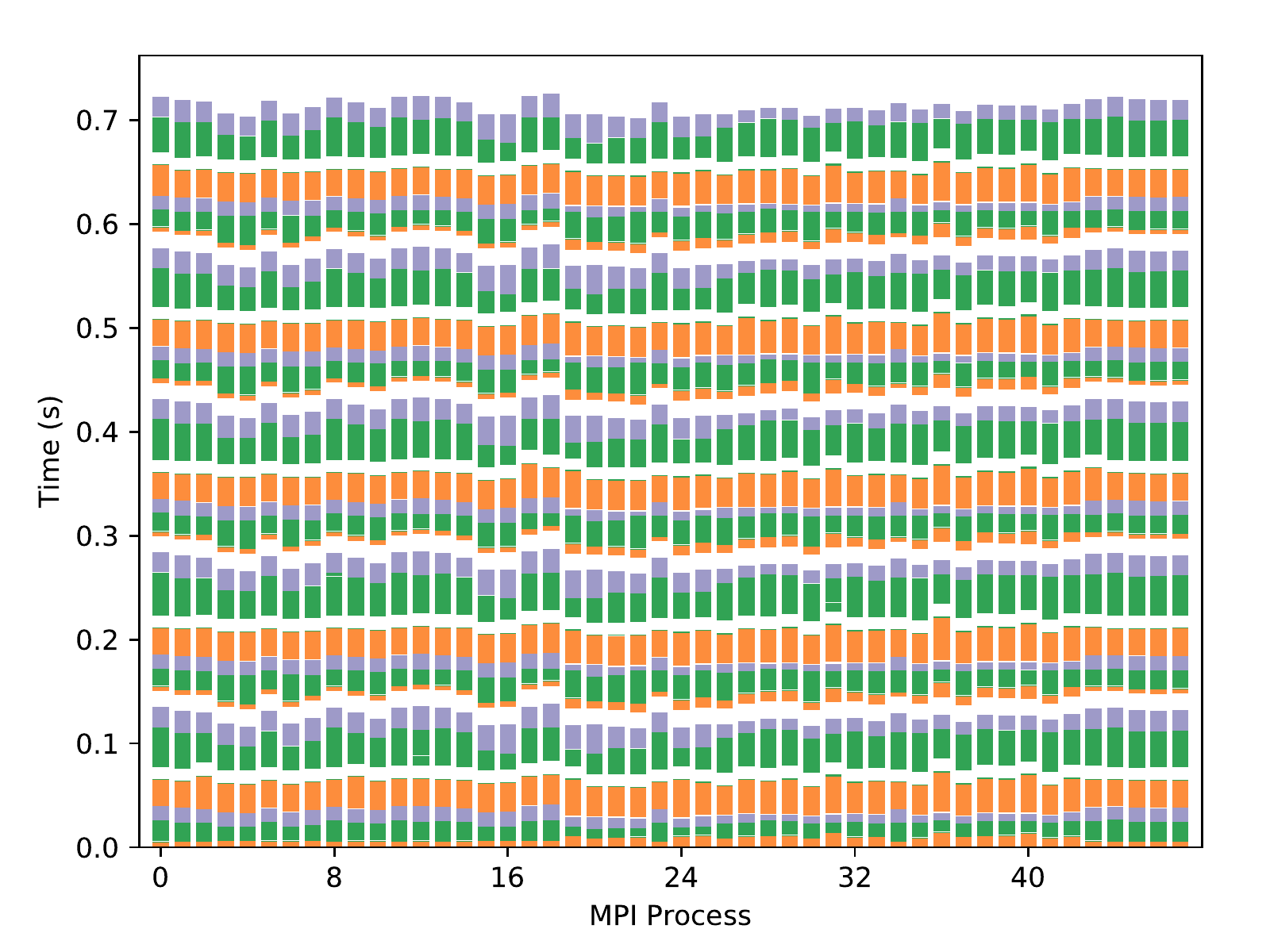}} \\
\subfloat[$p=3$, Moving]{\includegraphics[width=.5\textwidth]{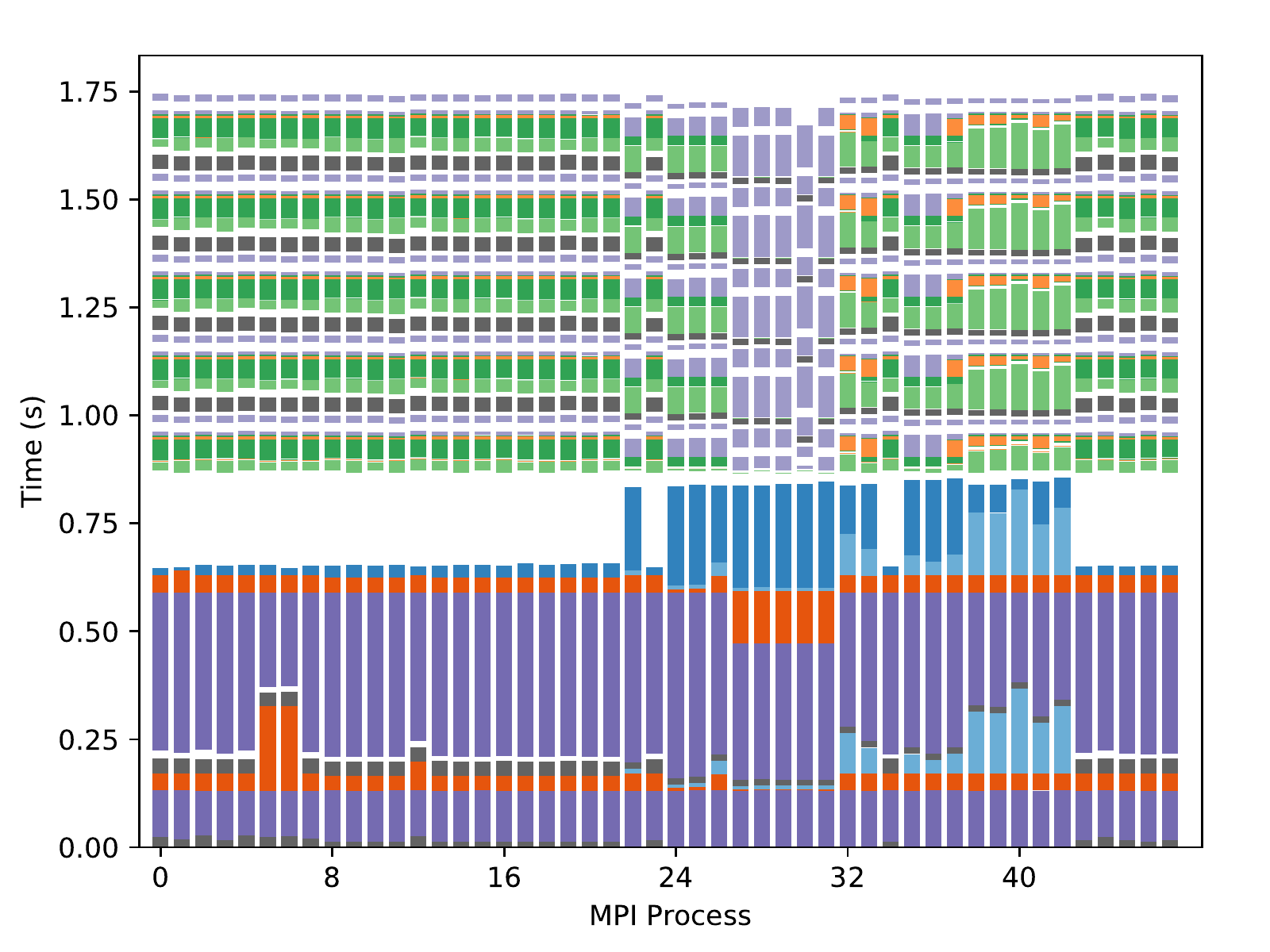}}
\subfloat[$p=4$, Moving]{\includegraphics[width=.5\textwidth]{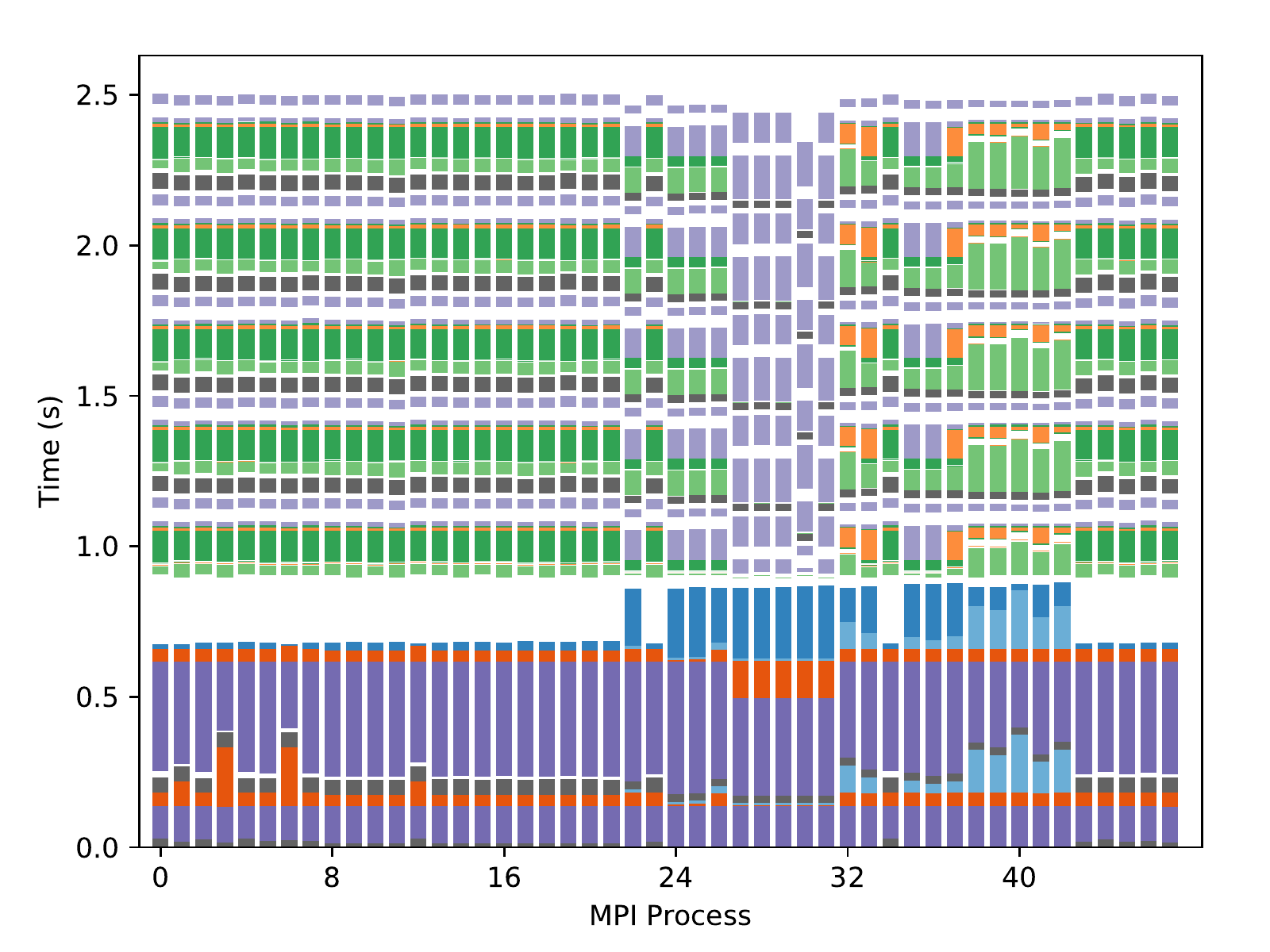}}
\caption{Timeline of overset grid work during one five-stage explicit time step over all ranks of the golf ball test case.  Top: Static overset grids; Bottom: Moving overset grids.  The prefix `DC' refers to work specific to our Direct Cut method; `TG' refers to standard procedures used within TIOGA; and `ZEFR' refers to work performed within the solver.}
\label{fig:timeline-gb}
\end{figure}

Detailed profiling results for both the static and moving cases for two polynomial orders are shown in Figure \ref{fig:timeline-gb}.  Similar to the profiling performed for the Taylor--Green test case, for the moving cases, the unblank procedure at the start of the time step is clearly distinguished from the Runge--Kutta residual calculation stages.  CPU-side preprocessing activities again take up most of the time of the unblank procedure (`TIOGA-PreProc', `DC-PreProc'); however in this case the hole cutting does take a noticeable amount of time.  It can also be seen that the ADT search and related operations within the point connectivity update procedure (`TIOGA-Point Conn') take up a substantial amount of time during each Runge--Kuta stage.

\begin{figure}
\centering
\includegraphics[width=.55\textwidth]{Pie-Legend-2.png} \\
\subfloat[$p=3$, Moving]{\includegraphics[height=.43\textwidth]{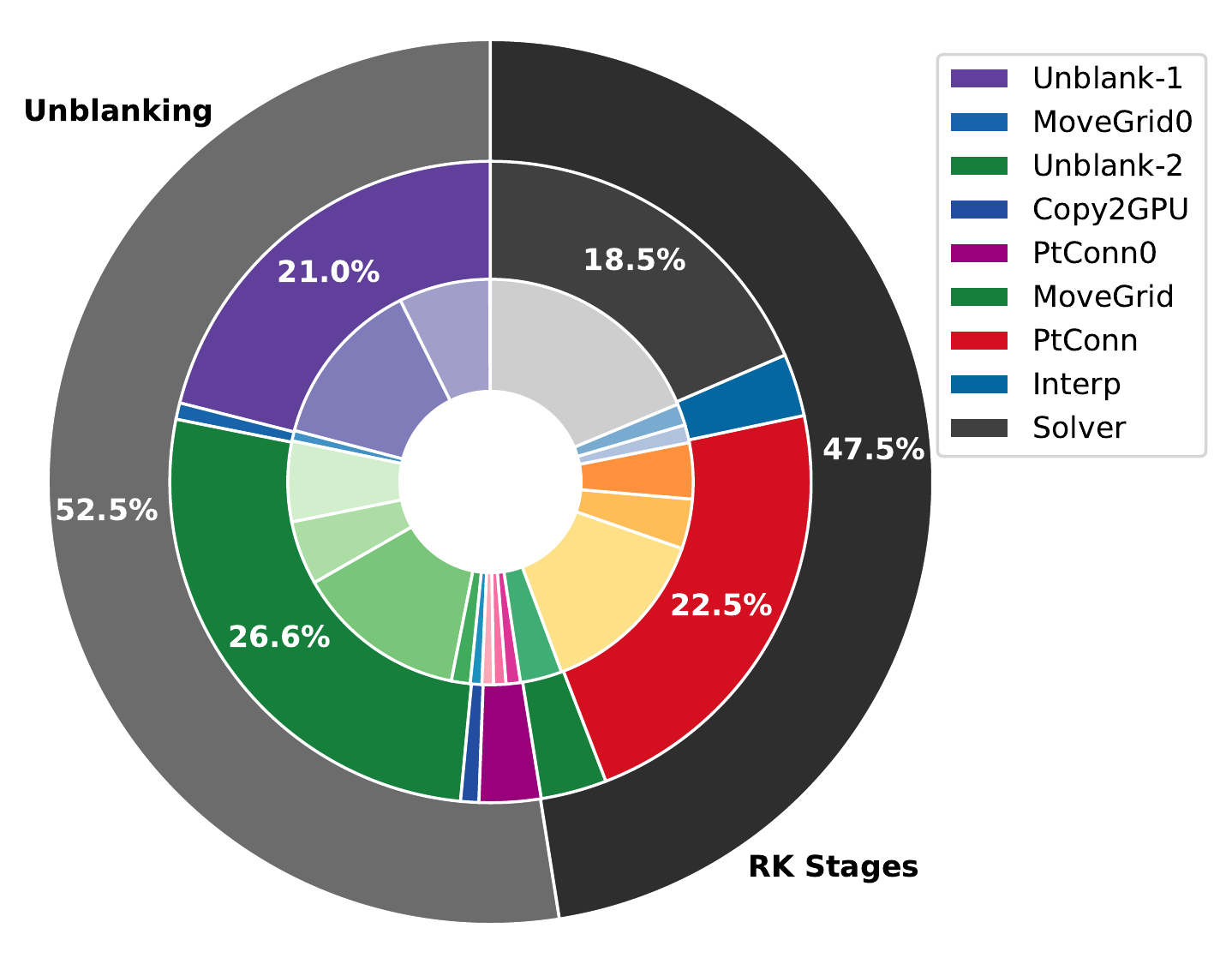}}
\subfloat[$p=4$, Moving]{\includegraphics[height=.43\textwidth]{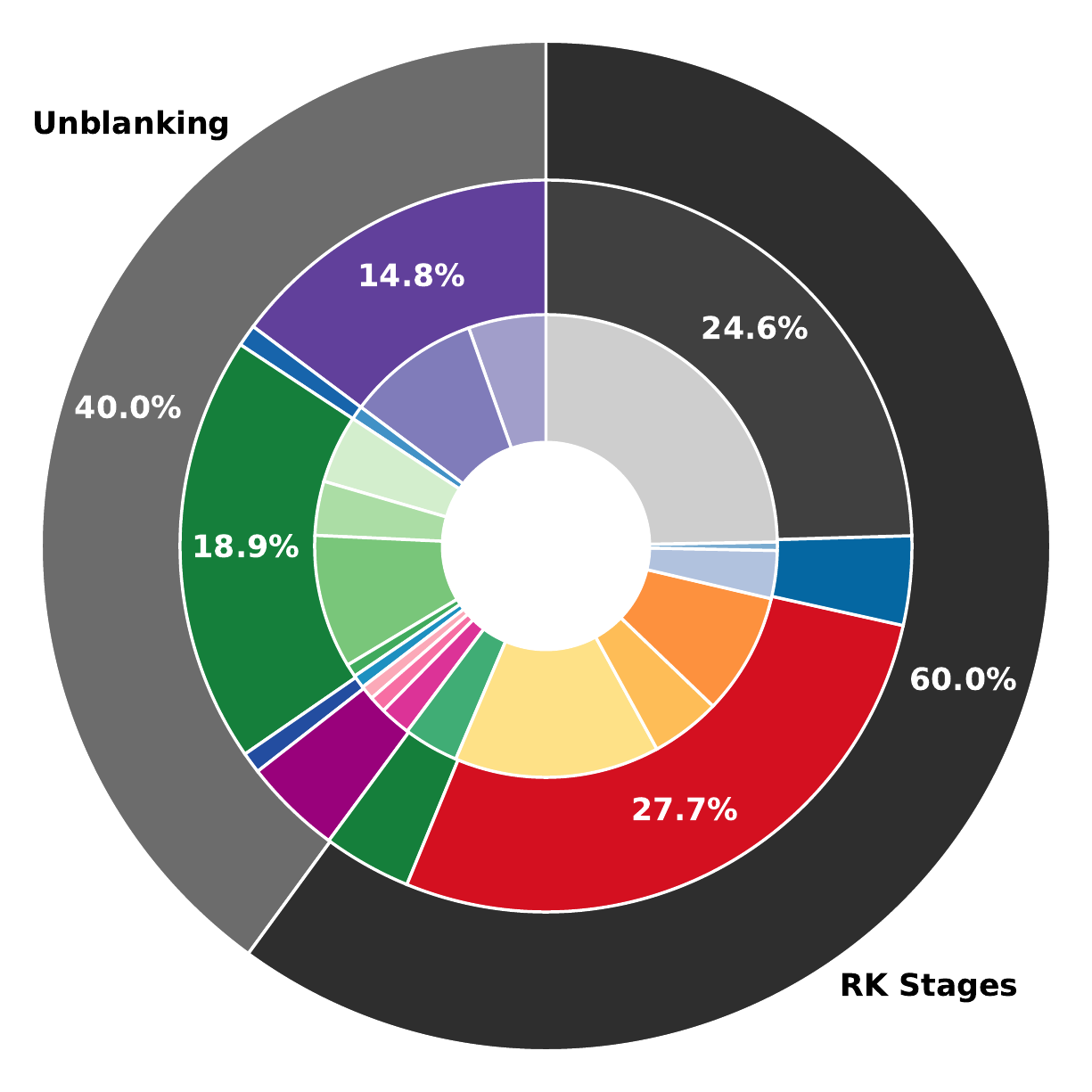}}
\caption{Breakdown of time spent on overset-related tasks for moving grids on GPUs (3-node, 48-GPU golf ball case).  Middle layer: High-level overset tasks; Inner layer: Lower-level subtasks for each slice of the middle layer.}
\label{fig:pie-gb}
\end{figure}

This is shown more clearly in Figure \ref{fig:pie-gb}.  In contrast to the relatively small and simple Taylor--Green test case, the golf ball domain connectivity takes up a considerably larger proportion of the total time, with the underlying numerical solver comprising just under a quarter of the total time for $p = 4$, and less for $p = 3$.  As was the case for the Taylor--Green simulation, the point connectivity procedure --- particularly the MPI communication of the fringe points --- is the single largest contributor to the total wall time.

\begin{figure}
\centering
\includegraphics[width=.6\textwidth]{Legend-H.pdf} \\ \vspace{-6pt}
\subfloat[$p=4$, Moving]{\includegraphics[width=.5\textwidth]{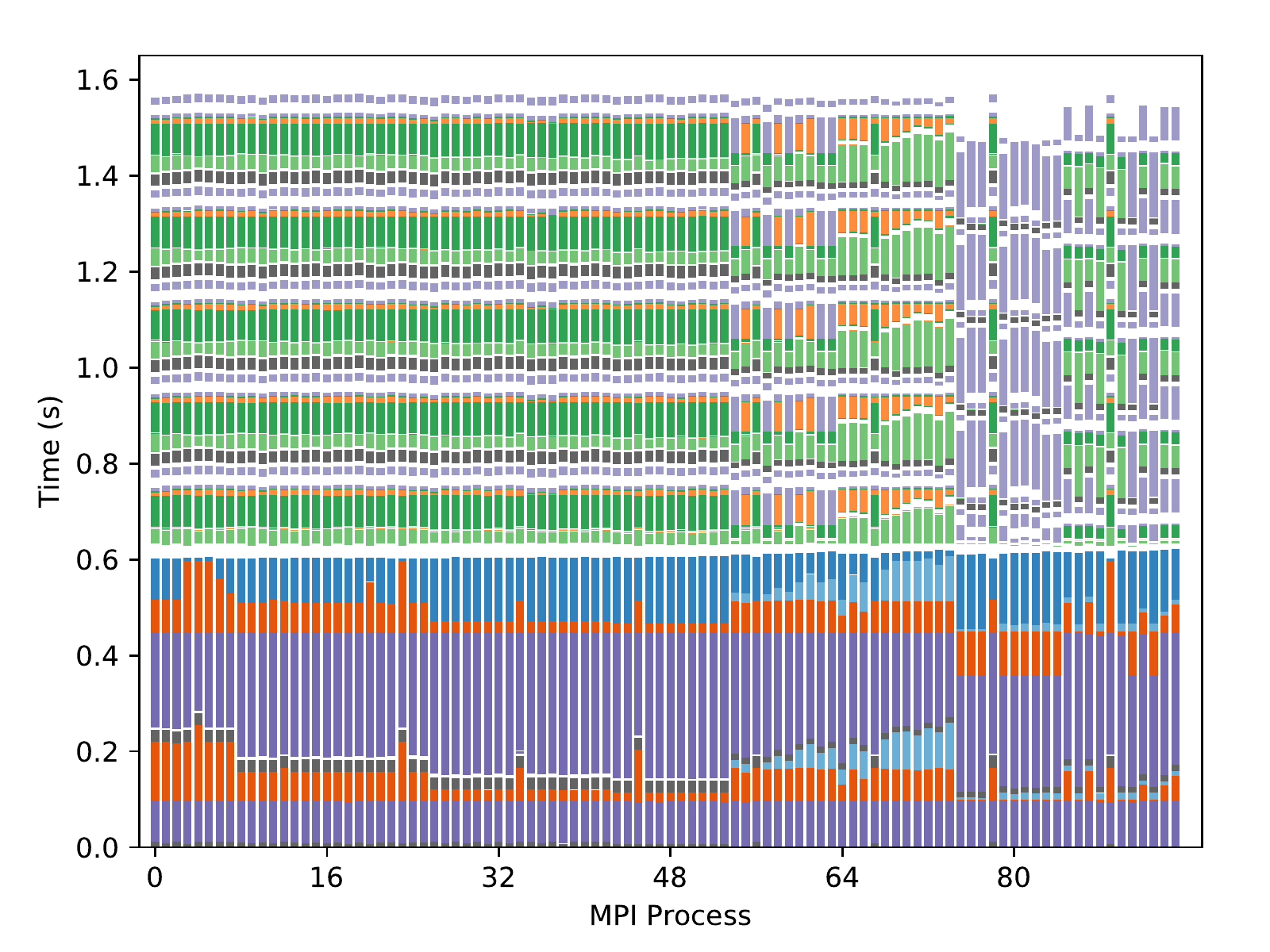}}
\subfloat[$p=5$, Moving]{\includegraphics[width=.5\textwidth]{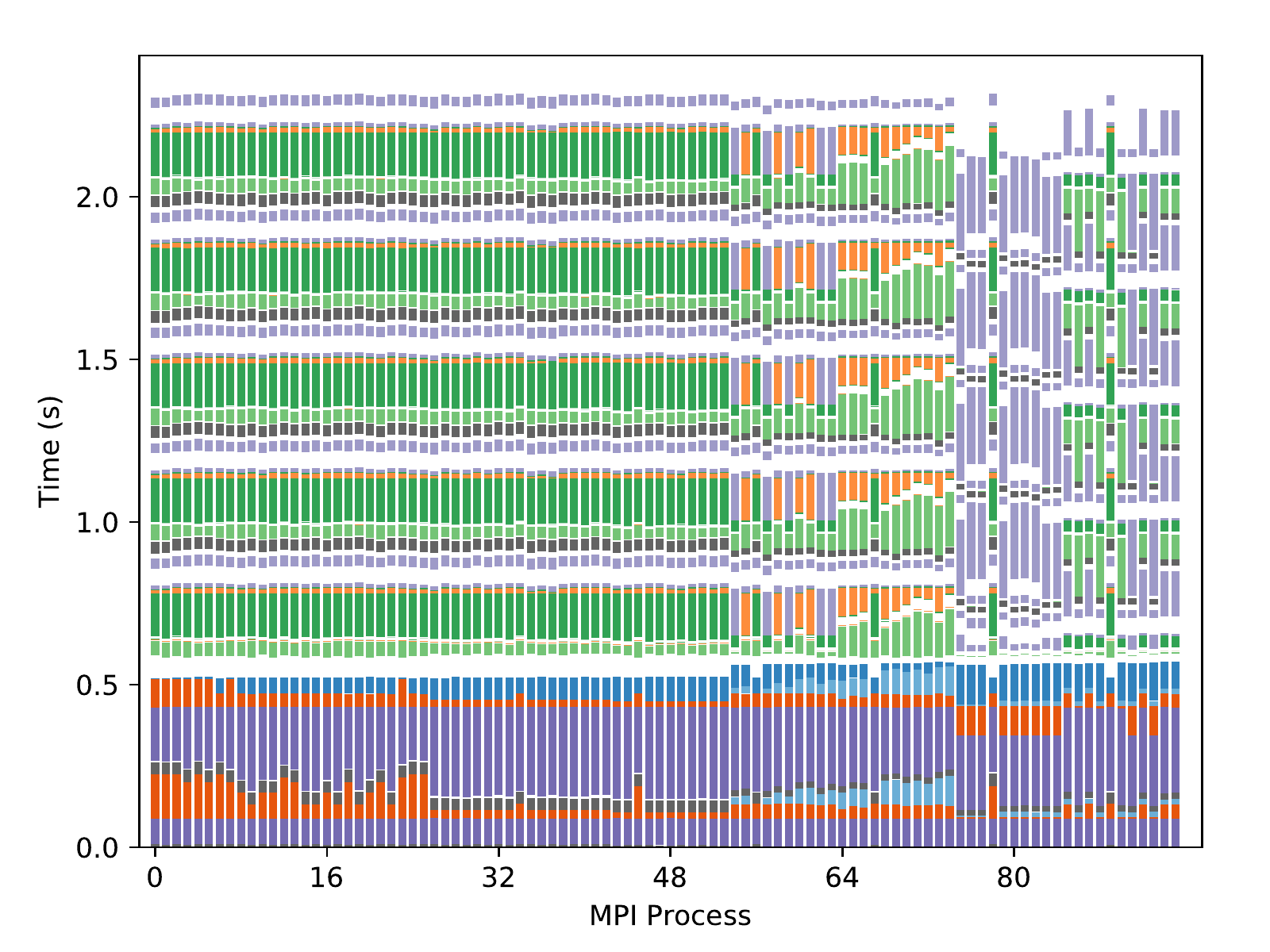}}
\caption{Timeline of overset operations for the golf ball case across 96 GPUs.}
\label{fig:timeline-gb-2}
\end{figure}

We can further test our method by scaling the case across double the number of computing nodes.  Figure \ref{fig:timeline-gb-2} shows the timelines of operations for the same moving-grid golf ball test case, but now across 96 GPUs (6 nodes on XStream) for $p=4$ and $p=5$.  Running the golf ball at $p=5$ was not possible on 48 GPUs due to requiring more than 576 GB of GPU memory, but here we can see that it runs quite well on 96 GPUs, with an efficiency comparable to that of the $p=4$ case on 48 GPUs.  The $p=4$ case also scales reasonably well; the next section will discuss quantitative performance metrics.

\subsection{Summary of Performance Results}

As a solver-agnostic, quantitative metric to compare performance between codes, Table \ref{tab:nsperdof} shows the average time per DOF required to compute the residual ($\mathbf{\nabla \cdot F}$) and all necessary connectivity and interpolation overhead once.  Since we have been using a 5-stage $4$th order Runge--Kutta time integration method, this is defined as $1/5$ of the time per DOF required to advance one complete time step.  We also compare against the same metric provided by Witherden et al. \cite{witherden15} for the PyFR code under equivalent conditions and on similar hardware (NVIDIA Tesla K40c).  Although PyFR and ZEFR are performing essentially identical numerical methods, the ZEFR single-grid performance metrics are slightly better due to the introduction of GiMMiK \cite{gimmik}, a library which greatly improves the performance of sparse matrix multiplications such as the ones which occur due to polynomial operations inside of tensor-product elements. These numbers have also been reported for the baseline (non-overset) ZEFR solver within the PhD thesis of Romero \cite{RomeroThesis}. 

Considering first the Taylor--Green case, we can see that for moving overset grids with linear hexahedra, the overhead involved is a factor of ${\sim}2.2-2.6$ over a single static grid, while the overhead for performing overset interpolation without grid motion and connectivity updates at every time step is only $35\%$.  The `worst-case' numbers are provided by the moving-overset calculations for the golf ball test case.  The case was first run over 3 nodes of the large computing cluster XStream (totaling 48 GPUs), and the body grid was represented with cubically curved hexahedra, increasing the amount of work required in the Parallel Direct Cut algorithm as well as the point connectivity update.  In this case, the cost for moving overset grid calculations relative to single-grid calculations becomes a factor of ${\sim} 3.8$ for $p=4$, or ${\sim} 5.2$ for $p=3$.  The increased apparent cost for $p=3$ vs. $p=4$ is because both cases are using the same grid, and hence the overset and geometry-related processing required for each time step is about the same between the two cases, but the $p=4$ case has about double the number of degrees of freedom, so the total cost is more evenly distributed.  The overhead for static overset grids, meanwhile, is at a relatively low 14\% for $p=4$, and 21\% for $p=3$.  

When the same case was run with double the number of MPI ranks (6 nodes, 96 GPUs), strong scaling efficiency of 79\% was achieved for $p=4$.  If the amount of work is increased by switching to 5th-order polynomials however (a 1.73x increase in the number of DOF vs $p=4$), then a weak scaling efficiency of 92\% is achieved, and is even more efficient than running $p=3$ on 48 ranks.  This again highlights an advantage of high order methods: high polynomial orders combined with relatively coarse grids (compared to a 2nd-order FV grid) can offer greatly improved efficiency in terms of mesh generation and mesh-related operations during the simulation.

\begin{table}
\centering
\caption{\label{tab:nsperdof}Time per Runge--Kutta stage per DOF in nanoseconds for various cases.  ZEFR used NVIDIA K80 GPUs and Intel Xeon E5-2680 v2 CPUs, while PyFR used NVIDIA K40 GPUs and Intel XeonE5-2697 v2 CPUs.}\vskip12pt
\begin{tabular}{l l c c c c c}
\toprule
 &  &  \multicolumn{5}{c}{Time per DOF / $10^{-9}s$} \\ \cmidrule{3-7}
 &  &  \multicolumn{3}{c}{NVIDIA GPUs} & \multicolumn{2}{c}{Intel CPUs} \\
 \cmidrule(lr){3-5} \cmidrule(lr){6-7}
Code  &  Case            & $p = 3$ & $p = 4$ & $p = 5$ & $p = 3$ & $p = 4$ \\ \cmidrule{1-2} \cmidrule(lr){3-5} \cmidrule(lr){6-7}
ZEFR       &  TGV (Base)      & \hphantom{0}4.28 & \hphantom{0}4.37 & -- & 340 & 370\\
(4 Ranks)  &  Static Overset  & \hphantom{0}5.78 & \hphantom{0}5.86 & -- & 360 & 380 \\
           &  Moving Overset  & 11.1\hphantom{0} & \hphantom{0}9.69 & -- & 450 & 500 \\ \cmidrule{2-2} \cmidrule(lr){3-5} \cmidrule(lr){6-7}
(29 Ranks) & Golf Ball (Base) & \hphantom{0}5.54 & \hphantom{0}5.67 & -- & -- & -- \\
(48 Ranks) &  Static Overset  & \hphantom{0}6.74 & \hphantom{0}6.46 & -- & -- & -- \\
           &  Moving Overset  & 28.7\hphantom{0} & 21.3\hphantom{0} & -- & -- & -- \\
(96 Ranks) &  Static Overset  & \hphantom{0}6.91 & \hphantom{0}7.09 & \hphantom{0}8.12 & -- & -- \\
           &  Moving Overset  & 36.2\hphantom{0} & 26.9\hphantom{0} & 23.2\hphantom{0} & -- & -- \\ \cmidrule{2-2} \cmidrule(lr){3-5} \cmidrule(lr){6-7}
PyFR       &  Cylinder       & \hphantom{0}4.88 & \hphantom{0}6.17 & -- & 332 & 383 \\ \bottomrule
\end{tabular}
\end{table}

\subsection{Comparison to CPU Performance}

For additional comparison, the right hand columns of Table \ref{tab:nsperdof} show the same absolute performance metric (wall time per DOF per RK stage) of the solver on Intel CPUs.  The CPU version of the code uses TIOGA's default implicit hole cutting method, modified slightly to handle artificial boundary interpolation for high-order solvers.  GiMMiK is again used to speedup the polynomial operations inherent to the FR method.  All times are normalized by the number of ranks used.  Therefore, to more directly compare metrics for an Intel Xeon E5-2680 v2 processor to a single logical K80 GPU, the numbers should be divided by the number of cores available on the chip, which in this case is 10.  Comparing the metrics CPU vs. GPU metrics in Table \ref{tab:nsperdof} suggests that approximately 8-10 Intel chips (totaling 80-100 available cores) must be used in order to match the performance of one physical K80 GPU (containing two logical GPUs), depending on the type of simulation being run.  Clearly, given that high-end GPUs cost less than 10x the price of a similarly high-end Intel processor and use far less than 10x the power, GPUs are a more efficient tool for complex fluid dynamic simulations of the type we have performed.

\section{Conclusions}
\label{S:conclusion}

In this paper we presented the Parallel Direct Cut Approach for overset methods.  This approach enables overset grids to be accurately and efficiently employed within the context of a GPU accelerated high-order discontinuous spectral element method.  The method is both robust and highly optimized.  Through a series of numerical experiments we have shown the ability of the parallel direct cut method to accurately simulate the challenging Taylor--Green vortex test case featuring a rotating inner grid.  In addition, we have also presented the first high order ILES simulations of a spinning golf ball at $Re = 150\,000$.  The results from these simulations demonstrate good agreement with both experimental and numerical data, and showcase the ability of the parallel direct cut approach to be employed  in the simulation of a complex three dimensional test case.  Near-term applications with the approach could include multicopters and other drones which operate at modest Reynolds numbers in reach of DNS or ILES simulations.

\section*{Acknowledgements}
\label{S:ack}

The authors would like to acknowledge the Army Aviation Development Directorate (AMRDEC) for providing funding for this research under the oversight of Roger Strawn, the Air Force Office of Scientific Research for their support under grant FA9550-14-1-0186 under the oversight of Jean-Luc Cambier, and Margot Gerritsen for access to the XStream GPU computing cluster, which is supported by the National Science Foundation Major Research Instrumentation program (ACI-1429830).  We would also like to thank Dr. Peter Eiseman for providing academic licensing to the GridPro meshing software and assisting with the creation of several golf ball grids.  Finally, we would like to thank Dr. Jay Sitaraman for his help in reviewing the initial draft of this paper and providing valuable feedback.

\bibliographystyle{elsarticle-num}
\bibliography{references.bib}


\clearpage
\begin{appendices}
\section{Parallel Direct Cut Algorithm}
\label{A:code}

\begin{algorithm}
\caption{\label{alg:cp0} Approximate Distance Calculation}
\begin{algorithmic}[1]
\State $F =$ current face ID
\State $E =$ current element ID
\State $\vec{B_e} =$ Extents of element OBB
\State $\bar{\bar{R}} =$ Axes of element OBB
\State $\bar{\bar{X_f}} = $ corner nodes of face $F$
\State $\bar{\bar{X_f}} = \bar{\bar{R}} \cdot \bar{\bar{X_f}}$ \Comment{Rotate face points to OBB axes}
\State $\vec{B_f} = $ AABB of $\bar{\bar{X_f}}$
\State $\vec{f_{cg}} =$ centroid of face $F$
\State $\vec{e_{cg}} =$ centroid of element $E$
\State $d_{xc} = ||\vec{f_{cg}} - \vec{e_{cg}}||_2$
\State $d_{b} = \Call{boundingBoxDist}{\vec{B_f},\vec{B_e}}$
\State $Distance(E,F) = .01 \cdot d_{xc} + d_{b}$
\end{algorithmic}
\end{algorithm}

\begin{algorithm}
\caption{\label{alg:bboxdix} Bounding-Box Distance Calculation}
\begin{algorithmic}[1]
\Function{boundingBoxDist}{$\vec{B_1}$,$\vec{B_2}$}
\State $D = 0$
\For{\forloop{i}{0}{3}}
  \If{$\vec{B}_{1,i+3} < \vec{B}_{2,i}$ or $\vec{B}_{2,i+3} < \vec{B}_{1,i}$}
    \State $d = \max ( \vec{B}_{2,i}-\vec{B}_{1,i+3}, \vec{B}_{1,i}-\vec{B}_{2,i+3} )$
    \State $D \pluseq d^2$
  \EndIf
\EndFor
\State return $\sqrt{D}$
\EndFunction
\end{algorithmic}
\end{algorithm}

\begin{algorithm}
\caption{\label{alg:cp1} $1^{st}$-Pass Distance Calculation}
\begin{algorithmic}[1]
\Function{cuttingPass1}{EleNodes,FaceNodes,outCorner,outDist}
\State $F =$ current face ID
\State $E =$ current element ID
\State $tol = 10^{-7}$
\State $\bar{\bar{X_e}} =$ corner nodes of element $E$
\State $\bar{\bar{X_f}} =$ corner nodes of face $F$

\State $\vec{f_{cg}} =$ centroid of face $F$

\State $D_{min} = $ \texttt{BIGVALUE}
\For{\forloop{i}{0}{7}}
  \State $D  = ||\overline{\overline{X_e[i,:]}} - \vec{f_{cg}}||_2$
  \If{$D < D_{min}$}
    \State $D_{min} = min(D,D_{min})$
    \State $c = i$
  \EndIf
\EndFor

\State $\vec{F}_{list} =$ IDs of faces touching corner $c$ (0 through 5)

\State $D_{min} =$ \texttt{BIGVALUE}
\For{\forloop{f}{0}{2}}:
  \State $fid = \vec{F}_{list}[f]$
  \For{\forloop{i}{0}{1}}:
    \State $\bar{\bar{TC}} =$ Nodes of sub-triangle $i$ of ele's face $fid$
    
    \For{\forloop{j}{0}{1}}:
      \State $\bar{\bar{TF}} =$ Nodes of sub-triangle $j$ of given 4-node face
      \State $D = \Call{triTriDistance}{TF, TC, tol}$
      \If{$D < D_{min}$}
        \State $D_{min} = D$
      \EndIf
    \EndFor
  \EndFor
\EndFor

\If{$D_{min} < tol$}
  \State $D_{min} = 0$
\EndIf

\State $outDist(E,F) = D_{min}$
\State $outCorner(E,F) = c$
\EndFunction
\end{algorithmic}
\end{algorithm}

\begin{algorithm}
\caption{\label{alg:cp2} $2^{nd}$-Pass Distance Calculation}
\begin{algorithmic}[1]
\Function{cuttingPass2}{EleNodes,FaceNodes,Corners,Dist,Vec}
\State $F =$ current face ID
\State $E =$ current element ID
\State $T =$ current sub-triangle of element to compare
\State $c =$ current corner of element $E$ to use
\State $T_C =$ nodes of triangle $T$ within element $E$ around corner $c$
\State $S_F =$ shape order of cutting boundary faces

\State $D_{min} =$ \texttt{BIGVALUE}
\State $\vec{dx}_{min} =$ (\texttt{BIGVALUE, BIGVALUE, BIGVALUE})

\For{$\forloop{M}{0}{S_F-1}$}
  \For{$\forloop{N}{0}{S_F-1}$}
    \For{$j = 0,1$}
      \State $T_F =$ points of sub-triangle $j$ of sub-quad $M,N$ of face $F$
      \State $D = \Call{triTriDistance}{T_F,T_C,\vec{dx},tol}$
      
      \If{$D < tol$}
        \State return 0
      \EndIf
      
      \If{$D < D_{min}$}
        \State $\vec{dx}_{min} = \vec{dx}$
        \State $D_{min} = D$
      \EndIf
    \EndFor
  \EndFor
\EndFor

\If{$D_{min} < tol$}
  \State $D_{min} = 0$
\EndIf

\State $Dist(E,F,T) = D_{min}$
\State $Vec(E,F,T) = \vec{dx}_{min}$
\EndFunction
\end{algorithmic}
\end{algorithm}

\end{appendices}


\end{document}